\documentclass{article}
\usepackage{graphicx}
\usepackage[letterpaper,top=2cm,bottom=2cm,left=3cm,right=3cm,marginparwidth=1.75cm]{geometry}

\usepackage{upgreek, color,amsfonts}
\usepackage{amsmath, hyperref,amssymb, bbm,color, comment,authblk}
\usepackage{tikz}
\usetikzlibrary{arrows.meta,calc}

\usepackage[numbers,sort&compress]{natbib}
\makeatletter
\def\BibitemShut#1{\unskip\csname bibitem#1\endcsname}
\makeatother

\numberwithin{equation}{section}

\newcommand{\dd}{\mathrm d}
\newcommand{\Tr}{\operatorname{Tr}}

\newcommand{\avg}[1]{\left\langle #1\right\rangle_{S^3}}

\title{Bilocal flows and universality with field-dependent regulators}
\author[1,2]{Alfio Bonanno}
\author[3]{Renata Ferrero}
\affil[1]{\small INAF – Osservatorio Astrofisico di Catania, Via S. Sofia 78, 95123 Catania, Italy}
\affil[2]{INFN, Sezione di Catania, Via S. Sofia 64, 95123 Catania, Italy}
\affil[3]{\small Institute for Theoretical Studies, ETH Zürich, 8006 Zürich, Switzerland}
\date{}

\begin{document}
\maketitle

\begin{abstract}
Many functional renormalization group flows employ a regulator that depends on the fields, as in proper-time and spectrally adjusted flows and in background-independent formulations of gauge theories and quantum gravity. Such a regulator generates additional terms in the flow, and their omission may spoil universality. We introduce bilocal flows, in which the effective action depends on the field and on the cutoff kernel as independent variables. The construction identifies the additional terms, and an exact sum rule shows that the first nontrivial coefficient of the anomalous dimension coincides with that of a fixed-regulator flow for every admissible regulator profile, so that universality is restored. For massless $\lambda\phi^4$ theory with a proper-time regulator, the reduced flow gives a profile-dependent anomalous dimension, while the completed flow gives the universal two-loop coefficient $C_\eta=1/6$ for every profile.
\end{abstract}

\tableofcontents

\section{Introduction}
\label{sec:introduction}

Functional flow equations with a one-loop trace structure provide a systematic framework  for studying quantum field theories across momentum scales, in the form of Wilson's exact renormalization group (RG) and of its later developments, notably the effective average action \cite{Wilson:1973jj,Wegner:1972ih,Polchinski:1983gv,Morris:1998da,Bagnuls:2000ae,Berges:2000ew,Polonyi:2001se,Pawlowski:2005xe,Gies:2006wv,Rosten:2010vm,Delamotte:2007pf,Aoki:2000wm,Dupuis:2020fhh}. Whether a given
flow is useful depends on the truncation, and also on whether the equation retains the universal perturbative information of the theory beyond leading order \cite{Falls:2017cze,Falls:2024noj,Falls:2026nuh}. For the Wetterich equation \cite{Wetterich:1992yh,Ellwanger:1993mw,Morris:1993qb} with a fixed regulator it does: scale and field derivatives are taken at fixed cutoff kernel, and the hierarchy of flow equations for the higher vertices follows from the effective action without further input. Perturbative consistency and the recovery of universal information from exact or functional RG equations have been analyzed in several formulations \cite{Papenbrock:1994kf,Bonini:1992vh,Tetradis:1993ts,Litim:2001ky,Litim:2002xm,Kopietz:2000bh,Codello:2013bra,Morris:2005tv}.

In many applications the regulator depends on the fields. In the background-field formulation of gauge theories and gravity \cite{Reuter:1993kw,Reuter:1996cp} the regulator is built from the background field $\bar\phi$ and held fixed while the fluctuation is varied, so the flow for the two-field functional $\Gamma_k[\phi;\bar\phi]$ is exact. This holds only as long as the two arguments are kept apart. The background-field approximation identifies $\bar\phi=\phi$ and differentiates the collapsed functional $\Gamma_k[\phi;\phi]$: the regulator then moves with the field, the collapsed derivatives differ from the fixed-background ones, and the split-Ward or Nielsen identity \cite{Nielsen:1975fs,Freire:2000bq,Pawlowski:2005xe,Safari:2015dva} measures the difference. Solving the modified split Ward identity together with the flow yields a background-independent flow in conformally reduced gravity \cite{Dietz:2015owa,Labus:2016lkh}, and constrains single-metric approximations \cite{Morris:2016spn}. In spectrally adjusted flows the regulator is moreover a function of the full inverse propagator, as introduced for Yang--Mills theory \cite{Gies:2002af} and used in gravity as the type III cutoff \cite{Codello:2008vh}. Proper-time flows belong to the same class
\cite{Schwinger:1951nm,Liao:1992fm,Bonanno:2000yp,Mazza:2001bp,Bonanno:2004sy,Bonanno:2019ukb,Giacometti:2025qyy}. Regulator optimization, derivative expansions and the dependence of truncated flows on cutoff choices have likewise been studied extensively \cite{Litim:2000ci,Litim:2001up,Litim:2001dt,Morris:1994ie,Morris:1994ki,Canet:2003qd,Pawlowski:2015mlf}.

A cutoff depending on the macroscopic fields has been introduced as a way to obtain a flow for a single gauge field or metric, compatible with analytic continuation, and acting on the fluctuations near the zeros of the full inverse propagator \cite{Wetterich:2016ewc,Wetterich:2020cxq,Wetterich:2024ivi}. The flow then no longer has an exact one-loop form, and the corrections can be organized as an expansion around a simplified flow equation \cite{Wetterich:2024ivi}.

The presence of additional terms has been noted before. With a spectrally adjusted cutoff the scale derivative of the regulator contains a term proportional to the scale derivative of the inverse propagator $\partial_t\Gamma_k^{(2)}$ \cite{Gies:2002af}, neglected in earlier work \cite{Reuter:1997gx}, and proper-time flows are in general not
exact \cite{Litim:2001hk,Litim:2002xm,deAlwis:2017ysy}. The universal two-loop coefficients have been recovered from proper-time flows for a particular prescription in which the vertices are generated locally \cite{Giacometti:2025qyy}. A systematic construction of all the terms that a field-dependent regulator adds at a given loop order has, however, not been given.

A field-dependent regulator distinguishes two operations that would otherwise coincide. One can either differentiate the effective action with respect to the field while holding the cutoff kernel fixed, or differentiate the functional obtained after the cutoff has been set equal to its field-dependent value. The two Hessians differ by a two-point function, the {\it mismatch tensor} $Q$, which is of order $\hbar$; it therefore drops out at one loop and enters at two loops \cite{Wetterich:2024ivi}. A reduced spectral flow can thus reproduce the one-loop running and fail at the next order. At two loops two additions are required: the $Q$ term, and the implicit scale dependence that the cutoff inherits from the running Hessian. With both included, the flow can be brought to the fixed-kernel two-loop form by absorbing a total scale derivative into the two-loop effective action. 

To keep track of the two operations we introduce a kernel-dependent effective average action $\Gamma[\phi,K]$, in which the field
$\phi$ and a symmetric two-point kernel $K$ are independent arguments. Any regulator prescription, whether fixed, field dependent or spectrally adjusted, then amounts to a rule $K=R_k[\phi]$ that assigns one kernel to each field
configuration and scale, and the mismatch tensor is
\begin{equation}
	Q_{AB}=\widehat\Gamma^{(2)}_{k,AB}-\Gamma^{(2)}_{k,AB},
	\label{eq:Q_intro}
\end{equation}
where the hat denotes the fixed-kernel Hessian. For a field-independent regulator, $Q$ vanishes identically, and the formulation reduces to the usual Wetterich equation.  The distinction between external and field-dependent cutoff structures is closely related to longstanding issues of background dependence and gauge symmetry in functional RG formulations \cite{Reuter:1993kw,Morris:1999px,Morris:2000fs,Litim:2002hj,Manrique:2009uh,Manrique:2010am,Wetterich:2016ewc,Falls:2020tmj,Ihssen:2026ucr}.

The sum of the two additions is fixed by an exact sum rule. It follows from an algebraic identity among the kernels that enter the flow, holds for every admissible regulator profile, and states that the reduced coefficient, the implicit scale term and the $Q$ term add up to the coefficient of the fixed-kernel flow. Since the fixed-kernel flow reproduces perturbation theory with regulated propagators, its two-loop coefficient is universal for any well-behaved regulator, and so is the completed one. We illustrate this for the two-loop anomalous dimension of massless $\lambda\phi^4$ theory in four dimensions, where $C_\eta=1/6$~\cite{Zinn-Justin:2002ecy}. With a proper-time regulator the reduced flow gives a coefficient that depends on the profile. The two additional terms have opposite signs, each exceeds in absolute value the deviation from $1/6$, and their sum restores $C_\eta=1/6$ for every profile.

The construction of Secs.~\ref{sec:bilocal_flow}--\ref{sec:general_completion} does not rely on the scalar example. It applies to any prescription $K=R_k[\phi]$, including the background-field identification in gauge theories and gravity, where the same two contributions first act at two loops; whether their omission affects the universal coefficients there is not settled by the present analysis.

\paragraph{Plan of the paper.}
The paper is divided into two parts.  Section~\ref{sec:bilocal_flow} sets up the kernel-dependent effective average action, defines the regulator prescription as a rule $K=R_k[\phi]$, and derives the exact flow that follows, together with the mismatch tensor $Q$ and its loop counting. In Section~\ref{sec:proper_time_vertex_hierarchies} we explain why a single one-loop trace does not determine a unique hierarchy of vertex flows, and construct two candidates: the fixed-kernel one and the one obtained by differentiating the spectral function directly. Section~\ref{sec:general_completion} contains the completion theorem and its proof. Here, we give its kernel form, and show that the three contributions recombine into the fixed-kernel coefficient for any projection satisfying the  hypotheses.

In the second part of the paper we illustrate the mechanism in massless $\lambda\phi^4$ theory. In Section~\ref{sec:scalar_setup} we compute the two-loop coefficient in the reduced and fixed-kernel hierarchies, evaluate the two missing completion terms, verify the general sum rule numerically, and compare different proper-time prescriptions. Finally, we conclude in Section \ref{sec:conclusions} and we give an outlook about future work. Technical derivations are collected in the appendices.

\paragraph{Terminology and notation.}
We summarize the main terms used throughout the paper.

\begin{center}
\begingroup
\small
\renewcommand{\arraystretch}{1.25}
\begin{tabular}{@{}p{0.27\linewidth}p{0.67\linewidth}@{}}
\hline
\textbf{Term} & \textbf{Meaning} \\
\hline

\multicolumn{2}{@{}l}{\emph{Bilocal formulation}} \\

Section and pullback
&
A section specifies the regulator $K=R_k[\phi]$.
The pullback is the action
$\Gamma_k[\phi]=\Gamma[\phi,R_k[\phi]]$.
\\

Spectral section
&
A regulator chosen as a function of the running Hessian,
$R_k=R_k(H_k)$, with $H_k=\Gamma_k^{(2)}$.
\\

Horizontal, vertical,
and along-section
&
Horizontal derivatives hold $K$ fixed. Along-section
derivatives also vary $K$ through $R_k[\phi]$.
The additional response to the kernel is called vertical.
\\

Mismatch tensor
&
$Q=\widehat\Gamma_k^{(2)}-\Gamma_k^{(2)}$,
the difference between the horizontal and
along-section Hessians.
\\

Exact and reduced
propagators
&
The exact propagator on the section is
$(\widehat\Gamma_k^{(2)}+R_k)^{-1}$.
The reduced propagator is
$G_k=(\Gamma_k^{(2)}+R_k)^{-1}$.
\\

\hline
\multicolumn{2}{@{}l}{\emph{Flow prescriptions}} \\

Fixed-kernel hierarchy
&
The vertex hierarchy obtained by taking field
derivatives at fixed regulator. For an external
regulator, this is the standard Wetterich hierarchy.
\\

Direct proper-time
hierarchy
&
The hierarchy obtained by differentiating the
spectral function $W(H_k)$ directly with respect
to the fields.
\\

Reduced and
completed flows
&
The reduced flow omits the vertical and implicit-scale
terms. The completed flow includes these contributions.
\\

Explicit and implicit
scale dependence
&
The explicit derivative changes $k$ at fixed $H_k$.
The implicit term arises because the Hessian in
$R_k(H_k)$ also runs with $k$.
\\

\hline
\multicolumn{2}{@{}l}{\emph{Kernels and vertices}} \\

Spectral kernel and
regulator reconstruction
&
$W$ is the function defining the spectral trace.
Reconstruction determines an external regulator
profile $r$ from $W=gs$.
\\

Single-scale insertion
&
$G_k S_k G_k$, where
$S_k=(\partial_tR_k)_{H_k}$.
It appears on the internal line carrying the
explicit scale derivative.
\\

Divided differences
&
$F^{[1]}(u,v)=[F(u)-F(v)]/(u-v)$,
with $F^{[1]}(u,u)=F'(u)$.
They arise when differentiating functions of
noncommuting operators.
\\

Bubble kernel
&
The kernel $\mathcal B(u,v)$ assigned to the
internal lines of a four-point bubble.
The fixed-kernel and direct proper-time
hierarchies give different kernels.
\\

One-loop source
&
The function $I_m(y)$ describing the momentum
dependence of the one-loop four-point bubble,
with $y=P^2/k^2$.
\\

\hline
\end{tabular}
\endgroup
\end{center}

\section{The kernel-dependent effective average action}
\label{sec:bilocal_flow}

When the regulator depends on the field, a field derivative acts both on the explicit field argument and on the regulator. To separate the two we treat the cutoff kernel as an independent variable and impose the regulator choice only afterwards.

Let $K_{AB} = K_{ab}(x,y)$ be an arbitrary symmetric bilocal kernel, i.e.\ a two-point function. We use condensed DeWitt notation: a collective index $A$ denotes both spacetime and internal labels, $x$ and $a$ respectively, and repeated indices are summed and integrated. We then define the kernel-dependent effective average action $\Gamma[\phi,K]$ by
\begin{equation}
	e^{-\Gamma[\phi,K]}
	=
	\int {\cal D}\hat\phi\,
	\exp\left[
	-S[\hat\phi]
	-\frac12(\hat\phi-\phi)^A K_{AB}(\hat\phi-\phi)^B
	+\Gamma_A[\phi,K](\hat\phi-\phi)^A
	\right].
	\label{eq:bilocal_definition}
\end{equation}
Here $\hat\phi$ denotes a generic  microscopic field or field multiplet, while $\phi$ is the corresponding macroscopic mean field. The source term proportional to 
\begin{equation}
\Gamma_A[\phi,K] \equiv \frac{\delta}{\delta \phi^{A}} \Gamma[\phi,K] 
\end{equation}
is fixed by the condition
\begin{equation}
	\langle \hat\phi^A-\phi^A\rangle =0 .
\end{equation}
In this enlarged description, $\phi$ and $K$ are independent coordinates. Unless stated otherwise, all functional derivatives of $\Gamma[\phi,K]$ are therefore taken at fixed $K$.

With $e^{\mathcal W[J,K]}=\int{\cal D}\hat\phi\,\exp[-S[\hat\phi]-\frac12\hat\phi^AK_{AB}\hat\phi^B+J_A\hat\phi^A]$ and $J_A=\Gamma_A+K_{AB}\phi^B$, the definition \eqref{eq:bilocal_definition} is equivalent to
\begin{equation}
	\Gamma[\phi,K]=-\mathcal W[J,K]+J_A\phi^A-\frac12\phi^AK_{AB}\phi^B ,
	\label{eq:Gamma_as_Legendre}
\end{equation}
with $\delta\mathcal W/\delta J_A=\phi^A$. Thus $\Gamma[\phi,K]$ is the Legendre transform with respect to the linear source at fixed bilocal source $K$, with the term $\frac12\phi K\phi$ subtracted. It depends on $k$ only through $K$; for a field-independent kernel $K=R_k$ it is the effective average action \cite{Wetterich:1992yh}, and for $K=0$ the ordinary one-particle-irreducible effective action. The two-particle-irreducible effective action \cite{Cornwall:1974vz} is obtained by a further Legendre transform with respect to $K$, which makes the propagator the independent variable. Here no such transform is taken, and $K$ remains an independent argument.

The response to a change of kernel is obtained by differentiating \eqref{eq:bilocal_definition} with respect to $K_{AB}$. This gives
\begin{equation}
	\frac{\delta \Gamma[\phi,K]}{\delta K_{AB}}
	=
	\frac12\,{\cal P}^{AB}[\phi,K],
	\label{eq:K_derivative_identity}
\end{equation}
where
\begin{equation}
	{\cal P}^{AB}[\phi,K]
	=
	\left(
	\Gamma^{(2)}[\phi,K]+K
	\right)^{-1\,AB}.
	\label{eq:bilocal_propagator}
\end{equation}
Thus ${\cal P}$ is the full connected two-point function in the presence of the bilocal kernel $K$.

If the kernel is allowed to depend on the RG time $t=\log k$, the chain rule gives the exact flow along an arbitrary trajectory $K_t$ in kernel space,
\begin{equation}
	\frac{\dd}{\dd t} \Gamma[\phi,K_t]
	=
	\frac12
	\Tr\left[
	{\cal P}[\phi,K_t] \, \partial_t K_t
	\right].
	\label{eq:exact_bilocal_flow}
\end{equation}
Equation~\eqref{eq:exact_bilocal_flow} holds for any trajectory $K_t$. Ordinary Wetterich flows correspond to an external choice $K_t=R_k$; field-dependent or spectrally adjusted prescriptions relate $K$ to $\phi$ and,
implicitly, to $\Gamma_k$, defined in \eqref{eq:reduced_action}. The prescriptions differ only in this choice and in how derivatives are taken after it.

An RG prescription assigns a kernel to each field configuration and scale, that is, it defines a section,
\begin{equation}
	K_{AB}=R_{k,AB}[\phi].
	\label{eq:section_definition}
\end{equation}

The scale-dependent effective action is obtained by evaluating $\Gamma[\phi,K]$ on this choice:
\begin{equation}
	\Gamma_k[\phi]
	=
	\Gamma[\phi,K]\big|_{K=R_k[\phi]} .
	\label{eq:reduced_action}
\end{equation}
We call $\Gamma_k$ the \emph{reduced action}, or the \emph{pullback} of $\Gamma[\phi,K]$ to the
section. A field variation of $\Gamma_k$ changes both arguments of
$\Gamma[\phi,K]$, the second through $R_k[\phi]$.

Figure~\ref{fig:bilocal_geometry_3d} illustrates the construction. The base
plane is the space of pairs $(\phi,K)$ and $\Gamma[\phi,K]$ is a surface above
it. A regulator prescription is a curve in the base plane, and the reduced
action $\Gamma_k[\phi]$ is the profile of the surface directly above that
curve. Moving in the field direction at fixed $K$ traces one path on the
surface; moving along the curve traces a different one.

\definecolor{sectionblue}{RGB}{28,82,140}
\definecolor{pullbackred}{RGB}{158,44,44}

\begin{figure}[tb]
\centering

\begin{tikzpicture}[
  x={(1.00cm,0cm)},
  y={(-0.72cm,-0.48cm)},
  z={(0cm,0.86cm)},
  >={Latex[length=2.1mm,width=1.5mm]},
  line cap=round,
  line join=round,
  font=\small,
  declare function={
    Rk(\x)     = 0.45 + 1.00*\x - 0.125*\x*\x + 0.0058*\x*\x*\x;
    dRk(\x)    = 1.00 - 0.25*\x + 0.0174*\x*\x;
    Gam(\x,\y) = 0.85 + 0.13*\x + 0.010*\x*\x
                      + 0.36*\y + 0.025*\y*\y + 0.030*\x*\y;
    Gk(\x)     = Gam(\x,Rk(\x));
  },
  surfacemesh/.style   = {black!30,line width=0.25pt},
  surfaceedge/.style   = {black!55,line width=0.5pt},
  framearrow/.style    = {->,black!75,line width=0.7pt},
  lbl/.style           = {inner sep=1.5pt, 
                          fill opacity=0.85,text opacity=1},
  tinylbl/.style       = {lbl,font=\footnotesize}
]

\def\Pmin{0.5}\def\Pmax{6.7}         
\def\Kmin{0.35}\def\Kmax{3.35}       
\def\PlaneP{7.2}\def\PlaneK{3.7}     
\def\Gax{4.9}                       
\def\phinode{2.6}                   

\pgfmathsetmacro\Knode{Rk(\phinode)}
\pgfmathsetmacro\Gnode{Gk(\phinode)}
\pgfmathsetmacro\dphi{1.35}
\pgfmathsetmacro\dK{dRk(\phinode)*\dphi}
\pgfmathsetmacro\dG{(Gk(\phinode+0.01)-Gk(\phinode-0.01))/0.02*\dphi}

\coordinate (Pbase) at (\phinode,\Knode,0);
\coordinate (Psurf) at (\phinode,\Knode,\Gnode);

\fill[black!4]
  (0,0,0) -- (\PlaneP,0,0) -- (\PlaneP,\PlaneK,0) -- (0,\PlaneK,0) -- cycle;

\foreach \k in {0.5,1.0,...,3.5}
  \draw[black!12,line width=0.2pt] (0,\k,0) -- (\PlaneP,\k,0);
\foreach \p in {0.5,1.0,...,7.0}
  \draw[black!12,line width=0.2pt] (\p,0,0) -- (\p,\PlaneK,0);

\draw[black!28,line width=0.4pt]
  (0,0,0) -- (\PlaneP,0,0) -- (\PlaneP,\PlaneK,0) -- (0,\PlaneK,0) -- cycle;
\draw[->,black!75] (0,0,0) -- ({\PlaneP+0.5},0,0) node[right]      {$\phi$};
\draw[->,black!75] (0,0,0) -- (0,{\PlaneK+0.4},0) node[below left] {$K$};
\draw[->,black!75] (0,0,0) -- (0,0,\Gax)          node[above]      {$\Gamma$};
\draw[sectionblue,line width=1.0pt]
  plot[domain=\Pmin:\Pmax,samples=60,variable=\x] (\x,{Rk(\x)},0);

\node[sectionblue,lbl,anchor=north west]
  at ({\Pmax+0.02},{Rk(\Pmax)-0.2},0) {$K=R_k[\phi]$};

\draw[framearrow] (Pbase) -- ({\phinode+\dphi},\Knode,0)
  node[tinylbl,anchor=south west,pos=1] {horizontal};

\draw[framearrow] (Pbase) -- (\phinode,{\Knode+0.95},0)
  node[tinylbl,anchor=east,pos=1] {vertical};

\draw[->,sectionblue,line width=0.9pt]
  (Pbase) -- ({\phinode+\dphi-0.1},{\Knode+\dK},-0.05)
  node[tinylbl,anchor=north west,pos=1] {along section};

\fill[sectionblue] (Pbase) circle (1.5pt);

\node[lbl,anchor=south] at (\phinode+0.2,{\Knode-0.74},0)
  {$\bigl(\phi,R_k[\phi]\bigr)$};

\path[fill=black!3,fill opacity=0.95]
  plot[domain=\Pmin:\Pmax,samples=60,variable=\x] (\x,\Kmin,{Gam(\x,\Kmin)})
  plot[domain=\Kmin:\Kmax,samples=30,variable=\t] (\Pmax,\t,{Gam(\Pmax,\t)})
  plot[domain=\Pmax:\Pmin,samples=60,variable=\x] (\x,\Kmax,{Gam(\x,\Kmax)})
  plot[domain=\Kmax:\Kmin,samples=30,variable=\t] (\Pmin,\t,{Gam(\Pmin,\t)})
  -- cycle;

\foreach \k in {0.95,1.55,2.15,2.75}
  \draw[surfacemesh]
    plot[domain=\Pmin:\Pmax,samples=50,variable=\x] (\x,\k,{Gam(\x,\k)});
\foreach \p in {1.74,2.98,4.22,5.46}
  \draw[surfacemesh]
    plot[domain=\Kmin:\Kmax,samples=30,variable=\t] (\p,\t,{Gam(\p,\t)});

\draw[surfaceedge]
  plot[domain=\Pmin:\Pmax,samples=60,variable=\x] (\x,\Kmin,{Gam(\x,\Kmin)});
\draw[surfaceedge]
  plot[domain=\Pmin:\Pmax,samples=60,variable=\x] (\x,\Kmax,{Gam(\x,\Kmax)});
\draw[surfaceedge]
  plot[domain=\Kmin:\Kmax,samples=30,variable=\t] (\Pmin,\t,{Gam(\Pmin,\t)});
\draw[surfaceedge]
  plot[domain=\Kmin:\Kmax,samples=30,variable=\t] (\Pmax,\t,{Gam(\Pmax,\t)});

\draw[black!45,line width=0.35pt] (0.95,2.75,4.12) -- (1.60,2.60,{Gam(1.60,2.60)});
\node[lbl,anchor=south west] at (0.95,2.85,4.12) {$\Gamma[\phi,K]$};

\foreach \p in {1.0,1.8,...,6.6}
  \draw[black!20,line width=0.3pt] (\p,{Rk(\p)},0) -- (\p,{Rk(\p)},{Gk(\p)});

\draw[pullbackred,line width=1.1pt]
  plot[domain=\Pmin:\Pmax,samples=60,variable=\x] (\x,{Rk(\x)},{Gk(\x)});

\node[pullbackred,lbl,anchor=west] at ({\Pmax+0.08},{Rk(\Pmax)-0.3},{Gk(\Pmax)})
  {$\Gamma_k[\phi]$};

\draw[->,pullbackred,line width=0.9pt]
  (Psurf) -- ({\phinode+\dphi},{\Knode+\dK},{\Gnode+\dG})
  node[tinylbl,anchor=south east,pos=1]
  {$\dfrac{\mathrm{d}\Gamma_k}{\mathrm{d}\phi}$};

\draw[black!55,densely dashed,line width=0.85pt] (Pbase) -- (Psurf);
\fill[black] (Psurf) circle (1.9pt);

\end{tikzpicture}

\caption{
Geometry of the bilocal formulation. The kernel-dependent effective average action
$\Gamma[\phi,K]$ is a surface over the $(\phi,K)$ configuration space.
A regulator prescription selects the section $K=R_k[\phi]$ (blue), whose
lift to the surface defines the reduced action
$\Gamma_k[\phi]=\Gamma\bigl[\phi,R_k[\phi]\bigr]$ (red). At a point of
the section, a field variation at fixed kernel is horizontal and a
variation of the kernel at fixed field is vertical; the tangent to the
section decomposes into these two components, so that
$\mathrm{d}/\mathrm{d}\phi=\partial_\phi+(\delta R_k/\delta\phi)\,\partial_K$.
Differentiation of the reduced action therefore carries both a
horizontal and a vertical contribution.
}
\label{fig:bilocal_geometry_3d}

\end{figure}
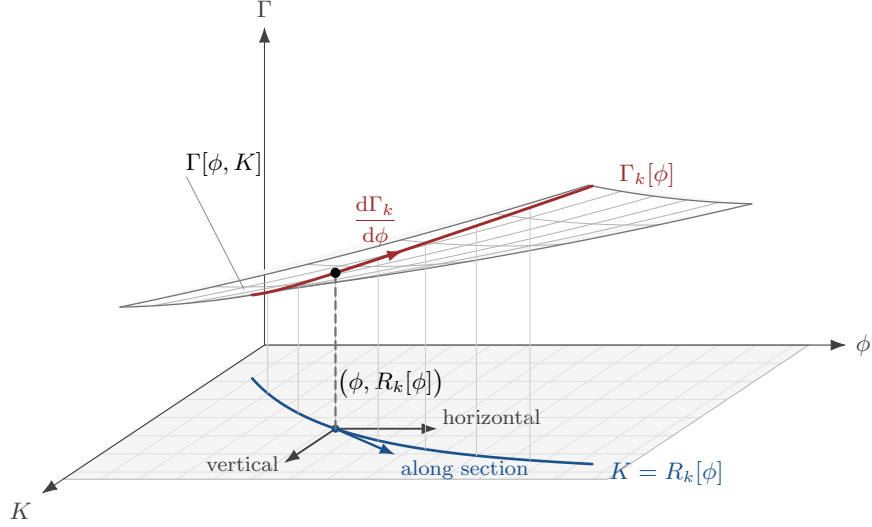

Evaluating the exact bilocal flow \eqref{eq:exact_bilocal_flow} on the section
gives
\begin{equation}
	\partial_t\Gamma_k[\phi]
	=
	\frac12
	\Tr\left[
	{\cal P}[\phi,R_k[\phi]]\,\partial_t R_k[\phi]
	\right].
	\label{eq:pulled_flow_raw}
\end{equation}
This is still exact, but ${\cal P}$ involves derivatives at fixed $K$, whereas one usually works with derivatives of $\Gamma_k$. We therefore distinguish two Hessians: the horizontal one, taken at fixed $K$ and then evaluated on the section,
\begin{equation}
	\widehat\Gamma^{(2)}_{k,AB}[\phi]
	:=
	\Gamma_{AB}[\phi,K]\big|_{K=R_k[\phi]} ,
	\label{eq:hat_hessian}
\end{equation}
and the reduced, or along-section one, which differentiates the reduced action itself,
\begin{equation}
	\Gamma^{(2)}_{k,AB}[\phi]
	:=
	\frac{\delta^2\Gamma_k[\phi]}{\delta\phi^A\delta\phi^B}.
	\label{eq:reduced_hessian}
\end{equation}
The latter also contains the response of $\Gamma$ in the $K$ direction, which we call \emph{vertical}. For a field-independent regulator the two Hessians coincide.

We encode this difference in the two-index mismatch tensor \cite{Wetterich:2024ivi}
\begin{equation}
	Q_{AB}[\phi]
	:=
	\widehat\Gamma^{(2)}_{k,AB}[\phi]
	-
	\Gamma^{(2)}_{k,AB}[\phi].
	\label{eq:Q_definition}
\end{equation}
By the chain rule,
\begin{equation}
	\Gamma^{(2)}_{k,AB}
	=
	\Gamma_{AB}
	+\frac{\delta\Gamma}{\delta K_{CD}}\,R_{k,CD,AB}
	+2\,\frac{\delta\Gamma_{(A}}{\delta K_{CD}}\,R_{k,CD,B)}
	+\frac{\delta^2\Gamma}{\delta K_{CD}\delta K_{EF}}\,
	R_{k,CD,A}R_{k,EF,B},
	\label{eq:pullback_chain_rule}
\end{equation}
where $R_{k,CD,A}=\delta R_{k,CD}/\delta\phi^A$ and parentheses denote symmetrization in $A$ and $B$. Comparing with \eqref{eq:hat_hessian} and
\eqref{eq:Q_definition},
\begin{equation}
	Q_{AB}
	=
	-\frac12\,{\cal P}^{CD}R_{k,CD,AB}
	-2\,\frac{\delta\Gamma_{(A}}{\delta K_{CD}}\,R_{k,CD,B)}
	-\frac{\delta^2\Gamma}{\delta K_{CD}\delta K_{EF}}\,
	R_{k,CD,A}R_{k,EF,B},
	\label{eq:Q_chain_rule}
\end{equation}
where the first term has been evaluated with the kernel derivative identity \eqref{eq:K_derivative_identity}. Every term carries at least one derivative of the section, which is why $Q$ vanishes identically for an external
regulator, and at least one derivative of $\Gamma$ with respect to $K$, which by \eqref{eq:K_derivative_identity} is a connected correlation function and is therefore of order $\hbar$.

In terms of $Q$ the propagator in \eqref{eq:pulled_flow_raw} is
\begin{equation}
	{\cal P}[\phi,R_k[\phi]]
	=
	\left(
	\Gamma_k^{(2)}[\phi]+R_k[\phi]+Q[\phi]
	\right)^{-1}.
	\label{eq:P_with_Q}
\end{equation}
Hence the exact pulled-back flow becomes
\begin{equation}
	\partial_t\Gamma_k[\phi]
	=
	\frac12
	\Tr\left[
	\left(
	\Gamma_k^{(2)}+R_k+Q
	\right)^{-1}
	\partial_tR_k
	\right].
	\label{eq:exact_pullback_flow}
\end{equation}
Equation \eqref{eq:exact_pullback_flow} is exact, but it is not closed in
$\Gamma_k[\phi]$: by \eqref{eq:Q_chain_rule}, $Q$ contains derivatives of
$\Gamma[\phi,K]$ in the $K$ direction, which are not functional derivatives of
$\Gamma_k$. The same information enters \eqref{eq:pulled_flow_raw} through
${\cal P}$, which requires $\Gamma[\phi,K]$ off the section. In principle, $Q$ can be determined by evolving the full $\Gamma[\phi,K]$, although this is generally impractical. For an external regulator, $Q$ vanishes identically.  In perturbation theory the $K$ dependence is known
explicitly: at one loop $\Gamma_1[\phi,K]=\frac12\Tr\log(H_0+K)+F[\phi]$, and
$Q_1$ follows directly, as shown in App.~\ref{app:determinant_Q1}. Beyond
one loop, a flow for $\Gamma_k[\phi]$ alone needs an approximation for $Q$, and
\eqref{eq:Q_chain_rule} shows how it is organized.

To display the relation to the commonly used reduced one-loop trace, define
\begin{equation}
	G_k[\phi]
	:=
	\left(
	\Gamma_k^{(2)}[\phi]+R_k[\phi]
	\right)^{-1}.
	\label{eq:G_definition}
\end{equation}
Expanding the exact propagator in powers of $Q$ gives
\begin{equation}
	\left(
	\Gamma_k^{(2)}+R_k+Q
	\right)^{-1}
	=
	G_k
	-
	G_k Q G_k
	+
	G_k Q G_k Q G_k
	-
	\cdots .
	\label{eq:P_expansion_Q}
\end{equation}
The flow can therefore be written as
\begin{equation}
	\partial_t\Gamma_k[\phi]
	=
	\frac12\Tr\left[G_k\,\partial_tR_k\right]
	-
	\frac12\Tr\left[G_kQG_k\,\partial_tR_k\right]
	+
	\frac12\Tr\left[G_kQG_kQG_k\,\partial_tR_k\right]
	-
	\cdots .
	\label{eq:flow_Q_expansion}
\end{equation}
The first term is the reduced trace; the others are vertical corrections
from the field dependence of the section. 

For a field-independent regulator,
\begin{equation}
	R_{k,AB}[\phi]=R_{k,AB},
	\qquad
	R_{k,AB,C}=0,
	\qquad
	R_{k,AB,CD}=0,
\end{equation}
one has $Q=0$, and \eqref{eq:exact_pullback_flow} reduces to the ordinary
fixed-regulator Wetterich form. For a field-dependent section, instead, $Q=O(\hbar)$, by the remark after
\eqref{eq:Q_chain_rule}, so the first term containing $Q$ in \eqref{eq:flow_Q_expansion} contributes at two loops.

This structure becomes explicit for a spectral section $R_k=R_k(H_k)$, with $H_k=\Gamma_k^{(2)}$, for which
\begin{equation}
\partial_tR_k(H_k)
=
(\partial_tR_k)_{H_k}
+
DR_k[H_k]\!\cdot\!\partial_tH_k .
\end{equation}
Keeping only the explicit scale dependence, together with the term of zeroth order in $Q$, gives the proper-time flow used in the next section. We will see in Sec.~\ref{sec:completion} that the neglected terms first enter at two-loop order and are those required to complete the flow.

\section{Proper time representation and vertex hierarchies}
\label{sec:proper_time_vertex_hierarchies}
Agreement at the level of the one-loop trace does not determine the flow of higher vertices. The two constructions considered below use the same spectral flow kernel and coincide on the tree-level spectrum at the symmetric field configuration, but differ under field differentiation. They therefore generate different flows for $\Gamma_k^{(2)}$, $\Gamma_k^{(4)}$, and higher vertices. We refer to this sequence of flow equations as the \emph{vertex hierarchy}.
We call the hierarchy obtained by differentiating the spectral trace $W(H_k)$ directly with respect to the fields the \emph{direct proper-time hierarchy}. It is the vertex hierarchy of the reduced proper-time flow, in which the vertical and implicit-scale contributions are not included.

We carry out the comparison in the spectral, or proper-time, representation, in which the flow is a scalar function evaluated on the running inverse propagator.

\label{subsec:spectral_form}

	Let $H$ be a positive operator. A proper-time flow can be written in the
	spectral form
	\begin{equation}
		\partial_t\Gamma_k^{\mathrm{PT}}
		=
		\frac12 \Tr\, W_k(H),
		\label{eq:PT_spectral_form}
	\end{equation}
	where the scalar function $W_k$ admits a Laplace representation
	\begin{equation}
		W_k(H)
		=
		\int_0^\infty \dd s\, \omega_k(s)\, e^{-sH}.
		\label{eq:W_laplace_general}
	\end{equation}
	Equivalently, define
	\begin{equation}
		\rho_k(s):=s\,\omega_k(s),
		\qquad
		W_k(H)=\int_0^\infty\frac{\dd s}{s}\,
		\rho_k(s)e^{-sH}.
		\label{eq:rho_omega_definition}
	\end{equation}
	The weights $\omega_k$ and $\rho_k$ specify the same coarse-graining
	scheme. We use the representation
	\eqref{eq:rho_omega_definition}. For the dimensionless spectral variable
	we write $W(u):=W_k(k^2u)$, so that $W_k(H)=W(H/k^2)$.

The same function $W$ can be used in two inequivalent ways. The first is the direct proper-time construction. In this case one identifies $H=\Gamma_k^{(2)}[\phi]$ in \eqref{eq:PT_spectral_form} and regards
	\begin{equation}
		\frac12\Tr\,W\left(\Gamma_k^{(2)}[\phi]/k^2\right)
	\end{equation}
	as the flow to be differentiated with respect to the fields. Since the Hessian itself is field dependent, the functional derivatives act on the full operator function $W(H)$. The resulting vertex hierarchy therefore depends not only on the propagator structure, but also on how the spectral function varies under changes of $H$.

	A different construction starts from the same scalar kernel $W$ but uses it only to reconstruct a cutoff profile of Wetterich type. We write
	\begin{equation}
		W(u)=g(u)s(u),
		\label{eq:W_as_gs_preview}
	\end{equation}
	with
	\begin{equation}
		g(u):=\frac{1}{u+r(u)},
		\qquad
		s(u):=2r(u)-2u r'(u).
		\label{eq:g_s_preview}
	\end{equation}
This relation reconstructs an external regulator $R_k(q^2)=k^2r(q^2/k^2)$
whose trace reproduces $W$. Since it is field independent, $Q=0$; we call
the resulting hierarchy the \emph{fixed-kernel  hierarchy}.

We begin with the fixed-kernel, or horizontal, construction and then derive its
direct proper-time counterpart.

\subsection{Differentiating at fixed kernel}
\label{sec:vertex_expansion_fixed_kernel}

The fixed-kernel hierarchy will serve as the reference against which the direct
proper-time hierarchy and the vertical terms are compared. We call the hierarchy obtained by differentiating at fixed regulator the \emph{fixed-kernel hierarchy}. For an external regulator, it coincides with the standard Wetterich hierarchy. In the bilocal formulation, the corresponding field derivatives are horizontal.

We start from
\begin{equation}
	\left(\partial_t\Gamma_k\right)_{\mathrm{hor}}
	=
	\frac12\Tr\left[
	G_k\,\partial_tR_k
	\right],
	\qquad
	G_k=
	\left(
	\Gamma_k^{(2)}+R_k
	\right)^{-1}.
	\label{eq:fixed_kernel_flow}
\end{equation}
Although $R_k$ may ultimately be chosen as a function of the running Hessian,
the defining rule of the fixed-kernel hierarchy is that it is held fixed under
functional differentiation,
\begin{equation}
	\frac{\delta R_k}{\delta\phi^A}=0,
	\qquad
	\frac{\delta\,\partial_tR_k}{\delta\phi^A}=0.
	\label{eq:fixed_R_derivatives}
\end{equation}
Thus all field dependence in the vertex expansion is generated by the Hessian
$\Gamma_k^{(2)}[\phi]$ inside the propagator. This is the same differentiation
rule as in the standard fixed-regulator hierarchy.

At one loop, the Hessian in the internal propagator may be replaced by its classical value,
\begin{equation}
	\Gamma^{(2)}[\phi,K]\;\longrightarrow\;S^{(2)}[\phi],
	\label{eq:oneloop_Hessian_replacement}
\end{equation}
which is independent of $K$. The kernel therefore enters the one-loop vertices only through the explicit regulator $R_k$ in \eqref{eq:fixed_kernel_flow}.

To generate the vertex hierarchy, we differentiate the propagator appearing in \eqref{eq:fixed_kernel_flow}. At fixed $R_k$,
\begin{equation}
	\delta G_k
	=
	-G_k\,\delta\Gamma_k^{(2)}\,G_k .
	\label{eq:delta_G}
\end{equation}
Inserting this in the variation of the flow,
\begin{equation}
\delta(\partial_t\Gamma_k)
=
\frac12\Tr\!\left[(\delta G_k)\,\partial_tR_k\right],
\end{equation}
and cyclically rearranging the trace gives the standard single-scale combination
\begin{equation}
G_k\,\partial_tR_k\,G_k
\end{equation}
on the differentiated internal line.
Introducing the vertex insertions
\begin{equation}
	\left(\Gamma^{(3)}_{k,A}\right)_{BC}
	:=
	\Gamma_{k,ABC},
	\qquad
	\left(\Gamma^{(4)}_{k,AB}\right)_{CD}
	:=
	\Gamma_{k,ABCD},
	\label{eq:vertex_insertions}
\end{equation}
one obtains
\begin{equation}
	\left(G_k\right)_{,A}
	=
	-G_k\,\Gamma^{(3)}_{k,A}\,G_k ,
	\label{eq:G_first_derivative}
\end{equation}
and
\begin{align}
	\left(G_k\right)_{,AB}
	&=
	G_k\,\Gamma^{(3)}_{k,A}\,G_k\,\Gamma^{(3)}_{k,B}\,G_k
	+
	G_k\,\Gamma^{(3)}_{k,B}\,G_k\,\Gamma^{(3)}_{k,A}\,G_k
	\nonumber\\
	&\quad
	-
	G_k\,\Gamma^{(4)}_{k,AB}\,G_k .
	\label{eq:G_second_derivative}
\end{align}
All products imply integration over intermediate DeWitt indices.

The first derivative of \eqref{eq:fixed_kernel_flow} yields the one-point flow
\begin{equation}
	\partial_t\Gamma_{k,A}
	=
	-\frac12
	\Tr\left[
	G_k\,\Gamma^{(3)}_{k,A}\,G_k\,\partial_tR_k
	\right].
	\label{eq:onepoint_flow_general}
\end{equation}
A second derivative gives
\begin{align}
	\partial_t\Gamma_{k,AB}
	&=
	\frac12
	\Tr\left[
	G_k\Gamma^{(3)}_{k,A}G_k\Gamma^{(3)}_{k,B}
	G_k\partial_tR_k
	\right]
	+
	\frac12
	\Tr\left[
	G_k\Gamma^{(3)}_{k,B}G_k\Gamma^{(3)}_{k,A}
	G_k\partial_tR_k
	\right]
	\nonumber\\
	&\quad
	-
	\frac12
	\Tr\left[
	G_k\Gamma^{(4)}_{k,AB}G_k\partial_tR_k
	\right].
	\label{eq:twopoint_flow_general}
\end{align}
This is the general two-point flow in the horizontal fixed-kernel hierarchy.
Its terms have the familiar one-loop topology: field derivatives insert
vertices on the internal propagator, while the scale derivative remains
localized on the regulator insertion.

Higher vertices follow by repeated differentiation. For example, the
four-point flow contains terms with two insertions of $\Gamma_k^{(4)}$, terms
with one insertion of $\Gamma_k^{(6)}$, and terms involving lower odd vertices
whenever allowed by the field content and symmetries. Schematically,
\begin{equation}
	\partial_t\Gamma_k^{(4)}
	=
	\left[\Gamma_k^{(4)}G_k\Gamma_k^{(4)}\right]
	+
	\left[\Gamma_k^{(6)}\right]
	+
	\left[\Gamma_k^{(3)}G_k\Gamma_k^{(5)}\right]
	+
	\left[\Gamma_k^{(3)}G_k\Gamma_k^{(3)}G_k\Gamma_k^{(4)}\right]
	+\cdots .
	\label{eq:fourpoint_flow_schematic}
\end{equation}
Each bracket denotes the corresponding trace with one insertion of
$G_k\partial_tR_kG_k$ on a differentiated internal line and ordinary
propagators on the remaining lines. The precise combinatorial coefficients and
index contractions depend on the theory under consideration.

The two-point flow and the bubble of this hierarchy are shown in the left
column of Fig.~\ref{fig:kernel_comparison}.

The two-point flow already displays the single-scale structure characteristic of the fixed-kernel hierarchy. In the direct proper-time hierarchy we will proceed to the four-point flow, where the bubble contribution introduces the kernel to be compared with the fixed-kernel result below.

\subsection{Differentiating the spectral function}
\label{sec:loop_hierarchy_direct_PT}

We now do the opposite. Instead of freezing a reconstructed profile, we take the spectral trace $\Tr\,W(H_k[\phi])$  and differentiate it, letting the derivatives act on the operator function itself. This is the prescription actually used in proper-time and spectrally adjusted calculations, and in the formalism of Sec.~\ref{sec:bilocal_flow} it is differentiation along the regulator section rather than at fixed kernel.

We write the flow in the form
\begin{equation}
	\partial_t\Gamma_k^{\mathrm{PT}}[\phi]
	=
	\frac{\hbar}{2}
	\int_0^\infty \frac{\dd s}{s}\,\rho_k(s)\,
	\Tr e^{-sH_k[\phi]},
	\qquad
	H_k[\phi]=\Gamma_k^{(2)}[\phi].
	\label{eq:PT_flow_rho}
\end{equation}
In this subsection, in Sec.~\ref{sec:hierarchy_difference} and in App.~\ref{app:direct_hierarchy}, $\hbar$ is displayed explicitly to keep track of the loop order. The function $\rho_k(s)$ is related to the spectral kernel $W_k$ by \eqref{eq:rho_omega_definition} at $H=H_k$. In this form the field derivatives act directly on the exponential.

We expand around the $\mathbb Z_2$-symmetric point $\phi=0$. Functional derivatives of $H_k[\phi]$ will be denoted by
\begin{equation}
	H_{k,A}
	:=
	\frac{\delta H_k}{\delta\phi^A},
	\qquad
	H_{k,AB}
	:=
	\frac{\delta^2 H_k}{\delta\phi^A\delta\phi^B},
\end{equation}
and similarly for higher derivatives. Since $H_k=\Gamma_k^{(2)}$, these derivatives correspond to higher vertices,
\begin{equation}
	H_{k,A}=\Gamma^{(3)}_{k,A},
	\qquad
	H_{k,AB}=\Gamma^{(4)}_{k,AB},
	\qquad
	H_{k,ABCD}=\Gamma^{(6)}_{k,ABCD}.
\end{equation}
At $\phi=0$, $\mathbb Z_2$ symmetry implies that all odd vertices vanish, and therefore
\begin{equation}
	H_{k,A}\big|_{\phi=0}=0.
	\label{eq:odd_vertices_vanish}
\end{equation}
Two field derivatives of the operator exponential produce a single insertion of $H_{k,AB}$ sandwiched between two exponentials, integrated over the proper-time position of the insertion. Terms with two odd insertions drop out by \eqref{eq:odd_vertices_vanish}. Cyclicity of the trace makes the integrand independent of the insertion point, so the integration  supplies a factor $s$, which cancels the $1/s$ in the measure of
\eqref{eq:PT_flow_rho}. The steps are written out in App.~\ref{app:direct_hierarchy}; the result is

\begin{equation}
	\left(\partial_t\Gamma^{(2)}_{k,AB}\right)_{\mathrm{PT}}
	=
	-\frac{\hbar}{2}
	\int_0^\infty \dd s\,\rho_k(s)\,
	\Tr\left[
	e^{-sH_k[0]}
	H_{k,AB}\big|_{\phi=0}
	\right].
	\label{eq:twopoint_flow_direct_hierarchy}
\end{equation}

In momentum space, for the scalar theory, this takes the form
\begin{equation}
	\left(\partial_t\Gamma_k^{(2)}(p)\right)_{\mathrm{PT}}
	=
	-\frac{\hbar}{2}
	\int_0^\infty \dd s\,\rho_k(s)
	\int\frac{\dd^4q}{(2\pi)^4}\,
	e^{-s\Gamma_k^{(2)}(q)}
	\Gamma_k^{(4)}(p,-p,q,-q).
	\label{eq:twopoint_flow_direct_momentum}
\end{equation}
Thus the two-point flow is generated by a four-point insertion inside the proper-time trace. Taking four functional derivatives gives one contact contribution and three bubble contributions. At $\phi=0$,
\begin{align}
	\left(\partial_t\Gamma^{(4)}_{k,ABCD}\right)_{\mathrm{PT}}
	&=
	-\frac{\hbar}{2}
	\int_0^\infty \dd s\,\rho_k(s)\,
	\Tr\left[
	e^{-sH_k[0]}
	H_{k,ABCD}\big|_{\phi=0}
	\right]
	\nonumber\\
	&\quad
	+
	\frac{\hbar}{2}
	\int_0^\infty \dd s\,\rho_k(s)
	\sum_{(AB)(CD)}
	\int_0^s \dd\alpha\,
	\Tr\left[
	e^{-(s-\alpha)H_k[0]}
	H_{k,AB}\big|_{\phi=0}
	e^{-\alpha H_k[0]}
	H_{k,CD}\big|_{\phi=0}
	\right].
	\label{eq:fourpoint_flow_direct_hierarchy}
\end{align}
Here the sum runs over the three unordered pairings of the external indices,
\begin{equation}
	(AB)(CD),
	\qquad
	(AC)(BD),
	\qquad
	(AD)(BC),
\end{equation}
and the summand is written for the first.

As in the two-point flow, the factor $1/s$ in the proper-time measure is canceled by the insertion integrals. For the bubble term, this follows from combining the two Duhamel orderings and using cyclicity of the trace; see App.~\ref{app:direct_hierarchy}.

The two-point flow and one bubble channel are shown in the right column of Fig.~\ref{fig:kernel_comparison}.

\subsection{Comparison of the two hierarchies}
\label{sec:hierarchy_difference}

Having derived the two vertex hierarchies, we now compare them at the level relevant for the two-loop anomalous dimension of Sec.~\ref{sec:scalar_setup}. For this purpose, the relevant difference first appears in the bubble contribution to the four-point flow. Although the two-point flows can already differ away from the symmetric configuration, that difference is momentum independent and therefore does not contribute to
the wave-function projection of Sec.~\ref{sec:fourpoint_formfactor}.

For one bubble channel, introduce the dimensionless spectral variables
\begin{equation}
	u=\frac{\Gamma^{(2)}_k(q)}{k^2},
	\qquad
	v=\frac{\Gamma^{(2)}_k(q+P)}{k^2},
	\label{eq:uv_internal_arguments}
\end{equation}
so that the bubble term of
\eqref{eq:fourpoint_flow_direct_hierarchy} can be written as
\begin{equation}
	\left(\partial_t\Gamma^{(4)}_k\right)_{\mathrm{bubble}}(P)
	=
	\frac{\hbar}{2}
	\sum_{\mathrm{pairings}}
	\int_\ell
	\mathcal V(\ell,P)\,
	\underbrace{
	\int_0^\infty \dd s\,\rho_k(s)
	\int_0^s \dd\alpha\,
	e^{-(s-\alpha)u-\alpha v}
	}_{\textstyle =:\;\mathcal B(u,v)}\,,
	\label{eq:bubble_host_equation}
\end{equation}
where
$\int_\ell:=\int\dd^4\ell/(2\pi)^4$, and
$\mathcal V(\ell,P)$ denotes the product of the two four-point vertices.
Once the measure, vertices, routing and pairing are fixed, the two
hierarchies differ only in the kernel $\mathcal B(u,v)$.

For the fixed-kernel hierarchy, choosing the internal line with spectral argument $u$ to carry the single-scale insertion gives
\begin{equation}
	\mathcal B_{\mathrm W}(u,v)
	=
	g(u)^2\,s(u)\,g(v)
	=
	g(u)\,W(u)\,g(v).
	\label{eq:KW_general}
\end{equation}
This is the usual structure $G\,\partial_tR\,G$ on one internal line,
together with an ordinary propagator on the other.

For the same orientation, the direct proper-time hierarchy instead gives
\begin{equation}
	\mathcal B_{\mathrm{PT}}(u,v)
	=
	W^{[2]}(u,v,u),
	\qquad
	W^{[2]}(u,v,u)
	:=
	\frac{W(v)-W(u)-(v-u)W'(u)}{(v-u)^2}.
	\label{eq:KPT_general}
\end{equation}
The kernel $\mathcal B_{\mathrm{PT}}$ in \eqref{eq:KPT_general} refers to one orientation, with the line of spectral argument $u$ distinguished. The reversed orientation is obtained by $u\leftrightarrow v$. Under the trace and loop integral the two give equal contributions after relabelling the internal momentum, so one may use $\mathcal B_{\mathrm{PT}}(u,v)=W^{[2]}(u,v,u)$ as the representative kernel. The ordering and normalization are derived explicitly in App.~\ref{app:direct_hierarchy}.

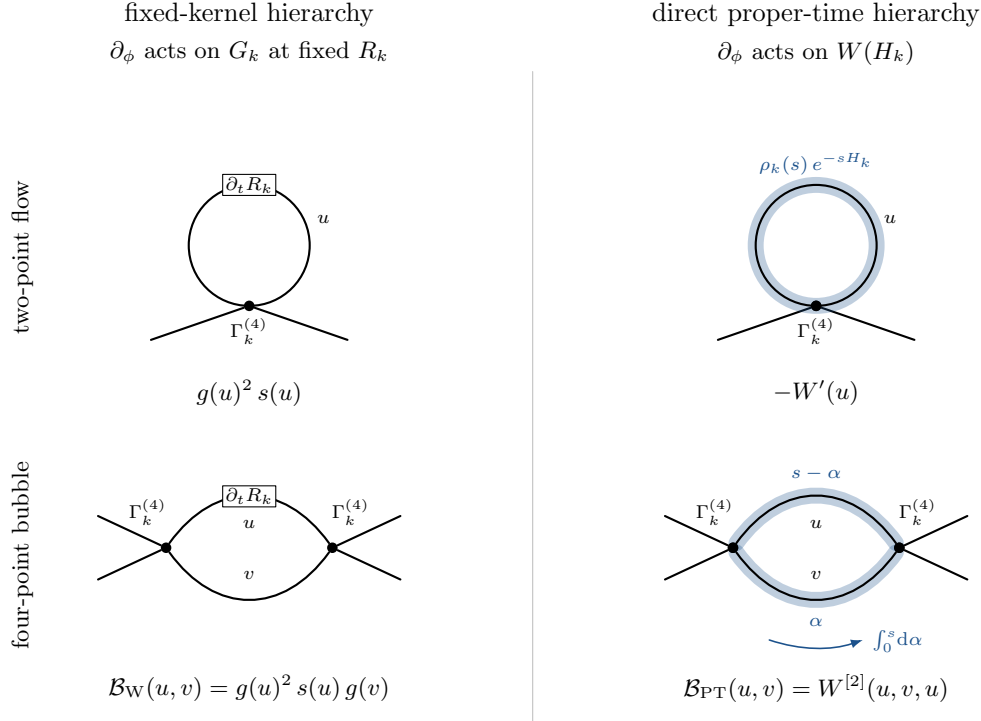
\begin{figure}[t]
\centering
\begin{tikzpicture}[
	line cap=round,
	line join=round,
	every node/.style={font=\normalsize},
	prop/.style={thick},
	ext/.style={thick},
	vtx/.style={circle,fill=black,inner sep=1.4pt},
	ins/.style={rectangle,draw,fill=white,inner sep=1pt,font=\scriptsize},
	band/.style={line width=6pt,sectionblue,opacity=0.28},
	ptlbl/.style={sectionblue,font=\scriptsize},
	title/.style={font=\normalsize},
	subtitle/.style={font=\small},
	kernellbl/.style={font=\small},
	rowlbl/.style={font=\small,rotate=90,anchor=center},
	speclbl/.style={font=\scriptsize}
]

\node[title]    at (0,7.05) {fixed-kernel hierarchy};
\node[subtitle] at (0,6.55) {$\partial_\phi$ acts on $G_k$ at fixed $R_k$};
\node[title]    at (7.5,7.05) {direct proper-time hierarchy};
\node[subtitle] at (7.5,6.55) {$\partial_\phi$ acts on $W(H_k)$};

\draw[black!25,line width=0.4pt] (3.75,6.3) -- (3.75,-2.3);

\node[rowlbl] at (-3.0,3.85) {two-point flow};
\node[rowlbl] at (-3.0,-0.1) {four-point bubble};

\begin{scope}[shift={(0,0)}]
	\node[vtx] (V) at (0,3.2) {};
	\draw[ext] (-1.3,2.75) -- (V) -- (1.3,2.75);
	\draw[prop] (0,4.0) circle[radius=0.8];
	\node[ins] at (0,4.8) {$\partial_tR_k$};
	\node[speclbl] at (0.98,4.35) {$u$};
	\node[below,inner sep=2pt] at (0,3.12) {\scriptsize $\Gamma_k^{(4)}$};
	\node[kernellbl] at (0,2.05) {$g(u)^2\,s(u)$};
\end{scope}

\begin{scope}[shift={(7.5,0)}]
	\node[vtx] (V) at (0,3.2) {};
	\draw[ext] (-1.3,2.75) -- (V) -- (1.3,2.75);
	\draw[band] (0,4.0) circle[radius=0.8];
	\draw[prop] (0,4.0) circle[radius=0.8];
	\node[vtx] at (0,3.2) {};
	\node[ptlbl] at (0,5.08) {$\rho_k(s)\,e^{-sH_k}$};
	\node[speclbl] at (0.98,4.35) {$u$};
	\node[below,inner sep=2pt] at (0,3.12) {\scriptsize $\Gamma_k^{(4)}$};
	\node[kernellbl] at (0,2.05) {$-W'(u)$};
\end{scope}

\begin{scope}[shift={(0,0)}]
	\node[vtx] (L1) at (-1.1,0) {};
	\node[vtx] (L2) at ( 1.1,0) {};
	\draw[prop] (L1) .. controls (-0.45,0.9) and (0.45,0.9) .. (L2);
	\draw[prop] (L1) .. controls (-0.45,-0.9) and (0.45,-0.9) .. (L2);
	\draw[ext] (-2.0,0.42) -- (L1);
	\draw[ext] (-2.0,-0.42) -- (L1);
	\draw[ext] (L2) -- (2.0,0.42);
	\draw[ext] (L2) -- (2.0,-0.42);
	\node[above left,inner sep=1pt]  at (-1.05,0.26) {\scriptsize $\Gamma_k^{(4)}$};
	\node[above right,inner sep=1pt] at (1.05,0.26)  {\scriptsize $\Gamma_k^{(4)}$};
	\node[ins] at (0,0.68) {$\partial_tR_k$};
	\node[speclbl] at (0,0.32)  {$u$};
	\node[speclbl] at (0,-0.36) {$v$};
	\node[kernellbl] at (0,-1.85) {$\mathcal B_{\mathrm W}(u,v)=g(u)^2\,s(u)\,g(v)$};
\end{scope}

\begin{scope}[shift={(7.5,0)}]
	\node[vtx] (R1) at (-1.1,0) {};
	\node[vtx] (R2) at ( 1.1,0) {};
	\draw[band] (R1) .. controls (-0.45,0.9) and (0.45,0.9) .. (R2);
	\draw[band] (R1) .. controls (-0.45,-0.9) and (0.45,-0.9) .. (R2);
	\draw[prop] (R1) .. controls (-0.45,0.9) and (0.45,0.9) .. (R2);
	\draw[prop] (R1) .. controls (-0.45,-0.9) and (0.45,-0.9) .. (R2);
	\draw[ext] (-2.0,0.42) -- (R1);
	\draw[ext] (-2.0,-0.42) -- (R1);
	\draw[ext] (R2) -- (2.0,0.42);
	\draw[ext] (R2) -- (2.0,-0.42);
	\node[vtx] at (-1.1,0) {};
	\node[vtx] at ( 1.1,0) {};
	\node[above left,inner sep=1pt]  at (-1.05,0.26) {\scriptsize $\Gamma_k^{(4)}$};
	\node[above right,inner sep=1pt] at (1.05,0.26)  {\scriptsize $\Gamma_k^{(4)}$};
	\node[ptlbl] at (0,0.98)  {$s-\alpha$};
	\node[ptlbl] at (0,-0.98) {$\alpha$};
	\node[speclbl] at (0,0.32)  {$u$};
	\node[speclbl] at (0,-0.36) {$v$};
	\draw[sectionblue,-{Latex[length=1.6mm]},line width=0.6pt]
		(-0.62,-1.22) .. controls (-0.2,-1.36) and (0.2,-1.36) .. (0.62,-1.22);
	\node[ptlbl,right] at (0.62,-1.24) {$\int_0^s\!\dd\alpha$};
	\node[kernellbl] at (0,-1.85) {$\mathcal B_{\mathrm{PT}}(u,v)=W^{[2]}(u,v,u)$};
\end{scope}

\end{tikzpicture}
\caption{Comparison of the two vertex hierarchies generated by the same
spectral kernel $W$, for the two-point flow (top) and for one channel and
orientation of the four-point bubble (bottom); $u$ and $v$ are the spectral
arguments of the internal lines. In the fixed-kernel hierarchy (left) the field
derivatives act on $G_k$ at fixed $R_k$, and the scale derivative stays on one
internal line. In the direct proper-time hierarchy (right), the derivatives act on the operator function $W(H_k)$. The blue shading indicates the proper-time evolution around the loop, with the two vertices dividing it into intervals of length $s-\alpha$ and $\alpha$ in the bubble diagram.. The two-point
kernels already differ at the symmetric point, but the difference is momentum
independent and does not enter the wave-function projection. The first
relevant difference is the bubble kernel, $\mathcal B_{\mathrm W}\neq\mathcal B_{\mathrm{PT}}$.}
\label{fig:kernel_comparison}
\end{figure}

Thus the two constructions start from the same one-point spectral kernel
$W$ and generate the same bubble topology, but they assign different
kernels to it:
\begin{equation}
	\mathcal B_{\mathrm W}(u,v)\neq \mathcal B_{\mathrm{PT}}(u,v)
\end{equation}
in general. This difference is what enters the two-loop calculation. Figure~\ref{fig:kernel_comparison} shows where the scale derivative acts in the two hierarchies.

 We will see
in Sec.~\ref{sec:general_completion} how the missing terms restore the
fixed-kernel result.

\section{Completion and universality}
\label{sec:general_completion}
\label{sec:completion}

We now show what has to be added to a reduced field-dependent flow, in
what sense the addition is complete, and how the three contributions
recombine into the fixed-kernel result.

\subsection{The three contributions at two loops}
\label{sec:three_contributions}

We consider a generic theory with a regulator that depends on the running Hessian. The exact pulled-back flow is given by \eqref{eq:exact_pullback_flow}; cf.~\cite{Wetterich:2024ivi}.

For a spectral regulator $R_k=R_k(H_k)$, with $H_k=\Gamma_k^{(2)}$, its scale
derivative has an explicit and an implicit part,
\begin{equation}
 \partial_tR_k(H_k)=S_k+U_k\!\cdot\!\partial_tH_k,
 \qquad
 S_k=(\partial_tR_k)_{H_k},\qquad U_k=DR_k[H_k].
 \label{eq:general_R_split}
\end{equation}
The first term, $S_k$ (not to be confused with the action $S$), changes the scale while holding $H_k$ fixed. The second is the
change induced because $H_k$ itself runs. The symbol $DR_k[H_k]$ is the
Fr\'echet derivative \cite{Daletskii:1965}: it is the linear response of the operator function
$R_k(H)$ to a small operator variation $X$,
\begin{equation}
 R_k(H+\epsilon X)=R_k(H)+\epsilon\,DR_k[H]\!\cdot\!X+o(\epsilon).
 \label{eq:frechet_definition}
\end{equation}
In an eigenbasis of $H$, this response is expressed through a divided
difference,
\begin{equation}
 \bigl(DF[H]\!\cdot\!X\bigr)_{ab}
 =F^{[1]}(\lambda_a,\lambda_b)X_{ab},
 \qquad
 F^{[1]}(u,v)=\frac{F(u)-F(v)}{u-v},
 \label{eq:frechet_divided_difference}
\end{equation}
with $F^{[1]}(u,u)=F'(u)$.

We expand
\begin{equation}
 \Gamma_k=S+\hbar\Gamma_{1,k}+\hbar^2\Gamma_{2,k}+\cdots,
 \qquad Q_k=\hbar Q_1+O(\hbar^2),
\end{equation}
and the two Hessians \eqref{eq:hat_hessian} and
\eqref{eq:reduced_hessian},
\begin{equation}
	\Gamma^{(2)}_k
	=
	H_0+\hbar H_1+O(\hbar^2),
	\qquad
	\widehat\Gamma^{(2)}_k
	=
	H_0+\hbar\widehat H_1+O(\hbar^2),
	\qquad
	H_0=S^{(2)},
	\label{eq:Hessian_loop_expansion}
\end{equation}
They share the leading term $H_0$: by \eqref{eq:Q_chain_rule} they differ by
terms in which a field derivative of the section multiplies a $K$-derivative
of $\Gamma$, which is a connected correlation function by
\eqref{eq:K_derivative_identity} and hence $O(\hbar)$, although the section
itself moves already at tree level. With $G_0=(H_0+R_k(H_0))^{-1}$, the
two-loop flow contains three pieces,
\begin{equation}
 \partial_t\Gamma_{2,k}
 =\frac12 D\Tr[GS]_{H_0}\!\cdot H_1
 +\frac12\Tr[G_0\,U_0\!\cdot\!\partial_tH_1]
 -\frac12\Tr[G_0Q_1G_0S_0].
 \label{eq:general_three_piece}
\end{equation}
\label{eq:twoloop_flow_three_terms}
The first is the two-loop part of the reduced flow. The second is the implicit scale term. The
third is the vertical term generated by $Q_1$. All three are of the same loop
order. Keeping only the first is therefore not a consistent two-loop
approximation.

With \eqref{eq:Hessian_loop_expansion}, the definition
\eqref{eq:Q_definition} of the mismatch tensor becomes a relation between the
one-loop Hessians,
\begin{equation}
	Q_1=\widehat H_1-H_1 .
	\label{eq:Q1_as_Hessian_difference}
\end{equation}
In practice, $Q_1$ is obtained by comparing the two vertex hierarchies of Sec.~\ref{sec:proper_time_vertex_hierarchies}, which determine the corresponding Hessians. In the scalar application, these are the one-loop four-point form factors of Sec.~\ref{sec:scalar_setup}. Their difference gives $Q_1$ in the crossed channels relevant for the projection (Sec.~\ref{appdetail:subsec:vertical_mismatch_two_loop}), and at the kernel level it is the difference between the fixed-kernel and along-section constructions. Both are determined by $W$: the along-section construction directly, the fixed-kernel one through the profile $r$ reconstructed from \eqref{eq:W_as_gs_preview}.

Using \eqref{eq:Q1_as_Hessian_difference} in
\eqref{eq:general_three_piece}, the vertical contribution can be split as
\begin{equation}
-\frac12\Tr[G_0Q_1G_0S_0]
=
-\frac12\Tr[G_0\widehat H_1G_0S_0]
+\frac12\Tr[G_0H_1G_0S_0].
\end{equation}
The two-loop flow therefore takes the form
\begin{equation}
 \partial_t\Gamma_{2,k}
 =
 -\frac12\Tr[G_0\widehat H_1G_0S_0]
 +\Delta_k,
 \label{eq:twoloop_flow_W_plus_Delta}
\end{equation}
where
\begin{equation}
 \Delta_k
 =
 \frac12D\Tr[GS]_{H_0}\!\cdot H_1
 +\frac12\Tr[G_0H_1G_0S_0]
 +\frac12\Tr[G_0U_0\!\cdot\!\partial_tH_1].
 \label{eq:Delta_definition_two_loop}
\end{equation}
Here the first and third terms in $\Delta_k$ are the reduced and implicit
contributions, while the middle term comes from the vertical term.

The Fr\'echet-derivative algebra of App.~\ref{app:frechet_cyclicity} gives
\begin{equation}
 \Delta_k
 =
 \frac12\partial_t
 \Tr[G_0\,U_0\!\cdot\!H_1].
 \label{eq:Delta_total_derivative}
\end{equation}
Hence
\begin{equation}
 \partial_t\Gamma_{2,k}
 =
 -\frac12\Tr[G_0\widehat H_1G_0S_0]
 +\frac12\partial_t
 \Tr[G_0\,DR_k[H_0]\!\cdot\!H_1].
 \label{eq:twoloop_final_structural}
\end{equation}
The first term is the usual fixed-kernel two-loop flow; the total scale
derivative can be absorbed into a redefinition of the two-loop effective
action. Defining
\begin{equation}
 \widetilde\Gamma_{2,k}
 :=
 \Gamma_{2,k}
 -\frac12\Tr[G_0\,DR_k[H_0]\!\cdot\!H_1],
 \label{eq:Gamma2_shifted_definition}
\end{equation}
one obtains
\begin{equation}
 \partial_t\widetilde\Gamma_{2,k}
 =
 -\frac12\Tr[G_0\widehat H_1G_0S_0].
 \label{eq:general_fixed_kernel_relation}
\end{equation}
This is the general completion of the reduced field-dependent flow.

\subsection{The kernel identity}
\label{sec:bivariate_identity}
\label{sec:general_universality}

The same completion can also be expressed in kernel form. Let $x$ be a dimensionless
spectral variable and let $r(x)$ be any admissible regulator profile. Define
\begin{equation}
 g(x):=\frac{1}{x+r(x)},\qquad
 s(x):=2r(x)-2x r'(x),\qquad
 W(x):=g(x)s(x),\qquad
 h(x):=g(x)^2s(x).
 \label{eq:kernel_definitions_general}
\end{equation}
Here $g$ is the regulated propagator profile, $s$ is its explicit scale
insertion, $W$ is the one-point flow kernel, and $h$ is the corresponding
single-scale outer kernel.
The profile $r$ is arbitrary and $W$ is defined from it, so the identity below
holds for any $r$. In Sec.~\ref{sec:proper_time_vertex_hierarchies} and in
the scalar application the logic is reversed: $W$ is given and
\eqref{eq:W_as_gs_preview} is an equation for $r$.

These definitions imply the pointwise identity
\begin{equation}
 \frac{\dd}{\dd x}\left[W(x)+2x\,g(x)r'(x)\right]=-h(x).
 \label{eq:kernel_identity_differential}
\end{equation}
Equivalently,
\begin{equation}
 W(x)+2x\,g(x)r'(x)+\int_0^x\dd x'\,h(x')=W(0)=2,
 \label{eq:kernel_identity_integrated}
\end{equation}
for $r(0)\neq0$ and $r'(0)$ finite. This is the radial form of the outer-kernel reorganization above. After differentiation, it relates the reduced kernel $W'$ and the implicit-scale kernel to the single-scale kernel $h=g^2s$. The vertical contribution, which acts on the inserted vertex rather than on the outer kernel, is considered separately below.

To obtain an explicit coefficient, the outer trace must reduce to a single radial
integral over $x$. This happens whenever the external projection leaves the outer loop with
a rotationally invariant measure, which is the case for the wave-function
projection used in Sec.~\ref{sec:fourpoint_formfactor} and, more generally, for projections that admit the required radial representation. We show here that in that situation the reduced and implicit contributions
combine by \eqref{eq:kernel_identity_differential} alone, and the vertical
contribution completes the fixed-kernel coefficient.

We now write the projected inserted vertex as a single function $A(x)$. Its
definition depends on the theory and on the observable; in the scalar
example $A=1-I-xI'$, with $I$ the one-loop four-point
source. The reduced and implicit contributions
of \eqref{eq:general_three_piece}, and the fixed-kernel reference, then differ
only in the kernel carried by the outer loop. The reduced spectral trace contributes
\begin{equation}
	C_{\mathrm{red}}
	=
	-\frac14\int_0^\infty\dd x\;W'(x)\,A(x),
	\label{eq:Cred_general_radial}
\end{equation}
the implicit scale term contributes
\begin{equation}
	C_{\mathrm{impl}}
	=
	\frac12\int_0^\infty\dd x\;
	\bigl[x\,g(x)r'(x)\bigr]\,A'(x),
	\label{eq:Cimpl_general_radial}
\end{equation}
and the fixed-kernel reference carries the single-scale outer kernel
$h=g^2s$,
\begin{equation}
	C[h;A]
	:=
	\frac14\int_0^\infty\dd x\;h(x)\,A(x).
	\label{eq:C_h_A_general}
\end{equation}

The first two now combine. Integrating \eqref{eq:Cimpl_general_radial} by
parts and using the kernel identity in the form
\begin{equation}
	\bigl[x\,g(x)r'(x)\bigr]'
	=
	-\tfrac12\bigl[h(x)+W'(x)\bigr],
	\label{eq:kernel_identity_derivative_form}
\end{equation}
which is \eqref{eq:kernel_identity_differential} rewritten, gives
\begin{equation}
	C_{\mathrm{red}}+C_{\mathrm{impl}}
	=
	C[h;A].
	\label{eq:Cred_plus_Cimpl_general}
\end{equation}
The reduced and implicit outer kernels have merged into the fixed-kernel outer
kernel, acting on the \emph{same} inserted vertex.

The inserted vertex, however, is not yet the fixed-kernel one. $A$ was generated along the
section, whereas the fixed-kernel reference requires its horizontal
counterpart $\widehat A$. The vertical term supplies that replacement: since
$Q_1=\widehat H_1-H_1$ is the difference between the fixed-kernel and
along-section one-loop vertices, inserting it into the same outer loop gives
\begin{equation}
	C_{\mathrm{vert}}
	=
	C[h;\widehat A]-C[h;A]
	=
	C_{\mathrm W}-C[h;A].
	\label{eq:Cvert_general_radial}
\end{equation}
Adding the three contributions, the intermediate object $C[h;A]$ cancels and
\begin{equation}
		C_{\mathrm{red}}+C_{\mathrm{impl}}+C_{\mathrm{vert}}
		=
		C_{\mathrm W}
	\label{eq:sum_rule_general}
\end{equation}
Equation \eqref{eq:sum_rule_general} is the kernel form of the completion
and one of the main results of this work.
The identity implies recombination with the fixed-kernel result for each reconstructed regulator profile $r$. At fixed kernel the flow reproduces perturbation theory with regulated propagators (App.~\ref{app:standard_twoloop}), and for a well-behaved regulator the $k$ dependence of the renormalized wave-function coefficient enters only through the ultraviolet logarithm, whose coefficient does not depend on $r$. The fixed-kernel coefficient is therefore universal, and by \eqref{eq:sum_rule_general} so is the completed one. The relation $C_{\mathrm{red}}+C_{\mathrm{impl}}=C[h;A]$ also assumes that the boundary term generated by the radial integration by parts vanishes. With $b=xgr'$, this requires $bA\to0$ at both endpoints. Finally, neither \eqref{eq:Cimpl_general_radial} nor \eqref{eq:Cvert_general_radial} can be assumed to give a small correction. Determining their size is therefore important for assessing the quantitative effect of the completion.

In the second part of the paper we construct $A$ explicitly for massless
$\lambda\phi^4$ and the $m$-family, evaluate the three coefficients for
$m=1,\ldots,8$, and check \eqref{eq:sum_rule_general} numerically.

\section{\texorpdfstring{Application: The two-loop anomalous dimension in massless $\lambda\phi^4$}{The two-loop anomalous dimension in massless lambda phi^4}}
\label{sec:scalar_setup}

Section~\ref{sec:general_completion} showed that the three contributions
recombine into the fixed-kernel coefficient. To see how large each of them is,
and how far the reduced flow deviates without them, we need a theory with an
independently known answer, a regulator family that can be deformed, and an
observable sensitive to the difference between the two hierarchies.

We use the universal two-loop coefficient of $\eta$ in massless
$\lambda\phi^4$ theory as the benchmark for this test. It is the first
quantity in this theory that responds to how the four-point vertex is
generated rather than to the one-loop trace alone: the two hierarchies of
Sec.~\ref{sec:proper_time_vertex_hierarchies} differ in the kernel they
assign to the four-point bubble, and it is the momentum dependence of that
bubble that the projection onto $p^2$ measures.

We first compute the two-loop coefficient in each hierarchy and exhibit their
difference. We then evaluate the implicit and vertical completion terms, verify
the sum rule \eqref{eq:sum_rule_general}, and compare the reduced flow with
other proper-time prescriptions.

We consider a single real, massless scalar field in four Euclidean
dimensions with scale dependent action
\begin{equation}
	\Gamma_k[\phi]
	=
	\int \dd^4x\,
	\left[
Z_k	\frac12 (\partial_\mu\phi)^2
	+
	\frac{\lambda_k}{4!}\phi^4
	\right].
	\label{eq:scalar_action}
\end{equation}
At tree level,
\begin{equation}
	\Gamma_k^{(2)}(p)=Z_kp^2,
	\qquad
	\Gamma_k^{(4)}=\lambda_k,
\end{equation}
while all odd vertices vanish and $\Gamma_k^{(6)}=0$.

We parametrize the momentum-dependent part of the two-point function as
\begin{equation}
	\Gamma_k^{(2)}(p)
	=
	Z_k p^2+\cdots ,
\end{equation}
and define the anomalous dimension by
\begin{equation}
	\eta_k=-\partial_t\ln Z_k .
	\label{eq:eta_definition}
\end{equation}
Its leading coefficient $C_\eta$ is defined by
\begin{equation}
	\eta_k
	=
	\frac{\lambda_k^2}{(16\pi^2)^2}\,C_\eta
	+O(\lambda_k^3),
	\qquad
	C_\eta^{\mathrm{univ}}=\frac16 .
	\label{eq:Ceta_normalization}
\end{equation}
The coefficient $C_\eta$ is obtained by projecting the two-point flow onto its $p^2$ term, which determines the flow of $Z_k$.

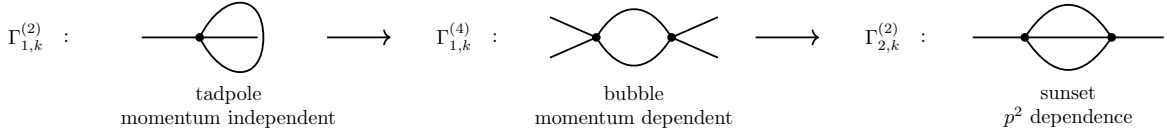
\begin{figure}[t]
	\centering
	\resizebox{\textwidth}{!}{
		\begin{tikzpicture}[
			scale=0.9,
			every node/.style={transform shape},
			prop/.style={thick},
			ext/.style={thick},
			vtx/.style={circle,fill=black,inner sep=1.5pt}
			]

			\node at (0,0) {$\Gamma_{1,k}^{(2)}$};
			\node at (0.7,0) {$:$};
			
			\begin{scope}[shift={(3.0,0)}]
				\node[vtx] at (0,0) {};
				
				\draw[ext] (-1.0,0) -- (0,0);
				\draw[ext] (0,0) -- (1.0,0);
				
				\draw[prop]
				(0,0)
				.. controls (0.4,0.8) and (1.1,0.8) ..
				(1.1,0)
				.. controls (1.1,-0.8) and (0.4,-0.8) ..
				(0,0);
			\end{scope}
			
			\node[align=center] at (3.5,-1.2)
			{ tadpole\\
				 momentum independent};

			\draw[->,thick] (5.2,0) -- (6.3,0);

			\node at (7.4,0) {$\Gamma_{1,k}^{(4)}$};
			\node at (8.1,0) {$:$};
			
			\begin{scope}[shift={(10.5,0)}]
				\node[vtx] at (-0.65,0) {};
				\node[vtx] at (0.65,0) {};
				
				\draw[prop]
				(-0.65,0)
				.. controls (-0.25,0.65) and (0.25,0.65) ..
				(0.65,0);
				
				\draw[prop]
				(-0.65,0)
				.. controls (-0.25,-0.65) and (0.25,-0.65) ..
				(0.65,0);
				
				\draw[ext] (-1.45,0.35) -- (-0.65,0);
				\draw[ext] (-1.45,-0.35) -- (-0.65,0);
				
				\draw[ext] (0.65,0) -- (1.45,0.35);
				\draw[ext] (0.65,0) -- (1.45,-0.35);
			\end{scope}
			
			\node[align=center] at (10.5,-1.2)
			{ bubble\\
				 momentum dependent};

			\draw[->,thick] (12.6,0) -- (13.7,0);

			\node at (14.8,0) {$\Gamma_{2,k}^{(2)}$};
			\node at (15.5,0) {$:$};
			
			\begin{scope}[shift={(18.0,0)}]
				\node[vtx] at (-0.75,0) {};
				\node[vtx] at (0.75,0) {};
				
				\draw[ext] (-1.65,0) -- (-0.75,0);
				\draw[ext] (0.75,0) -- (1.65,0);
				
				\draw[prop]
				(-0.75,0)
				.. controls (-0.25,0.75) and (0.25,0.75) ..
				(0.75,0);
				
				\draw[prop]
				(-0.75,0)
				.. controls (-0.25,0) and (0.25,0) ..
				(0.75,0);
				
				\draw[prop]
				(-0.75,0)
				.. controls (-0.25,-0.75) and (0.25,-0.75) ..
				(0.75,0);
			\end{scope}
			
			\node[align=center] at (18.0,-1.2)
			{ sunset\\
				 $p^2$ dependence};
			
		\end{tikzpicture}
	}
	
	\caption{
		Perturbative origin of the two-loop anomalous dimension in massless
		$\lambda\phi^4$ theory. The one-loop two-point flow gives a
		momentum-independent tadpole. Momentum dependence first appears in the
		one-loop four-point bubble. Reinserting this form factor into the two-point
		flow generates the two-loop sunset topology responsible for the wave-function
		renormalization.
	}
	\label{fig:PT_twoloop_eta}
\end{figure}

We expand the vertices in powers of $\hbar$,
\begin{equation}
	\Gamma_k^{(2)}
	=
	S^{(2)}
	+\hbar\,\Gamma^{(2)}_{1,k}
	+\hbar^2\,\Gamma^{(2)}_{2,k}
	+O(\hbar^3).
	\label{eq:loop_expansion_two_point}
\end{equation}
Similarly,
\begin{equation}
	\Gamma_k^{(4)}
	=
	S^{(4)}
	+\hbar\,\Gamma^{(4)}_{1,k}
	+\hbar^2\,\Gamma^{(4)}_{2,k}
	+O(\hbar^3).
	\label{eq:loop_expansion_four_point}
\end{equation}
The mechanism is the same in both hierarchies. At one loop the two-point flow is a tadpole: the four-point
vertex is a constant, so the loop carries no external momentum and the result
has no $p^2$ piece. This renormalizes the mass. Momentum
dependence enters one order later, through the one-loop four-point bubble,
which is a  function of the momentum flowing through the channel.
Feeding that momentum-dependent vertex back into the two-point flow produces
the two-loop sunset diagram, and the $p^2$ term of the sunset is what $\eta$
measures. Figure~\ref{fig:PT_twoloop_eta} shows the three steps.

\subsection{The regulator family and its reconstruction}
\label{subsec:typeC_family}
To test the dependence on the regulator we use a standard one-parameter family of spectrally adjusted cutoff
profiles, labelled by a shape parameter $m$ and known in the proper-time
literature as type C \cite{Liao:1992fm,Bonanno:2000yp,Mazza:2001bp}. We call
it the \emph{$m$-family} below. The parameter $m$ sets the shape of the
proper-time kernel and interpolates between two familiar limits: $m=1$ is the mass cutoff underlying the Callan--Symanzik flow \cite{Callan:1970yg,Symanzik:1970rt,Alexandre:2000eg}, and
$m\to\infty$ is the exponential kernel with a Wilsonian reading. In
dimensionful form, we choose the kernel such that
	\begin{equation}
		W_{k,m}(z)
		=
		2\left(
		\frac{m k^2}{z+m k^2}
		\right)^m ,
		\qquad m>0.
		\label{eq:Wkm_dimensionful}
	\end{equation}
Equivalently, the profile $W_{k,m}$ admits the proper-time representation
\begin{equation}
	W_{k,m}(z)
	=
	\int_0^\infty \dd s\,
	\omega_{k,m}(s)\,e^{-sz},
	\label{eq:Wkm_laplace}
\end{equation}
with weight
\begin{equation}
	\omega_{k,m}(s)
	=
	2\,\frac{(m k^2)^m}{\Gamma(m)}\,
	s^{m-1}e^{-s m k^2}.
	\label{eq:typeC_weight}
\end{equation}
Thus the corresponding $\rho$ weight is positive,
\begin{equation}
	\rho_{k,m}(s)
	=
	2\frac{(m k^2)^m}{\Gamma(m)}
	s^m e^{-s m k^2}.
	\label{eq:typeC_rho_weight}
\end{equation}
Using
\begin{equation}
	\int_0^\infty \dd s\,
	s^{m-1}e^{-s(z+m k^2)}
	=
	\frac{\Gamma(m)}{(z+m k^2)^m},
\end{equation}
one recovers \eqref{eq:Wkm_dimensionful}; this form shows how $m$ shapes
the distribution of proper times.

For the calculations below it is more convenient to work with the dimensionless spectral variable
\begin{equation}
	u=\frac{z}{k^2},
	\label{eq:u_dimensionless}
\end{equation}
in terms of which the family reduces to
\begin{equation}
	W_m(u)
	=
	2\left(
	\frac{m}{u+m}
	\right)^m .
	\label{eq:Wm_typeC}
\end{equation}
The dependence on the proper-time profile is therefore encoded entirely in the single shape parameter $m$, and the subsequent flow equations can be written directly in terms of $W_m$ and its derivatives.

We use $m=1,2,3$ for the explicit calculations and extend the evaluation to $m\le8$ in Sec.~\ref{subsec:explicit_checks_and_numerics}. For the first three,
\begin{align}
	W_1(u) &= \frac{2}{u+1},
	\label{eq:W1_typeC}\\
	W_2(u) &= \frac{8}{(u+2)^2},\\
	W_3(u) &= \frac{54}{(u+3)^3}.
\end{align}
The case $m=1$ serves as a control: the reconstructed cutoff is a
field-independent mass term, so $Q$ and the implicit correction vanish
and the two hierarchies coincide. The cases $m=2$ and $m=3$ are the two
smallest nondegenerate integers and admit closed-form reconstructed
profiles.

The along-section hierarchy requires only the spectral profile. The
fixed-kernel hierarchy of Sec.~\ref{sec:vertex_expansion_fixed_kernel}
requires in addition an external regulator profile $r$, carrying no field
dependence, whose ordinary trace reproduces the same $W_m$. It follows from the
reconstruction condition \eqref{eq:W_as_gs_preview},
$W=gs$, specialized to the $m$-family. Written out, it gives
\begin{equation}
	\frac{
		r_m(u)-u r_m'(u)
	}{
		u+r_m(u)
	}
	=
	\left(
	\frac{m}{u+m}
	\right)^m .
	\label{eq:rm_ode_mixed_simplified}
\end{equation}

Equation \eqref{eq:rm_ode_mixed_simplified} is a first-order ordinary
differential equation for $r_m$. It is
solved subject to a boundary condition, fixed by requiring $r_m$ to
regulate the infrared in the usual way, $r_m(u)\to$ const as $u\to0$, and to
switch off in the ultraviolet. The solutions for $m=2$ and $m=3$ are worked
out in Sec.~\ref{app:typeC_checks}.

Two properties of this construction are used below. First, a solution
need not exist for every profile; only where an external $r$ reproduces
$W$ is there a Wetterich flow to compare with.
Second, once obtained, $r_m$ is held fixed throughout the horizontal
hierarchy. Functional derivatives act on vertices and propagators and never on
$r_m$ itself, which is why $Q$ vanishes for that hierarchy.

\subsection{From the one-loop four-point vertex to the anomalous dimension}
\label{sec:fourpoint_formfactor}

The momentum dependence relevant for $\eta$ first appears in the one-loop
four-point vertex. For a single channel with momentum $P=p_i+p_j$ and
$y=P^2/k^2$, the contact term vanishes since $S^{(6)}=0$, while the bubble gives
\begin{equation}
    \partial_t\Gamma^{(4)}_{1,k,\mathrm{1ch}}(P^2)
    =
    \frac{\lambda_k^2}{2(4\pi)^2}
    \int_0^1 \dd a\,
    W_m\!\left(a(1-a)y\right).
    \label{eq:oneloop_fourpoint_Wm}
\end{equation}
The proper-time evaluation leading to \eqref{eq:oneloop_fourpoint_Wm} is given
in App.~\ref{app:proper_time_fourpoint}.

Using the profile \eqref{eq:Wm_typeC}, we define
\begin{equation}
    I_m(y)
    :=
    \int_0^1 \dd a\,
    \left[
        1+\frac{a(1-a)}{m}y
    \right]^{-m},
    \label{eq:Im_from_flow_new}
\end{equation}
so that
\begin{equation}
    \partial_t\Gamma^{(4)}_{1,k,\mathrm{1ch}}(P^2)
    =
    \frac{\lambda_k^2}{(4\pi)^2}I_m(y).
    \label{eq:fourpoint_flow_Im_final}
\end{equation}
At zero channel momentum $I_m(0)=1$ for every $m$, and summing the three
channels gives
\begin{equation}
    \partial_t\lambda_k
    =
    3\,\frac{\lambda_k^2}{(4\pi)^2},
    \label{eq:betalambda_check_section6}
\end{equation}
the universal one-loop coefficient. This does not constrain the momentum
dependence of the four-point vertex: the profile dependence is confined to the nonlocal part $I_m(y)-I_m(0)$, which enters the two-loop
wave-function projection.

To obtain the two-loop wave-function flow, the momentum-dependent part of the
one-loop four-point vertex must first be integrated in the RG scale. For one
channel we write
\begin{equation}
    \Gamma^{(4)}_{1,k,\mathrm{1ch}}(P^2)
    =
    \Gamma^{(4)}_{1,k,\mathrm{1ch}}(0)
    +
    \frac{\lambda_k^2}{(4\pi)^2}\,f_m(y),
    \qquad
    y=\frac{P^2}{k^2}.
    \label{eq:one_channel_formfactor_definition}
\end{equation}
The local term contributes to the running of $\lambda_k$, whereas $f_m$
contains the momentum dependence relevant for $\eta$. Integrating the
subtracted one-loop flow at fixed physical momentum gives
\begin{equation}
    f_m'(y)
    =
    \frac{1-I_m(y)}{2y}.
    \label{eq:fprime_m_from_Im}
\end{equation}
The derivation, including the subtraction convention, is given in
App.~\ref{app:angular_projection}.

The integrated one-loop vertex is then inserted into the outer two-point
flow. At two-loop order its momentum-dependent part is
\begin{equation}
    \left(\partial_t\Gamma^{(2)}_{2,k}(p)\right)_{\mathrm{PT}}
    =
    -\frac12
    \int_0^\infty \dd s\,\rho_k(s)
    \int\frac{\dd^4q}{(2\pi)^4}\,
    e^{-sq^2}
    \Gamma^{(4)}_{1,k}(p,-p,q,-q)
    +\cdots .
    \label{eq:twoloop_twopoint_integrated_fourpoint}
\end{equation}
The omitted terms are momentum independent in the massless theory and do not
contribute to $\eta$. Of the three channels of the four-point subdiagram, the
zero-momentum channel is local, while the two crossed channels give equal
contributions. Choosing $P=p+q$ and writing $x=q^2/k^2$ and $y=p^2/k^2$, the
wave-function projection uses
\begin{equation}
    \left.
    \frac{\partial}{\partial y}
    \avg{
        F\!\left(x+y+2\sqrt{xy}\,c\right)
    }
    \right|_{y=0}
    =
    F'(x)+\frac{x}{2}F''(x).
    \label{eq:angular_projection_identity_new}
\end{equation}
The channel routing and the angular expansion are collected in
App.~\ref{app:angular_projection}.

Combining \eqref{eq:fprime_m_from_Im} with this projection gives the form used
throughout the remainder of the calculation,
\begin{equation}
    C[\mathcal B]
    =
    \frac14
    \int_0^\infty \dd x\,\kappa_0(x)
    \left[
        1-I(x)-xI'(x)
    \right].
    \label{eq:C_general_I_new}
\end{equation}
Here $I(x)$ is the one-loop four-point source generated by the bubble kernel
$\mathcal B$, while $\kappa_0(x)$ is the kernel carried by the outer two-point loop. A
general representation of $I$ in terms of $\mathcal B$ is given in
App.~\ref{app:bubble_kernel_representation}. For the fixed-kernel hierarchy,
$\kappa_0=h=g^2s$. The relation of \eqref{eq:C_general_I_new} to the standard
two-loop sunset contribution is shown in App.~\ref{app:standard_twoloop}.

\subsection{Two-loop coefficients from the two hierarchies}
\label{sec:two_loop_coefficients}

We now specialize the two kernels of Sec.~\ref{sec:hierarchy_difference} to
the $m$-family and insert them into \eqref{eq:C_general_I_new}. The
diagrammatic organization is the same in both hierarchies; the difference is
the kernel assigned to the one-loop four-point bubble and to the outer loop.

For the direct proper-time hierarchy, the bubble kernel is
\begin{equation}
    \mathcal B_{\mathrm{PT}}^{(m)}(u,v)
    =
    W_m^{[2]}(u,v,u),
    \label{eq:KPT_direct_coefficient}
\end{equation}
as obtained in Sec.~\ref{sec:loop_hierarchy_direct_PT}. Together with the
one-loop source $I_m$ of \eqref{eq:Im_from_flow_new}, the projection gives
\begin{equation}
    C_{\mathrm{PT}}^{(m)}
    =
    \frac12
    \int_0^\infty \dd x\,
    \left(
        \frac{m}{m+x}
    \right)^{m+1}
    \left[
        1-I_m(x)-xI_m'(x)
    \right].
    \label{eq:CPT_general}
\end{equation}
For representative values,
\begin{equation}
    C_{\mathrm{PT}}^{(1)}=\frac16,
    \qquad
    C_{\mathrm{PT}}^{(2)}\simeq0.150405,
    \qquad
    C_{\mathrm{PT}}^{(\infty)}\simeq0.127836.
    \label{eq:CPT_minfinity}
\end{equation}
Thus the reduced direct hierarchy agrees with the universal coefficient at
the mass-cutoff point $m=1$, but becomes profile dependent for $m>1$.

We now move to the fixed-kernel  hierarchy.
For the regulator reconstructed from the same $W_m$, the fixed-kernel bubble
is instead
\begin{equation}
    \mathcal B_{\mathrm W}^{(m)}(u,v)
    =
    g_m(u)^2s_m(u)g_m(v)
    =
    g_m(u)W_m(u)g_m(v).
    \label{eq:KW_mixed}
\end{equation}
Defining $h_m(u):=g_m(u)^2s_m(u)$, the corresponding one-loop source is
\begin{equation}
    I_{\mathrm W}^{(m)}(y)
    =
    \int_0^\infty \dd x\,x\,h_m(x)\,
    \avg{
        g_m\!\left(
        x+y+2\sqrt{xy}\,c
        \right)
    }.
    \label{eq:IW_mixed}
\end{equation}
Its normalization $I_{\mathrm W}^{(m)}(0)=1$ is shown in
App.~\ref{app:bubble_kernel_representation}. Equation \eqref{eq:C_general_I_new} then gives
\begin{equation}
    C_{\mathrm W}^{(m)}
    =
    \frac14
    \int_0^\infty \dd x\,h_m(x)
    \left[
        1-I_{\mathrm W}^{(m)}(x)
        -x\frac{\dd I_{\mathrm W}^{(m)}}{\dd x}
    \right].
    \label{eq:CW_mixed_Iprime}
\end{equation}
For the reconstructed profiles $m=2$ and $m=3$,
\begin{equation}
    C_{\mathrm W}^{(2)}
    =
    C_{\mathrm W}^{(3)}
    =
    \frac16
\end{equation}
within numerical accuracy. The profile-specific kernels and intermediate
integrations are collected in App.~\ref{app:typeC_checks}.

The two constructions agree at the level of the undifferentiated
one-loop trace and reproduce the same local one-loop running of the quartic
coupling. They differ only in the generation of nonlocal vertices. Although
both kernels are built from the same profile $W_m$, for $m>1$ they lead to
different two-loop coefficients: the direct proper-time hierarchy is profile
dependent, whereas the reconstructed fixed-kernel hierarchy gives the
universal value $1/6$. At $m=1$, where the reconstructed cutoff is field
independent, the two kernels coincide; the explicit check is given in
App.~\ref{app:typeC_checks}.

\subsection{Completion of the reduced flow}
\label{sec:restoring_universality}

The fixed-kernel result provides the horizontal reference for the completion.
We now evaluate, for the same scalar theory and $m$-family, the two terms
not included in the reduced spectral flow: the implicit scale contribution and the
vertical contribution generated by the mismatch tensor $Q$. This realizes the
general argument of Sec.~\ref{sec:general_universality} in the concrete
two-loop projection and also determines the size of the individual pieces.

\subsubsection*{The implicit scale term}

	The implicit-scale term in \eqref{eq:twoloop_flow_three_terms} is
	\begin{equation}
		\left(\partial_t\Gamma_{2,k}\right)_{\mathrm{impl}}
		=
		\frac12
		\Tr\left[
		G_0\,U_0\cdot\partial_tH_1
		\right].
		\label{eq:Timpl_definition}
	\end{equation}
	At the symmetric point of the massless theory the odd vertices vanish, so
	the momentum-dependent part of its two-point projection is obtained by
	letting both external derivatives act on $\partial_tH_1$. The insertion is
	then diagonal in the loop momentum, so the Fr\'echet derivative $U_0$
	reduces to the ordinary derivative of the profile,
	\begin{equation}
		U_0\big|_{\mathrm{diag}}
		=
		r_m'(x),
		\qquad
		x=\frac{q^2}{k^2},
		\label{eq:U0_diagonal}
	\end{equation}
	and the outer kernel of the implicit term is
	\begin{equation}
		\kappa_{\mathrm{impl}}(x)
		=
		x\,g_m(x)\,r_m'(x).
		\label{eq:kimpl_definition}
	\end{equation}
	The inserted object is the scale derivative of the one-loop four-point
	function at fixed momenta. In the scaling decomposition used in
	Sec.~\ref{sec:fourpoint_formfactor}, the local part is the running of
	$\lambda_k$ and does not contribute to the wave-function projection, while
	each momentum-carrying channel contributes the subtracted flow form factor
	\begin{equation}
		\partial_t\big|_{P}\,f_{\mathrm{PT}}^{(m)}\!\left(\frac{P^2}{k^2}\right)
		=
		-2y\,\frac{\dd f_{\mathrm{PT}}^{(m)}}{\dd y}
		=
		I_m(y)-1 .
		\label{eq:flow_formfactor_insertion}
	\end{equation}
	Applying the projection identity
	\eqref{eq:angular_projection_identity_new} to $\Phi(y)=I_m(y)-1$, and
	using that \eqref{eq:Timpl_definition} enters the two-loop flow with the
	opposite sign relative to the reference structure
	$-\frac12\Tr[G\,\Gamma^{(4)}_1G\,\partial_tR]$ that defines the
	normalization of \eqref{eq:C_general_I_new}, one obtains
	\begin{equation}
		C_{\mathrm{impl}}^{(m)}
		=
		-
		\int_0^\infty \dd x\;
		x\,g_m(x)\,r_m'(x)
		\left[
		I_m'(x)+\frac{x}{2}\,I_m''(x)
		\right].
		\label{eq:Cimpl_formula}
	\end{equation}
	Introducing
	\begin{equation}
		A_m(x)
		:=
		1-I_m(x)-x\,I_m'(x)
		=
		\frac{\dd}{\dd x}
		\left[
		x\left(1-I_m(x)\right)
		\right],
		\qquad
		I_m'+\frac{x}{2}I_m''=-\frac12A_m',
		\label{eq:A_definition}
	\end{equation}
	this can be written compactly as
	\begin{equation}
		C_{\mathrm{impl}}^{(m)}
		=
		\frac12
		\int_0^\infty \dd x\;
		x\,g_m(x)\,r_m'(x)\,A_m'(x).
		\label{eq:Cimpl_Aform}
	\end{equation}
	For $m=1$ the reconstructed profile is the constant mass cutoff,
	$r_1(u)=1$, so $r_1'=0$ and
	\begin{equation}
		C_{\mathrm{impl}}^{(1)}=0
		\label{eq:Cimpl_m1}
	\end{equation}
	identically: the Callan--Symanzik-type cutoff carries no implicit scale
	dependence.

\subsubsection*{The vertical mismatch term}
\label{appdetail:subsec:vertical_mismatch_two_loop}

The vertical contribution was obtained in
Sec.~\ref{sec:three_contributions}. At two loops it is
\begin{equation}
    \left(\partial_t\Gamma_{2,k}\right)_{\mathrm{vert}}
    =
    -\frac12
    \Tr\left[
        G_0\,Q_1\,G_0\,S_0
    \right],
    \label{eq:vertical_two_loop_explicit}
\end{equation}
where, from \eqref{eq:Q1_as_Hessian_difference},
$
Q_1=\widehat H_1-H_1
$.
For the scalar two-point projection, the outer single-scale kernel is
therefore the fixed-kernel one,
\begin{equation}
    h_m(x)=g_m(x)^2s_m(x),
\end{equation}
while the difference lies in the inserted one-loop vertex.

Its four-point part is
\begin{equation}
    Q_1^{(4)}
    =
    \widehat\Gamma^{(4)}_{1,k}
    -
    \Gamma^{(4)}_{1,k},
    \label{eq:Q1_fourpoint_difference}
\end{equation}
namely the difference between the fixed-kernel and along-section one-loop
four-point functions. Strictly speaking, two further field derivatives of the
two-point mismatch do not coincide identically with the four-point mismatch.
At the symmetric point,
\begin{equation}
    \left(Q^{(2)}_{AB}\right)_{,CD}
    =
    Q^{(4)}_{ABCD}
    +\frac{\delta\Gamma_{AB}}{\delta K_{EF}}\,R_{k,EF,CD},
\end{equation}
since the terms with a single derivative of the section vanish there. The
additional term contains the classical insertion at coincident points and
carries zero channel momentum. It therefore drops from the $p^2$ projection,
so the two definitions agree for the crossed channels relevant below.

The local part of the mismatch is momentum independent and drops out of the
projection. For each momentum-carrying channel, the remaining difference is
encoded in the two subtracted form factors,
\begin{equation}
    D_m(y)
    :=
    f_{\mathrm W}^{(m)}(y)
    -
    f_{\mathrm{PT}}^{(m)}(y),
    \qquad
    D_m'(y)
    =
    \frac{I_m(y)-I_{\mathrm W}^{(m)}(y)}{2y},
    \qquad
    D_m(0)=0,
    \label{eq:D_difference_formfactor}
\end{equation}
where $I_{\mathrm W}^{(m)}$ is the fixed-kernel one-loop source
\eqref{eq:IW_mixed}. Since the two hierarchies share the same universal
logarithm, $D_m$ is analytic at $y=0$. For the reconstructed $m=2$ and $m=3$
profiles it approaches a constant rather than zero at large $y$; in the
integration by parts below the complete boundary term nevertheless vanishes,
because the source difference is weighted by the decaying outer kernel.

The projection identity gives
\begin{equation}
    C_{\mathrm{vert}}^{(m)}
    =
    \frac14
    \int_0^\infty \dd x\;
    h_m(x)
    \left[
        \left(I_m-I_{\mathrm W}^{(m)}\right)(x)
        +
        x\left(I_m-I_{\mathrm W}^{(m)}\right)'(x)
    \right],
    \label{eq:Cvert_formula}
\end{equation}
or, after integration by parts,
\begin{equation}
    C_{\mathrm{vert}}^{(m)}
    =
    -\frac14
    \int_0^\infty \dd x\;
    x h_m'(x)
    \left[
        I_m(x)-I_{\mathrm W}^{(m)}(x)
    \right].
    \label{eq:Cvert_byparts}
\end{equation}
Thus the vertical contribution measures the difference between inserting the
along-section and fixed-kernel one-loop form factors into the same Wetterich
outer loop. At $m=1$ the two hierarchies coincide,
$I_{\mathrm W}^{(1)}=I_1$, and therefore $C_{\mathrm{vert}}^{(1)}=0$.

\subsubsection*{Recombination of the three contributions}

The two completion terms combine with the reduced contribution exactly as
predicted by the kernel identity. Using the notation $C[h;A]$ introduced in
Sec.~\ref{sec:bivariate_identity}, specialized here to $h_m$ and $A_m$, one has
\begin{equation}
    C_{\mathrm{PT}}^{(m)}+C_{\mathrm{impl}}^{(m)}
    =
    C[h_m;A_m],
    \label{eq:CPT_plus_Cimpl}
\end{equation}
while the vertical contribution replaces the along-section one-loop vertex by
its fixed-kernel counterpart,
\begin{equation}
    C_{\mathrm{vert}}^{(m)}
    =
    C_{\mathrm W}^{(m)}-C[h_m;A_m].
    \label{eq:Cvert_as_difference}
\end{equation}
Therefore
\begin{equation}
    C_{\mathrm{PT}}^{(m)}
    +C_{\mathrm{impl}}^{(m)}
    +C_{\mathrm{vert}}^{(m)}
    =
    C_{\mathrm W}^{(m)}.
    \label{eq:sum_rule_exact}
\end{equation}
This is \eqref{eq:sum_rule_general} realized in the scalar theory: the outer
kernels are those of the $m$-family and the three coefficients recombine
exactly as the general argument requires.

\subsection{Explicit check of the sum rule}
\label{subsec:explicit_checks_and_numerics}

We now verify the kernel identity and the completed coefficient explicitly across
several nontrivial members of the $m$-family.

The recombination test does not by itself determine the  proper-time one-loop source $I_m$,
since the same source also enters the vertical contribution. The fixed-kernel
coefficient $C_{\mathrm W}$ and the $m=1$ case provide consistency checks on
the calculation, but they do not independently determine the mismatch.
An independent computation of $Q_1$ is given in
App.~\ref{app:determinant_Q1}.

\paragraph{\texorpdfstring{Profile $m=2$.}{Profile m=2.}}
For
\begin{equation}
    W_2(u)=\frac{8}{(u+2)^2},
\end{equation}
the reconstructed regulator is
\begin{equation}
    r_2(u)
    =
    (u+2)\exp\!\left[-\frac{2}{u+2}\right]-u .
    \label{eq:r2_solution}
\end{equation}
The coefficients for this profile are collected in
\eqref{eq:decomposition_table}.

\paragraph{\texorpdfstring{Profile $m=3$.}{Profile m=3.}}
\label{sec:m3_reconstruction}

For
\begin{equation}
    W_3(u)=\frac{54}{(u+3)^3},
\end{equation}
the reconstructed regulator is
\begin{equation}
    r_3(u)
    =
    (u+3)
    \exp\!\left[-\frac{3(2u+9)}{2(u+3)^2}\right]-u .
    \label{eq:r3_solution}
\end{equation}
The coefficients for this profile are also collected in
\eqref{eq:decomposition_table}.

\paragraph{Completed two-loop coefficient.}
	\label{subsec:profile_independent_coefficient}

	Evaluating \eqref{eq:Cimpl_formula} and \eqref{eq:Cvert_byparts} for the
	reconstructed profiles gives
	\begin{equation}
		\begin{array}{c|ccc|c}
			m &
			C_{\mathrm{PT}}^{(m)} &
			C_{\mathrm{impl}}^{(m)} &
			C_{\mathrm{vert}}^{(m)} &
			\text{sum}
			\\[0.3em]
			\hline
			\rule{0pt}{1.1em}
			1 & 1/6 & 0 & 0 & 1/6
			\\
			2 & 0.150405 & -0.026829 & \phantom{-}0.043091 & 0.166667
			\\
			3 & 0.143635 & -0.033516 & \phantom{-}0.056548 & 0.166667
			\\
			{4} & {0.139959} & {-0.036228} & {\phantom{-}0.062936} & {0.166667}
			\\
			{5} & {0.137659} & {-0.037628} & {\phantom{-}0.066636} & {0.166667}
			\\
			{6} & {0.136088} & {-0.038464} & {\phantom{-}0.069042} & {0.166667}
			\\
			{7} & {0.134949} & {-0.039011} & {\phantom{-}0.070729} & {0.166667}
			\\
			{8} & {0.134084} & {-0.039395} & {\phantom{-}0.071977} & {0.166667}
		\end{array}
		\label{eq:decomposition_table}
	\end{equation}
	in agreement with \eqref{eq:sum_rule_exact} to the numerical accuracy of the
	quadratures. We evaluate the radial momentum integral over $x=q^2/k^2$ using 216 Gauss--Legendre integration points. At each radial point, the angular integral over $c=\cos\theta$ is evaluated using 80 Gauss--Gegenbauer integration points, with the four-dimensional angular weight $\sqrt{1-c^2}$.  The total agrees with $1/6$ to a
	residual below $10^{-11}$ for both $m=2$ and $m=3$, and below $10^{-13}$ for
	$m=4,\ldots,8$, and the algebraic
	identity $C_{\mathrm{PT}}^{(m)}+C_{\mathrm{impl}}^{(m)}=C[h_m;A_m]$
	underlying \eqref{eq:CPT_plus_Cimpl} holds to better than $10^{-13}$;
	convergence was checked across multiple radial orders and both residuals
	are consistent with pure quadrature error. For $m>3$ we use the closed
	form of the reconstruction valid for every integer $m$,
	$u+r_m(u)=(u+m)\exp[-\sum_{j=2}^{m}m^{j-1}/((j-1)(u+m)^{j-1})]$, which
	reduces to \eqref{eq:r3_solution} for $m=3$.

    The implicit contribution is negative and the vertical contribution
positive. For every profile considered, both are larger in absolute
value than their sum and therefore partially cancel. Including only
one of them gives a larger error than the reduced flow itself.
The net correction $1/6-C_{\mathrm{PT}}^{(m)}$ is $0.016262$ for $m=2$,
$0.023032$ for $m=3$, and $0.032583$ for $m=8$.
At $m=1$, the regulator reduces to a field-independent mass cutoff
and both corrections vanish. Their magnitudes increase over the
sampled values $m=1,\ldots,8$, while $C_{\mathrm{PT}}^{(m)}$
decreases towards the limiting value
$C_{\mathrm{PT}}^{(\infty)}$ in \eqref{eq:CPT_minfinity}.
This trend is observed numerically and does not establish
monotonicity for arbitrary $m$. We next compare this result with the local-vertex prescription of
Ref.~\cite{Giacometti:2025qyy}.

\subsection{Comparison of different prescriptions}
\label{sec:other_prescriptions}

\label{subsec:local_projection_k2s}

The comparison above used the full momentum dependence of the one-loop
four-point form factor when inserting it into the outer two-point flow.
For the reduced proper-time equation, however, this is not the only possible
projection. One may instead expand the inserted four-point vertex locally in
the channel momentum before carrying out the outer proper-time integral.

This alternative is useful because it is closely related to the prescription
used in Ref.~\cite{Giacometti:2025qyy}. In the local projection, the
momentum-dependent one-loop four-point vertex is replaced by its slope at
vanishing channel momentum. For the $m$-family this gives
\begin{equation}
	C_{\mathrm{loc}}^{(m)}
	=
	\frac{m}{6(m-1)},
	\qquad
	m>1,
	\label{eq:Cloc_result}
\end{equation}
and therefore
\begin{equation}
	C_{\mathrm{loc}}^{(m)}
	\longrightarrow
	\frac16
	\qquad
	(m\to\infty).
	\label{eq:Cloc_minfinity}
\end{equation}
The local projection thus reproduces $1/6$ at the exponential endpoint, whereas
the full momentum-dependent projection derived above reproduces the universal
coefficient at the Callan--Symanzik point $m=1$.

The origin of this behaviour is seen in the proper-time representation. The normalized measure $\mu_m$ (defined in App.~\ref{app:local_projection}) becomes sharply
localized as $m$ increases,
\begin{equation}
	\langle s\rangle_{\mu_m}
	=
	\frac{1}{k^2},
	\qquad
	\frac{\mathrm{Var}_{\mu_m}(s)}
	{\langle s\rangle_{\mu_m}^2}
	=
	\frac{1}{m}.
	\label{eq:gamma_measure_moments_main}
\end{equation}
Hence, in the limit $m\to\infty$, the proper time selected by the outer
kernel collapses onto
\begin{equation}
	s=\frac{1}{k^2}.
	\label{eq:k2_s_limit_main}
\end{equation}
Only in this limit can the scale $k$ appearing in the locally expanded
inner vertex be identified unambiguously with the proper-time scale carried
by the outer loop.

The two projections therefore display complementary dependence on the shape
parameter:
\begin{equation}
	\begin{array}{c|cccc}
		& m=1 & m=2 & m=3 & m=\infty
		\\[0.2em]
		\hline
		C_{\mathrm{PT}}^{(m)}\ \text{(full)}
		& 1/6 & 0.1504 & 0.1436 & 0.1278
		\\
		C_{\mathrm{loc}}^{(m)}\ \text{(local)}
		& \text{---} & 1/3 & 1/4 & 1/6
	\end{array}
	\label{eq:projection_comparison_table}
\end{equation}
The full projection retains the complete channel-momentum dependence of the
one-loop bubble, whereas the local projection discards this information
before the outer integration is performed.

The two coefficients in \eqref{eq:projection_comparison_table} are
projections of the same reduced hierarchy, so both neglect the implicit and the
vertical term. At the exponential endpoint $C_{\mathrm{loc}}^{(m)}$ reaches $1/6$ while
$C_{\mathrm{PT}}^{(m)}$, which keeps the channel dependence that the local
projection discards, does not. The information discarded by the local
projection is therefore numerically equal, at that endpoint, to the terms the
reduced hierarchy does not consider. Away from it the two quantities are unrelated:
\begin{equation}
	\begin{array}{c|ccc}
		& m=2 & m=3 & m\to\infty
		\\[0.2em]
		\hline
		\rule{0pt}{1.1em}
		C_{\mathrm W}-C_{\mathrm{PT}}^{(m)}
		\ \text{(vertical and implicit contributions)}
		& 0.0163 & 0.0230 & 0.0388
		\\
		C_{\mathrm{loc}}^{(m)}-C_{\mathrm{PT}}^{(m)}
		\ \text{(local-vertex approximation)}
		& 0.1829 & 0.1064 & 0.0388
	\end{array}
	\label{eq:two_errors_comparison}
\end{equation}
They differ by an order of magnitude at $m=2$ and coincide only in the
limit. The agreement of the local projection with the universal coefficient
at the exponential endpoint thus results from a cancellation between the
 completion terms and the discarded channel dependence; it does not indicate
that the local projection becomes exact there. The statement concerns $C_{\mathrm{loc}}^{(m)}$ as constructed in
App.~\ref{app:local_projection}.

This projection dependence is another indication that the reduced
proper-time equation does not by itself determine a unique higher-loop
hierarchy. Different ways of iterating the same one-loop trace lead
to different finite-$m$ coefficients, even though particular prescriptions
may reproduce the universal result at special endpoints of the regulator
family. The detailed derivation of \eqref{eq:Cloc_result}, together with the
proper-time interpretation of the identification $s=1/k^2$, is given in
App.~\ref{app:local_projection}.

Relative to $C_{\mathrm W}=1/6$, the reduced proper-time coefficient is
smaller by $9.8\%$ for $m=2$, $13.8\%$ for $m=3$, and $23.3\%$ as
$m\to\infty$; see \eqref{eq:projection_comparison_table}. These deviations
are specific to massless $\lambda\phi^4$ and the wave-function projection
used here. For $m>1$, the size of the completion terms depends on the theory
and observable.

\section{Conclusions}
\label{sec:conclusions}

A flow equation whose cutoff depends on the running inverse propagator is
not, as it stands, complete at two loops. In this work we have shown
explicitly that two additional terms must be included: the vertical term
generated by the mismatch tensor $Q$, and the implicit scale dependence that
the cutoff inherits from the running Hessian.

Writing the effective action as
$\Gamma[\phi,K]$ with the cutoff kernel as an independent argument shows
that a regulator prescription is a choice of curve $K=R_k[\phi]$ in
the space of pairs $(\phi,K)$, and that a field derivative can be taken either at fixed
$K$ or along the curve. The two disagree by the mismatch tensor $Q$. Since $Q$
starts at order $\hbar$, one-loop results are unaffected, but two-loop results
are not.

A proper-time representation fixes the one-loop trace but does not uniquely determine the hierarchy of vertex flows.  Freezing a reconstructed
regulator and differentiating propagators gives the ordinary single-scale
structure. Differentiating the spectral function directly gives divided
differences instead. Both reproduce the same one-loop trace at
the symmetric configuration $\phi=0$ and the same one-loop running of $\lambda$; they
assign different kernels to the internal lines of the one-loop bubble, and
the anomalous dimension $\eta$ is sensitive to this difference.

We show this for the two-loop anomalous dimension in massless $\lambda\phi^4$ theory. The direct proper-time hierarchy reproduces the one-loop running of
$\lambda$   but  then yields a two-loop coefficient that depends
on the shape parameter of the regulator family: $0.150405$ at $m=2$ and
$0.143635$ at $m=3$ instead of $1/6$. For every $m$ considered, $m=1,\ldots,8$, the fixed-kernel hierarchy, using the regulator $r_m$ reconstructed from the corresponding spectral profile $W_m$, gives the universal coefficient $1/6$. Adding the implicit scale and vertical contributions to the reduced proper-time result recovers the same value.

The two corrections are individually non-negligible and have opposite signs. As shown in Sec.~\ref{subsec:explicit_checks_and_numerics}, including either correction alone gives a larger deviation from $1/6$ than the reduced flow without corrections, for every value of $m$ considered. Both terms are therefore needed to recover the universal coefficient.  The sum rule \eqref{eq:sum_rule_general} is not restricted to the scalar theory. It follows from the kernel identity of Sec.~\ref{sec:bivariate_identity} for any admissible regulator profile, provided the projection has the required radial form and the boundary terms vanish. The completed coefficient therefore coincides with the fixed-kernel one and is universal. The calculations for $m=1,\ldots,8$ verify this explicitly.

It is important to emphasize that agreement with the universal coefficient at a given order does not necessarily verify every step of the calculation. The sum rule confirms that the three contributions recombine, but it does not independently check the one-loop four-point form factor, which cancels between the reduced and vertical terms (Sec.~\ref{subsec:explicit_checks_and_numerics}). Similarly, the local proper-time projection reproduces $1/6$ at $m\to\infty$ because its approximation compensates for the omitted contributions (Sec.~\ref{sec:other_prescriptions}).

The exact bilocal formulation provides a framework for organizing the
contributions induced by a field-dependent regulator order by order in the
loop expansion. Extending the explicit two-loop completion developed here
to higher orders requires the corresponding higher-loop kernel responses
and a consistent treatment of the implicit spectral section. We leave this
extension for future work.

Finally, in gauge theories, the background-field formulation introduces an additional complication. In the background-field formulation the identification of the background with the mean field is a section in the sense of Sec.~\ref{sec:bilocal_flow}, and the cutoff becomes field dependent. The part of the mismatch generated by the regulator is again of order $\hbar$. At one loop the flow therefore integrates to the regulated one-loop determinant evaluated on the section, whatever the field dependence of $R_k$, and the one-loop coefficient of the beta function is reproduced by every such prescription. At this order the prescriptions differ only in non-universal quantities: at finite $k$ the one-loop Hessians differ by $Q_1$, as the one-loop form factors of Sec.~\ref{sec:restoring_universality} illustrate in the scalar case, while the coefficients of the ultraviolet logarithms and the limit $k\to0$ coincide. The agreement of the one-loop beta function is thus not a test of how the field dependence of the regulator is treated. The first coefficient that is sensitive to it is the two-loop one, for which a manifestly gauge-invariant calculation is available \cite{Morris:2005tv}. In a gauge theory the omission may then produce a  discrepancy in the two-loop beta function, or reside in quantities that no universality protects. The operator decomposition of Sec.~\ref{sec:three_contributions} applies there as well; extending the recombination argument of Sec.~\ref{sec:general_universality} requires a suitable regulator reconstruction and a projection onto gauge-theory observables.

\section*{Acknowledgements}
We are grateful to Kevin Falls for many discussions in the early stages of
this work and for a careful critical reading of the manuscript, whose
comments improved both the arguments and their presentation. We thank Oleg
Melichev for suggesting the comparison with the standard two-loop effective
action of App.~\ref{app:standard_twoloop}.

Large language models (ChatGPT and Claude) were used to assist with the
revision of the text and with the checking of symbolic algebra and numerical
code. All equations, derivations and numerical results were verified by the
authors, who take full responsibility for the content.

\appendix

\section{Vertex hierarchy of the direct proper-time flow}
\label{app:direct_hierarchy}

This appendix collects the operator algebra behind
\eqref{eq:twopoint_flow_direct_hierarchy} and
\eqref{eq:fourpoint_flow_direct_hierarchy}. The derivation does not depend on the details of the theory or the cutoff profile. It uses only cyclicity of the trace and the vanishing of odd vertices at the $\mathbb Z_2$-symmetric point. Throughout this appendix we set $k=1$, so that $W_k=W$ and the eigenvalues $u,v$ are the spectral variables of Sec.~\ref{sec:hierarchy_difference}.

Taking two functional derivatives of the exponential and evaluating at the symmetric point gives
\begin{equation}
	\left.
	\frac{\delta^2 e^{-sH_k[\phi]}}
	{\delta\phi^A\,\delta\phi^B}
	\right|_{\phi=0}
	=
	-
	\int_0^s \dd\alpha\,
	e^{-(s-\alpha)H_k[0]}
	H_{k,AB}\big|_{\phi=0}
	e^{-\alpha H_k[0]} .
	\label{eq:second_variation_exp_Z2}
\end{equation}
Terms containing two insertions of $H_{k,A}$ and $H_{k,B}$ vanish at this point because of \eqref{eq:odd_vertices_vanish}. By cyclicity of the trace, the $\alpha$-integration produces a factor $s$, which cancels the factor $1/s$ in the measure of \eqref{eq:PT_flow_rho}. The two-point flow therefore becomes
\begin{equation}
\left(\partial_t\Gamma^{(2)}_{k,AB}\right)_{\mathrm{PT}}
	=
	-\frac{\hbar}{2}
	\int_0^\infty \dd s\,\rho_k(s)\,
	\Tr\left[
	e^{-sH_k[0]}
	H_{k,AB}\big|_{\phi=0}
	\right],
	\label{eq:twopoint_flow_appendix}
\end{equation}
which is \eqref{eq:twopoint_flow_direct_hierarchy}.

For the bubble, the second Duhamel formula gives two ordered insertion
integrals. Combining them by cyclicity of the trace yields
\begin{equation}
	\begin{split}
		&\int_0^s \dd\alpha_1
		\int_0^{\alpha_1}\dd\alpha_2\,
		\Tr\left[\cdots\right]
		+
		(AB)\leftrightarrow(CD)
		\\
		&\qquad =
		s
		\int_0^s \dd\alpha\,
		\Tr\left[
		e^{-(s-\alpha)H_k[0]}
		H_{k,AB}\big|_{\phi=0}
		e^{-\alpha H_k[0]}
		H_{k,CD}\big|_{\phi=0}
		\right].
	\end{split}
	\label{eq:duhamel_orderings_combined}
\end{equation}
The resulting factor $s$ cancels the factor $1/s$ in the proper-time measure
of \eqref{eq:PT_flow_rho}.

The combined expression must be distinguished from the individual
ordered terms. Consider one ordering, with insertions $X$ and $Y$,
and eigenstates
\begin{equation}
    H_k[0]|n\rangle=u|n\rangle,
    \qquad
    H_k[0]|n'\rangle=v|n'\rangle.
\end{equation}
The coefficient of $X_{nn'}Y_{n'n}$ is
\begin{equation}
    I_2(s;u,v)
    =
    \int_0^s\dd s_1
    \int_0^{s-s_1}\dd s_2\,
    e^{-s_1u-s_2v-(s-s_1-s_2)u}.
    \label{eq:ordered_three_segment_simplex}
\end{equation}
The three intervals add up to $s$. Writing $\Delta=v-u$,
integration gives
\begin{equation}
    I_2(s;u,v)
    =
    \frac{s e^{-su}}{\Delta}
    -
    \frac{e^{-su}-e^{-sv}}{\Delta^2}.
    \label{eq:ordered_three_segment_integral}
\end{equation}

For this ordering we retain the original proper-time measure
$\rho_k(s)\,\dd s/s$. From the definition of $W$,
\begin{equation}
    \begin{aligned}
    W'(H)
    &=
    -\int_0^\infty\dd s\,\rho_k(s)e^{-sH},
    \\
    M_0(H)
    &:=
    \int_0^\infty\dd s\,\rho_k(s)e^{-sH}
    =
    -W'(H).
    \end{aligned}
    \label{eq:proper_time_M0_identity}
\end{equation}
It follows that
\begin{equation}
    \begin{aligned}
    \int_0^\infty\frac{\dd s}{s}\,
    \rho_k(s)I_2(s;u,v)
    &=
    -\frac{W'(u)}{\Delta}
    -
    \frac{W(u)-W(v)}{\Delta^2}
    \\
    &=
    \frac{W(v)-W(u)-\Delta W'(u)}{\Delta^2}
    =
    W^{[2]}(u,v,u).
    \end{aligned}
    \label{eq:ordered_simplex_to_W2}
\end{equation}
The reversed ordering gives $W^{[2]}(v,u,v)$.

Combining the orderings as in \eqref{eq:duhamel_orderings_combined}
gives instead
\begin{equation}
    \begin{aligned}
    \int_0^\infty\dd s\,\rho_k(s)
    \int_0^s\dd\alpha\,
    e^{-(s-\alpha)u-\alpha v}
    &=
    \frac{W'(v)-W'(u)}{v-u}
    \\
    &=
    W^{[2]}(u,v,u)+W^{[2]}(v,u,v).
    \end{aligned}
    \label{eq:combined_simplex_symmetric_kernel}
\end{equation}

For an unordered pairing, this sum carries the factor $1/2$
in \eqref{eq:fourpoint_flow_direct_hierarchy}. Under the trace
and loop integral, the two terms give equal contributions
after exchanging the internal lines. Hence
\begin{equation}
    \begin{aligned}
    \int \mathcal B_{\mathrm{pair}}
    &=
    \frac12\int\!
    \left[
        W^{[2]}(u,v,u)+W^{[2]}(v,u,v)
    \right]\\
    &=\int W^{[2]}(u,v,u).
    \end{aligned}
    \label{eq:oriented_pairing_equivalence}
\end{equation}
The fixed-kernel half-sum of $g(u)^2s(u)g(v)$ and
$g(v)^2s(v)g(u)$ likewise gives $g(u)^2s(u)g(v)$ after integration.

Only when the insertions commute with $H_k[0]$ can the
combined proper-time expression be reduced further to
$s e^{-sH_k[0]}XY$, with measure $\rho_k(s)s\,\dd s$.
The oriented representative in \eqref{eq:KPT_general} is therefore
justified under the trace and loop integral, not by discarding an ordering
at the operator level.

\section{Fr\'echet-derivative algebra for the completion theorem}
\label{app:frechet_cyclicity}
We derive \eqref{eq:Delta_total_derivative} from
\eqref{eq:Delta_definition_two_loop},

\begin{equation}
\Delta_k=\frac12 D\,\Tr[GS]_{H_0}\cdot H_1
+\frac12\Tr[G_0H_1G_0S_0]+\frac12\Tr[G_0U_0\cdot\partial_tH_1].
\end{equation}

To evaluate its operator variation, write
	\begin{equation}
		Y(H)=H+R_k(H),
		\qquad
		G(H)=Y(H)^{-1}.
		\label{eq:Y_G_H_functions}
	\end{equation}
	For an arbitrary variation $X$ of $H$,
	\begin{equation}
		DY[H]\cdot X
		=
		X+U[H]\cdot X,
		\label{eq:DY_variation}
	\end{equation}
	where
	\begin{equation}
		U[H]=DR_k[H].
	\end{equation}
	Moreover,
	\begin{equation}
		DG[H]\cdot X
		=
		-G(H)\,
		\left(DY[H]\cdot X\right)
		G(H).
		\label{eq:DG_variation}
	\end{equation}
	Therefore
	\begin{align}
		D(GS)[H]\cdot X
		&=
		\left(DG[H]\cdot X\right)S(H)
		+
		G(H)\left(DS[H]\cdot X\right)
		\nonumber\\
		&=
		-G(H)\,X\,G(H)S(H)
		-
		G(H)\left(U[H]\cdot X\right)G(H)S(H)
		\nonumber\\
		&\quad
		+
		G(H)\left(DS[H]\cdot X\right).
		\label{eq:DGS_variation}
	\end{align}
	Setting $X=H_1$ and evaluating at $H=H_0$, we obtain
	\begin{align}
		D\,\Tr[GS]_{H_0}\cdot H_1
		&=
		\Tr\left[
		-G_0H_1G_0S_0
		-
		G_0\left(U_0\cdot H_1\right)G_0S_0
		+
		G_0\left(DS_0\cdot H_1\right)
		\right].
		\label{eq:DTrGS_explicit}
	\end{align}
	Substitution into \eqref{eq:Delta_definition_two_loop} gives
	\begin{align}
		\Delta_k
		&=
		\frac12\Tr\left[
		-G_0H_1G_0S_0
		-
		G_0\left(U_0\cdot H_1\right)G_0S_0
		+
		G_0\left(DS_0\cdot H_1\right)
		\right]
		\nonumber\\
		&\quad
		+
		\frac12\Tr\left[
		G_0H_1G_0S_0
		\right]
		+
		\frac12\Tr\left[
		G_0U_0\cdot\partial_tH_1
		\right].
		\label{eq:Delta_before_cancellation}
	\end{align}
	The first and fourth terms cancel exactly. Hence
	\begin{equation}
		\Delta_k
		=
		\frac12\Tr\left[
		-G_0\left(U_0\cdot H_1\right)G_0S_0
		+
		G_0\left(DS_0\cdot H_1\right)
		+
		G_0U_0\cdot\partial_tH_1
		\right].
		\label{eq:Delta_after_cancellation}
	\end{equation}

	Now note that
	\begin{equation}
		\left(\partial_tG_0\right)_{\mathrm{expl}}
		=
		-G_0S_0G_0,
		\label{eq:dtG_expl}
	\end{equation}
	because $G_0=(H_0+R_k(H_0))^{-1}$ and the explicit scale derivative of
	$H_0$ vanishes. Also,
	\begin{equation}
		\left(\partial_tU_0\right)_{\mathrm{expl}}
		=
		DS_0.
		\label{eq:dtU_expl}
	\end{equation}
	The second identity follows from
	\begin{equation}
		U_0=DR_k[H_0],
		\qquad
		S_0=\left(\partial_tR_k\right)_{\mathrm{expl}}\big|_{H_0},
	\end{equation}
	and from commuting the explicit scale derivative with the Fr\'echet derivative
	with respect to $H$.

	Using \eqref{eq:dtG_expl} and \eqref{eq:dtU_expl}, we compute
	\begin{align}
		\partial_t
		\Tr\left[
		G_0\,U_0\cdot H_1
		\right]
		&=
		\Tr\left[
		\left(\partial_tG_0\right)_{\mathrm{expl}}
		U_0\cdot H_1
		\right]
		+
		\Tr\left[
		G_0\left(\partial_tU_0\right)_{\mathrm{expl}}\cdot H_1
		\right]
		\nonumber\\
		&\quad
		+
		\Tr\left[
		G_0\,U_0\cdot\partial_tH_1
		\right]
		\nonumber\\
		&=
		\Tr\left[
		-G_0S_0G_0\left(U_0\cdot H_1\right)
		+
		G_0\left(DS_0\cdot H_1\right)
		+
		G_0U_0\cdot\partial_tH_1
		\right].
		\label{eq:dt_total_trace}
	\end{align}
	By cyclicity of the trace,
	\begin{equation}
		\Tr\left[
		-G_0S_0G_0\left(U_0\cdot H_1\right)
		\right]
		=
		\Tr\left[
		-G_0\left(U_0\cdot H_1\right)G_0S_0
		\right].
		\label{eq:cyclic_first_term}
	\end{equation}
	Comparison with \eqref{eq:Delta_after_cancellation} gives
	\begin{equation}
		\Delta_k
		=
		\frac12
		\partial_t
		\Tr\left[
		G_0\,U_0\cdot H_1
		\right].
		\label{app:eq:Delta_total_derivative}
	\end{equation}

\section{Technical details of the scalar two-loop calculation}
\label{app:twoloop_projection}

\subsection{Proper-time evaluation of the one-loop four-point form factor}
\label{app:proper_time_fourpoint}

For one momentum channel, the bubble term of the along-section hierarchy gives
\begin{equation}
    \partial_t\Gamma^{(4)}_{1,k,\mathrm{1ch}}(P^2)
    =
    \frac{\lambda_k^2}{2}
    \int_0^\infty \dd s\,\rho_k(s)
    \int_0^s \dd\alpha
    \int\frac{\dd^4\ell}{(2\pi)^4}\,
    e^{-(s-\alpha)\ell^2-\alpha(\ell+P)^2}.
    \label{eq:oneloop_fourpoint_flow_start}
\end{equation}
The factor $\lambda_k^2$ comes from the two classical four-point vertices; the
contact term proportional to $S^{(6)}$ vanishes for the classical
$\lambda\phi^4$ action. Setting $\alpha=sa$, with $0\leq a\leq1$, gives
\begin{equation}
    (s-\alpha)\ell^2+\alpha(\ell+P)^2
    =
    s\left[(\ell+aP)^2+a(1-a)P^2\right],
    \label{eq:complete_square_fourpoint}
\end{equation}
and the Gaussian integral is
\begin{equation}
    \int\frac{\dd^4\ell}{(2\pi)^4}e^{-s(\ell+aP)^2}
    =
    \frac{1}{(4\pi s)^2}.
    \label{eq:gaussian_loop_integral}
\end{equation}
Since $\dd\alpha=s\,\dd a$, Eq.~\eqref{eq:oneloop_fourpoint_flow_start}
becomes
\begin{equation}
    \partial_t\Gamma^{(4)}_{1,k,\mathrm{1ch}}(P^2)
    =
    \frac{\lambda_k^2}{2(4\pi)^2}
    \int_0^\infty\frac{\dd s}{s}\,\rho_k(s)
    \int_0^1\dd a\,e^{-s a(1-a)P^2}.
    \label{eq:oneloop_fourpoint_after_gaussian}
\end{equation}
With $y=P^2/k^2$ and the definition of $W_m$ in
Eq.~\eqref{eq:Wm_typeC}, the proper-time integral gives
\begin{equation}
    \partial_t\Gamma^{(4)}_{1,k,\mathrm{1ch}}(P^2)
    =
    \frac{\lambda_k^2}{2(4\pi)^2}
    \int_0^1\dd a\,W_m\!\left(a(1-a)y\right),
\end{equation}
which is Eq.~\eqref{eq:oneloop_fourpoint_Wm} of the main text.

\subsection{Form-factor integration and angular projection}
\label{app:angular_projection}

We collect here the steps needed in Sec.~\ref{sec:fourpoint_formfactor}.
Subtracting the zero-momentum part of the one-loop four-point flow gives
\begin{equation}
    \partial_t f_m(y)\big|_{P^2}
    =
    I_m(y)-1.
    \label{eq:subtracted_formfactor_flow_m}
\end{equation}
At fixed physical channel momentum $P^2$, $y=P^2/k^2$ obeys
\begin{equation}
    \partial_t y\big|_{P^2}=-2y,
    \label{eq:dt_y_fixed_P}
\end{equation}
so that
\begin{equation}
    \partial_t f_m(y)\big|_{P^2}
    =
    -2y f_m'(y).
    \label{eq:canonical_scaling_formfactor_new}
\end{equation}
Combining these equations yields \eqref{eq:fprime_m_from_Im}. With the
subtraction convention $f_m(0)=0$,
\begin{equation}
    f_m(y)
    =
    \int_0^y \dd z\,\frac{1-I_m(z)}{2z}.
    \label{eq:fm_integrated}
\end{equation}
This is the integration of the subtracted flow at fixed physical momentum,
rewritten in terms of the dimensionless variable $y$; it is not an integration
to an endpoint of the RG trajectory.

For the external configuration $(p,-p,q,-q)$, the four-point subdiagram has
three channels,
\begin{equation}
    0,\qquad p+q,\qquad p-q.
    \label{eq:channels_ppqq_new}
\end{equation}
The zero-momentum channel is local, while the two crossed channels give equal
contributions after angular integration. Choosing $P=p+q$, define
\begin{equation}
    x=\frac{q^2}{k^2},
    \qquad
    y=\frac{p^2}{k^2},
    \qquad
    c=\cos\theta=\frac{p\cdot q}{|p||q|},
    \label{eq:x_y_c_definitions_new}
\end{equation}
so that
\begin{equation}
    v:=\frac{P^2}{k^2}
    =x+y+2\sqrt{xy}\,c.
    \label{eq:v_from_channel_momentum_new}
\end{equation}
The inserted form factor is therefore $f(v)$.

In four dimensions the normalized angular average is
\begin{equation}
    \avg{A(c)}
    =
    \frac{2}{\pi}
    \int_{-1}^{1}\dd c\,\sqrt{1-c^2}\,A(c).
    \label{eq:S3_average_new}
\end{equation}
Writing $v=x+\delta v$ with
$\delta v=y+2\sqrt{xy}\,c$, and expanding a smooth function $F(v)$ to the
order required for a derivative with respect to $y$, gives
\begin{equation}
    F(v)
    =
    F(x)+F'(x)\,\delta v
    +\frac12F''(x)(\delta v)^2+O(y^{3/2},y^2).
\end{equation}
Using
\begin{equation}
    \avg{c}=0,
    \qquad
    \avg{c^2}=\frac14,
    \label{eq:angular_moments_S3}
\end{equation}
one obtains \eqref{eq:angular_projection_identity_new}.

Before eliminating the integrated form factor, the projected coefficient may
be written as
\begin{equation}
    C[\mathcal B]
    =
    \int_0^\infty \dd x\,x\,\kappa_0(x)
    \left[
        f'(x)+\frac{x}{2}f''(x)
    \right].
    \label{eq:C_general_kernel_new}
\end{equation}
Using $f'(x)=[1-I(x)]/(2x)$ gives
\begin{equation}
    f''(x)
    =
    -\frac{I'(x)}{2x}
    -\frac{1-I(x)}{2x^2},
\end{equation}
and hence
\begin{equation}
    f'(x)+\frac{x}{2}f''(x)
    =
    \frac{1}{4x}\left[1-I(x)-xI'(x)\right].
\end{equation}
Substitution in \eqref{eq:C_general_kernel_new} yields the master formula
\eqref{eq:C_general_I_new} used in the main text.

\subsection{Kernel representation of the one-loop form factor}
\label{app:bubble_kernel_representation}

The same one-loop source used in Sec.~\ref{sec:fourpoint_formfactor} can be
represented directly in terms of the bubble kernel appearing in the vertex
hierarchy. Let
\begin{equation}
    x=\frac{q^2}{k^2},
    \qquad
    y=\frac{P^2}{k^2},
    \qquad
    c=\cos\theta=\frac{P\cdot q}{|P||q|},
    \label{eq:x_y_c_for_kernel_source}
\end{equation}
and
\begin{equation}
    v
    :=
    \frac{(q+P)^2}{k^2}
    =
    x+y+2\sqrt{xy}\,c .
    \label{eq:v_for_kernel_source}
\end{equation}
For a general one-loop bubble kernel $\mathcal B(x,v)$, define
\begin{equation}
    I(y)
    =
    \int_0^\infty \dd x\,x\,
    \avg{
        \mathcal B\!\left(x,x+y+2\sqrt{xy}\,c\right)
    } .
    \label{eq:I_general_kernel_new}
\end{equation}
The normalization is chosen such that
\begin{equation}
    I(0)=1.
    \label{eq:I_general_normalization_new}
\end{equation}
In the direct proper-time hierarchy,
\begin{equation}
    \mathcal B(x,v)
    =
    \mathcal B_{\mathrm{PT}}^{(m)}(x,v)
    =
    W_m^{[2]}(x,v,x).
    \label{eq:KPT_kernel_section6}
\end{equation}
To verify that the kernel representation reproduces $I_m(y)$, we use
the Schwinger representation
\begin{equation}
    W_m(z)
    =
    \frac{2m^m}{\Gamma(m)}
    \int_0^\infty \dd t\,t^{m-1}e^{-t(m+z)}.
    \label{eq:Wm_schwinger_kernel_appendix}
\end{equation}
The integral representation of the second divided difference gives
\begin{align}
    W_m^{[2]}(u,v,u)
    &=
    \int_0^1\dd\xi\int_0^{1-\xi}\dd\zeta\,
    W_m''\!\left((1-\xi)u+\xi v\right)
    \nonumber\\
    &=
    \frac{2m^m}{\Gamma(m)}
    \int_0^\infty\dd t\,t^{m+1}e^{-mt}
    \int_0^1\dd\xi\,(1-\xi)
    e^{-t[(1-\xi)u+\xi v]}.
    \label{eq:hermite_genocchi_kernel_appendix}
\end{align}
For $u=x$ and $v=x+y+2\sqrt{xy}\,c$, introduce the dimensionless
momenta $\widehat q=q/k$ and $\widehat P=P/k$. Completing the square,
\begin{equation}
    (1-\xi)\widehat q^2+\xi(\widehat q+\widehat P)^2
    =
    (\widehat q+\xi\widehat P)^2
    +\xi(1-\xi)\widehat P^2.
\end{equation}
The radial integral and the four-dimensional angular average
\eqref{eq:S3_average_new} then give
\begin{equation}
    \int_0^\infty\dd x\,x\,
    \avg{e^{-t[(1-\xi)x+\xi v]}}
    =
    \frac{1}{t^2}e^{-t\xi(1-\xi)y}.
    \label{eq:gaussian_kernel_reduction_appendix}
\end{equation}
The factor $t^{-2}$ cancels the $t^2$ from the second divided
difference. Since the integrands are non-negative for $m>0$ and
Euclidean $y\geq0$, the order of integration can be exchanged.
Substituting into \eqref{eq:I_general_kernel_new} gives
\begin{align}
    I(y)
    &=
    \frac{2m^m}{\Gamma(m)}
    \int_0^1\dd\xi\,(1-\xi)
    \int_0^\infty\dd t\,t^{m-1}
    e^{-t[m+\xi(1-\xi)y]}
    \nonumber\\
    &=
    \frac{m^m}{\Gamma(m)}
    \int_0^1\dd\xi
    \int_0^\infty\dd t\,t^{m-1}
    e^{-t[m+\xi(1-\xi)y]}
    \nonumber\\
    &=
    \int_0^1\dd\xi\,
    \left[1+\frac{\xi(1-\xi)}{m}y\right]^{-m}
    =I_m(y).
    \label{eq:kernel_to_Im_appendix}
\end{align}
In the second line we used the symmetry
$\xi\leftrightarrow1-\xi$, which allows $(1-\xi)$ to be replaced
by $1/2$ under the integral. Thus the kernel representation agrees
with \eqref{eq:Im_from_flow_new} for every $m>0$ and $y\geq0$.

For explicit checks at $m=2$ and $m=3$, let
$a_m=x+m$, $d=v-x$, and $b_m=v+m=a_m+d$.
The corresponding kernels are
\begin{align}
    W_2^{[2]}(x,v,x)
    &=
    \frac{8(3a_2+2d)}{a_2^3b_2^2}
    =
    \frac{8}{a_2^3}
    \left(\frac{a_2}{b_2^2}+\frac{2}{b_2}\right),
    \nonumber\\
    W_3^{[2]}(x,v,x)
    &=
    \frac{54(6a_3^2+8a_3d+3d^2)}{a_3^4b_3^3}
    =
    \frac{54}{a_3^4}
    \left(\frac{a_3^2}{b_3^3}
    +\frac{2a_3}{b_3^2}+\frac{3}{b_3}\right).
    \label{eq:kernel_explicit_m2_m3_appendix}
\end{align}
At coincident arguments, these reduce to
$24/(x+2)^4$ and $324/(x+3)^5$, respectively. Their integrals
satisfy
\begin{equation}
    \int_0^\infty\dd x\,x\,\frac{24}{(x+2)^4}
    =
    \int_0^\infty\dd x\,x\,\frac{324}{(x+3)^5}
    =1,
\end{equation}
in agreement with $I_m(0)=1$.

The channel normalization uses the oriented representative justified in
App.~\ref{app:direct_hierarchy}.

The fixed-kernel source \eqref{eq:IW_mixed} has the same normalization.
Using $s_m=2r_m-2xr_m'$, its integrand at zero external
momentum is a total derivative,
\begin{equation}
    xg_m(x)^3s_m(x)
    =
    \frac{\dd}{\dd x}\left[x^2g_m(x)^2\right].
    \label{eq:IW_normalization_identity}
\end{equation}
Therefore
\begin{equation}
    I_{\mathrm W}^{(m)}(0)
    =
    \left[x^2g_m(x)^2\right]_0^\infty
    =
    1,
\end{equation}
where the boundary values follow from the infrared regularity
of $r_m$ and $g_m(x)\sim 1/x$ as $x\to\infty$.

\subsection{Relation to the standard two-loop effective action}
\label{app:standard_twoloop}

We relate the form-factor construction of
Sec.~\ref{sec:fourpoint_formfactor} to the standard two-loop expression for
the one-particle irreducible effective action. In a fixed-kernel formulation,
with the propagator
\begin{equation}
    G_0=\left(S^{(2)}+R_k\right)^{-1},
\end{equation}
the one-loop action is
\begin{equation}
    \Gamma_{1,k}
    =
    \frac12\Tr\log G_0^{-1},
    \label{eq:Gamma1_logPk_relation}
\end{equation}
up to a $k$-independent constant. Taking two functional derivatives gives
\begin{equation}
    \Gamma_{1,k;AB}
    =
    \frac12\Tr\left[
        G_0 S^{(4)}_{AB}
        -
        G_0 S^{(3)}_B G_0 S^{(3)}_A
    \right].
    \label{eq:Gamma1_second_derivative_relation}
\end{equation}
The two-loop flow can then be written as
\begin{equation}
    \partial_t\Gamma_{2,k}
    =
    -\frac12\Tr\left[
        G_0\Gamma_{1,k}^{(2)}G_0\,\partial_tR_k
    \right]
    =
    \frac12\Tr\left[
        (\partial_tG_0)\Gamma_{1,k}^{(2)}
    \right],
    \label{eq:Gamma2_flow_relation}
\end{equation}
where
\begin{equation}
    \partial_tG_0
    =
    -G_0(\partial_tR_k)G_0 .
\end{equation}
Substituting \eqref{eq:Gamma1_second_derivative_relation} into
\eqref{eq:Gamma2_flow_relation} and integrating in $t$ reconstructs the usual
regulated two-loop effective action,
\begin{equation}
    \Gamma_{2,k}
    =
    \frac18\,S^{(4)}_{ABCD}\,G_0^{AB}G_0^{CD}
    -
    \frac1{12}\,
    S^{(3)}_{ABC}\,G_0^{AD}G_0^{BE}G_0^{CF}\,
    S^{(3)}_{DEF},
    \label{eq:standard_two_loop_EA_regulated_relation}
\end{equation}
up to $k$-independent terms. The first term is the double-bubble diagram,
whereas the second term is the sunset diagram.

In the symmetric massless $\lambda\phi^4$ theory,
\begin{equation}
    S^{(3)}[0]=0.
\end{equation}
Thus the sunset term vanishes if
\eqref{eq:standard_two_loop_EA_regulated_relation} is evaluated directly at
$\phi=0$. However, the anomalous dimension is extracted from the two-point
function, so one must first take two further functional derivatives and only
then set the background to zero. These derivatives act on the two cubic
vertices in the sunset and produce two quartic vertices,
\begin{equation}
    \left.
    \frac{\delta S^{(3)}}{\delta\phi}
    \right|_{\phi=0}
    =
    S^{(4)}[0].
\end{equation}
Therefore the momentum-dependent part of the two-loop self-energy is the
sunset topology. In the flow hierarchy the same object appears as an outer
single-scale loop with the integrated one-loop four-point vertex inserted,
\begin{equation}
    \partial_t\Gamma^{(2)}_{2,k}(p)
    =
    -\frac12
    \Tr\left[
        G_0\,
        \Gamma^{(4)}_{1,k}(p,-p,\cdot,\cdot)\,
        G_0\,\partial_tR_k
    \right]
    +\cdots .
    \label{eq:twopoint_flow_as_sunset_relation}
\end{equation}
The wave-function projection is the extraction of the coefficient of $p^2$ in
this two-point flow,
\begin{equation}
    \eta_k
    =
    -\partial_t\ln Z_k,
    \qquad
    Z_k
    =
    \left.
    \frac{\partial\Gamma_k^{(2)}(p)}{\partial p^2}
    \right|_{p^2=0}.
    \label{eq:eta_projection_relation}
\end{equation}
This identifies the form-factor insertion with the sunset contribution
to the wave-function flow.

\section{\texorpdfstring{Explicit calculations for $m=2$ and $m=3$}{Explicit calculations for m=2 and m=3}}
\label{app:typeC_checks}

In this appendix we report the profile-specific kernel formulas and intermediate
integrations used in the checks of Sec.~\ref{sec:two_loop_coefficients}.

At the mass-cutoff point $m=1$,
$W_1(u)=2/(u+1)$, $g_1(u)=1/(u+1)$ and $s_1=2$, so that
\begin{equation}
    \mathcal B_{\mathrm W}^{(1)}(u,v)
    =
    \frac{2}{(u+1)^2(v+1)}
    =
    \mathcal B_{\mathrm{PT}}^{(1)}(u,v).
    \label{eq:m1_kernels_equal}
\end{equation}
This is the explicit kernel-level form of the degeneracy at $m=1$.

For the fixed-kernel coefficient it is convenient in the numerical evaluation
to integrate \eqref{eq:CW_mixed_Iprime} by parts,
\begin{equation}
    C_{\mathrm W}^{(m)}
    =
    \frac14
    \int_0^\infty \dd x\,
    \left[
        h_m(x)
        +xh_m'(x)I_{\mathrm W}^{(m)}(x)
    \right],
    \label{eq:CW_mixed_byparts}
\end{equation}
where the boundary terms vanish for the profiles considered below.

	We now apply the fixed-kernel construction to the first non-trivial proper-time
	kernel,
	\begin{equation}
		W_2(u)
		=
		\frac{8}{(u+2)^2}.
		\label{appdetail:eq:W2_mixed_section}
	\end{equation}
	The reconstruction equation is
	\begin{equation}
		\frac{r_2(u)-u r_2'(u)}{u+r_2(u)}
		=
		\left(\frac{2}{u+2}\right)^2 .
		\label{appdetail:eq:r2_reconstruction_equation}
	\end{equation}
	The solution normalized by $u+r_2(u)\sim u$ for $u\to\infty$ is
	\begin{equation}
		r_2(u)
		=
		(u+2)\exp\left[-\frac{2}{u+2}\right]-u .
		\label{appdetail:eq:r2_solution}
	\end{equation}
	Thus
	\begin{equation}
		g_2(u)
		=
		\frac{1}{u+r_2(u)}
		=
		\frac{\exp\left[\frac{2}{u+2}\right]}{u+2},
		\label{appdetail:eq:g2_solution}
	\end{equation}
	and
	\begin{equation}
		s_2(u)
		=
		2r_2(u)-2u r_2'(u)
		=
		\frac{8\exp\left[-\frac{2}{u+2}\right]}{u+2}.
		\label{appdetail:eq:s2_solution}
	\end{equation}
	One checks that
	\begin{equation}
		g_2(u)s_2(u)
		=
		\frac{8}{(u+2)^2}
		=
		W_2(u).
		\label{appdetail:eq:W2_check}
	\end{equation}

	The single-scale factor entering the fixed-kernel four-point bubble is
	\begin{equation}
		h_2(u)
		:=
		g_2(u)^2s_2(u)
		=
		\frac{8\exp\left[\frac{2}{u+2}\right]}{(u+2)^3}.
		\label{appdetail:eq:h2_solution}
	\end{equation}
	Therefore the fixed-kernel bubble kernel is
	\begin{equation}
		\mathcal B_{\mathrm W}^{(2)}(u,v)
		=
		g_2(u)^2s_2(u)g_2(v)
		=
		\frac{
			8\exp\left[\frac{2}{u+2}+\frac{2}{v+2}\right]
		}{
			(u+2)^3(v+2)
		}.
		\label{appdetail:eq:KW_m2}
	\end{equation}

	The direct proper-time kernel is, from \eqref{eq:kernel_explicit_m2_m3_appendix},
	\begin{equation}
		\mathcal B_{\mathrm{PT}}^{(2)}(u,v)
		=
		W_2^{[2]}(u,v,u)
		=
		\frac{8(u+2v+6)}{(u+2)^3(v+2)^2}.
		\label{appdetail:eq:KPT_m2_again}
	\end{equation}
	Thus
	\begin{equation}
		\mathcal B_{\mathrm W}^{(2)}(u,v)
		\neq
		\mathcal B_{\mathrm{PT}}^{(2)}(u,v).
		\label{appdetail:eq:m2_kernels_different}
	\end{equation}

	Using \eqref{appdetail:eq:h2_solution} in the fixed-kernel formulas
	\eqref{eq:IW_mixed} and \eqref{eq:CW_mixed_byparts}, we obtain
	\begin{equation}
		C_{\mathrm W}^{(2)}
		=
		\frac16 ,
		\label{appdetail:eq:CW_m2_result}
	\end{equation}
	with a numerical residual of $1.4\times10^{-13}$ at the quadrature orders of
	Sec.~\ref{subsec:explicit_checks_and_numerics}.

	By contrast, the  direct proper-time hierarchy gives
	\begin{equation}
		C_{\mathrm{PT}}^{(2)}
		\simeq
		0.150405 ,
		\label{appdetail:eq:CPT_m2_reminder}
	\end{equation}
	as found in Sec.~\ref{sec:two_loop_coefficients}.

	We now repeat the same construction for
	\begin{equation}
		W_3(u)
		=
		2\left(\frac{3}{u+3}\right)^3
		=
		\frac{54}{(u+3)^3}.
		\label{appdetail:eq:W3_mixed_section}
	\end{equation}
	The reconstruction equation is
	\begin{equation}
		\frac{r_3(u)-u r_3'(u)}{u+r_3(u)}
		=
		\left(\frac{3}{u+3}\right)^3 .
		\label{appdetail:eq:r3_reconstruction_equation}
	\end{equation}
	The solution normalized by $u+r_3(u)\sim u$ for $u\to\infty$ is
	\begin{equation}
		r_3(u)
		=
		(u+3)
		\exp\left[
		-\frac{3(2u+9)}{2(u+3)^2}
		\right]
		-u .
		\label{appdetail:eq:r3_solution}
	\end{equation}
	Under the rescaling $u=mx$ the
	reconstruction equation \eqref{eq:rm_ode_mixed_simplified} coincides with
	the adapted-cutoff equation, Eq.~(18) of~\cite{Wetterich:2024ivi},
	with $\gamma=1$ and $n=m$. In particular, $r_3$ is, up to this
	rescaling, the exact solution given there for $n=3$,
	$x+r=(x+1)\,e^{-z-z^2/2}$ with $z=1/(1+x)$, and $r_2$
	corresponds to the solution $x+r=(x+1)\,e^{-z}$ of the $n=2$ case.
	The profiles reconstructed here from the frozen spectral kernel are
	therefore identical to the cutoffs adapted to the simplified functional flow
	equation \cite{Wetterich:2024ivi}. With this profile one proceeds as for
	$m=2$:
	\begin{equation}
		g_3(u)
		=
		\frac{1}{u+r_3(u)}
		=
		\frac{
			\exp\left[
			\frac{3(2u+9)}{2(u+3)^2}
			\right]
		}{u+3},
		\label{appdetail:eq:g3_solution}
	\end{equation}
	and
	\begin{equation}
		s_3(u)
		=
		2r_3(u)-2u r_3'(u)
		=
		\frac{54}{(u+3)^2}
		\exp\left[
		-\frac{3(2u+9)}{2(u+3)^2}
		\right].
		\label{appdetail:eq:s3_solution}
	\end{equation}
	One checks that
	\begin{equation}
		g_3(u)s_3(u)
		=
		\frac{54}{(u+3)^3}
		=
		W_3(u).
		\label{appdetail:eq:W3_check}
	\end{equation}

	The single-scale factor entering the fixed-kernel four-point bubble is
	\begin{equation}
		h_3(u)
		:=
		g_3(u)^2s_3(u)
		=
		\frac{54}{(u+3)^4}
		\exp\left[
		\frac{3(2u+9)}{2(u+3)^2}
		\right].
		\label{appdetail:eq:h3_solution}
	\end{equation}
	Therefore the fixed-kernel bubble kernel is
	\begin{equation}
		\mathcal B_{\mathrm W}^{(3)}(u,v)
		=
		g_3(u)^2s_3(u)g_3(v).
		\label{appdetail:eq:KW_m3_compact}
	\end{equation}
	Explicitly,
	\begin{equation}
		\mathcal B_{\mathrm W}^{(3)}(u,v)
		=
		\frac{54}{(u+3)^4(v+3)}
		\exp\left[
		\frac{3(2u+9)}{2(u+3)^2}
		+
		\frac{3(2v+9)}{2(v+3)^2}
		\right].
		\label{appdetail:eq:KW_m3}
	\end{equation}

	The corresponding direct proper-time kernel is
	\begin{equation}
		\mathcal B_{\mathrm{PT}}^{(3)}(u,v)
		=
		W_3^{[2]}(u,v,u),
		\label{appdetail:eq:KPT_m3}
	\end{equation}
	which differs from \eqref{appdetail:eq:KW_m3}.

	Using \eqref{appdetail:eq:h3_solution} in
	\eqref{eq:IW_mixed} and \eqref{eq:CW_mixed_byparts}, we obtain
	\begin{equation}
		C_{\mathrm W}^{(3)}
		=
		\frac16 ,
		\label{appdetail:eq:CW_m3_result}
	\end{equation}
	with a numerical residual of $1.8\times10^{-14}$ at the quadrature orders of
	Sec.~\ref{subsec:explicit_checks_and_numerics}.
These numerical values agree with the benchmark $1/6$ for both
reconstructed profiles; the decomposition into completion terms is given
in \eqref{eq:decomposition_table}.

\section{Local proper-time projection}
\label{app:local_projection}

The local prescription of Sec.~\ref{sec:other_prescriptions}
retains the slope of the inserted four-point vertex at zero channel
momentum before performing the outer proper-time integral.

For a single channel, the slope of the integrated one-loop four-point vertex
at vanishing channel momentum is
\begin{equation}
	\left.
	\partial_{P^2}
	\Gamma^{(4)}_{1,k,\mathrm{1ch}}(P)
	\right|_{P=0}
	=
	\frac{\lambda_k^2}
	{12(16\pi^2)}
	\frac{1}{k^2},
	\label{eq:local_slope_scale_k}
\end{equation}
up to a scale-independent contribution, which does not affect the running.

If this local slope is inserted directly into the outer flow at the same RG
scale $k$, the remaining proper-time integral contains the moment
\begin{equation}
	\int_0^\infty \dd s\,
	\frac{\rho_k^{(m)}(s)}{s^2}
	=
	2k^2\,
	\frac{m}{m-1},
	\qquad
	m>1.
	\label{eq:s2_moment}
\end{equation}
Including the outer two-point-flow prefactor, the three channel assignments,
the Gaussian loop factor, and the sign in
$\eta_k=-\partial_t\ln Z_k$, combining
\eqref{eq:local_slope_scale_k} with \eqref{eq:s2_moment} gives, in the
normalization used for $C_{\mathrm{PT}}^{(m)}$,
\begin{equation}
	C_{\mathrm{loc}}^{(m)}
	=
	\frac{m}{6(m-1)}
	=
	\frac16
	\left(
	1+\frac{1}{m-1}
	\right).
	\label{eq:Cloc_result_appendix}
\end{equation}
The local coefficient is therefore not defined at $m=1$, where
\eqref{eq:s2_moment} diverges, and approaches the universal value only for
$m\to\infty$,
\begin{equation}
	C_{\mathrm{loc}}^{(m)}
	=
	\frac16
	+
	O\!\left(\frac1m\right).
	\label{eq:Cloc_large_m}
\end{equation}

The special role of the exponential limit can be understood from the
proper-time measure. Define
\begin{equation}
	\dd\mu_m(s)
	:=
	\frac12\,
	\frac{\dd s}{s}\,
	\rho_k^{(m)}(s).
	\label{eq:mu_m_definition}
\end{equation}
For the $m$-family this is a normalized Gamma distribution. Its first two
moments are
\begin{equation}
	\langle s\rangle_{\mu_m}
	=
	\frac{1}{k^2},
	\qquad
	\frac{\mathrm{Var}_{\mu_m}(s)}
	{\langle s\rangle_{\mu_m}^2}
	=
	\frac{1}{m}.
	\label{eq:gamma_measure_moments}
\end{equation}
Thus
\begin{equation}
	\dd\mu_m(s)
	\longrightarrow
	\dd s\,
	\delta\!\left(
	s-\frac{1}{k^2}
	\right)
	\qquad
	(m\to\infty).
	\label{eq:mu_delta_limit}
\end{equation}

At finite $m$, the outer proper-time integral samples a distribution of
proper times around $1/k^2$. The scale appearing in the locally expanded
inner vertex is nevertheless the single RG scale $k$. These are therefore
not identical notions of scale. In the exponential limit the distribution
collapses, and the two may be identified:
\begin{equation}
	\frac{1}{k^2}
	\;\longleftrightarrow\;
	s.
	\label{eq:k2_s_identification}
\end{equation}
At finite $m$, the relative width of the sampled proper times is
\begin{equation}
	\frac{\sqrt{\mathrm{Var}_{\mu_m}(s)}}
	{\langle s\rangle_{\mu_m}}
	=
	\frac{1}{\sqrt m}.
\end{equation}

If one formally replaces $1/k^2$ in
\eqref{eq:local_slope_scale_k} by the proper time $s$ selected by the outer
kernel, the remaining integral involves only the normalization
\begin{equation}
	\int_0^\infty
	\frac{\dd s}{s}\,
	\rho_k^{(m)}(s)
	=
	2,
	\label{eq:rho_normalization_local}
\end{equation}
and the universal coefficient $1/6$ follows. At finite $m$, however, this
replacement neglects the spread of proper times in
\eqref{eq:gamma_measure_moments}. Retaining that spread gives instead the
finite-$m$ result \eqref{eq:Cloc_result_appendix}.

The formal replacement selects the exponential endpoint, whereas the
full projection retains the channel dependence throughout the outer loop.
Their finite-$m$ comparison is given in Sec.~\ref{sec:other_prescriptions}.

The direct proper-time hierarchy does not contain the vertical and implicit-scale contributions. As shown in Sec.~\ref{sec:restoring_universality}, restoring these terms recovers the fixed-kernel two-loop coefficient. Whether a similar completion can reconcile other proper-time prescriptions remains to be established.

\section{The one-loop mismatch from the determinant}
\label{app:determinant_Q1}

The vertical contribution in Sec.~\ref{sec:restoring_universality} is
expressed through the difference $I_m-I_{\mathrm W}^{(m)}$. Since the same
difference enters the sum rule, recombination alone does not provide an
independent check of the direct source $I_m$. Here we calculate
$Q_1=\widehat H_1-H_1$ directly from the one-loop effective action and
project its crossed-channel four-point part onto the anomalous dimension.
The determinant supplies the mismatch and its derivatives; the flow sources
are used only for the numerical comparisons at the end of this appendix.

\subsection{Fixed-kernel and pulled-back Hessians}

At one loop, the kernel derivative identity
\eqref{eq:K_derivative_identity} gives, with $H_0[\phi]=S^{(2)}[\phi]$,
\begin{equation}
    \frac{\delta\Gamma_1[\phi,K]}{\delta K}
    =\frac12\bigl(H_0[\phi]+K\bigr)^{-1}.
    \label{eq:det_dGammadK}
\end{equation}
Integrating with respect to $K$ at fixed $\phi$ yields
\begin{equation}
    \Gamma_1[\phi,K]=\frac12\Tr\log\bigl(H_0[\phi]+K\bigr)+F[\phi].
    \label{eq:det_Gamma1}
\end{equation}
The response identity fixes the $K$ dependence, but does not determine
the functional $F[\phi]$, which is fixed by the normalization or
renormalization prescription. We use the same ultraviolet regularization
and subtraction prescription for both Hessians. Since $F$ is independent
of $K$, its field derivatives are identical before and after the pullback
and cancel from $Q_1$; no assumption that $F=0$ is required.

To obtain $\widehat H_1$, we differentiate at fixed $K$ and then set
$K=R_k(H_0)$. For $H_1$, we make this substitution before taking the
field derivatives. The latter gives
\begin{equation}
    \frac12\Tr\log\bigl(H_0+R_k(H_0)\bigr)
    =\frac12\Tr L\bigl(H_0/k^2\bigr)+\text{const},
    \qquad
    L(x)=\log Y(x),\quad Y(x)=x+r(x),
    \label{eq:det_L}
\end{equation}
where $R_k(k^2x)=k^2r(x)$. We use $g(x)=1/Y(x)$ and
$J(x)=L'(x)=Y'(x)/Y(x)$.

In massless $\lambda\phi^4$ theory, write
$H_0=-\partial^2+V$, where $V=\lambda_k\phi^2/2$.
The four-point vertex at $\phi=0$ comes from the term quadratic in $V$.
At fixed $K$, with $G=(-\partial^2+R_k(-\partial^2))^{-1}$,
\begin{equation}
    \frac12\Tr\log\bigl(G^{-1}+V\bigr)
    =\frac12\Tr\log G^{-1}+\frac12\Tr(GV)
    -\frac14\Tr(GVGV)+O(V^3).
    \label{eq:det_fixed_expansion}
\end{equation}
For momentum eigenstates with dimensionless eigenvalues $u$ and $v$,
the quadratic term is
\[
-\frac14\sum_{u,v}g(u)g(v)\,V_{uv}V_{vu}/k^4.
\]

After the pullback, set $A=-\partial^2/k^2$ and $B=V/k^2$ and expand
$\Tr L(A+B)$. Since $A$ and $B$ do not generally commute, the
variation of $J(A)$ cannot be replaced by $J'(A)B$. Its matrix
elements are
\begin{equation}
    \bigl(\delta J\bigr)_{uv}=J^{[1]}(u,v)\,B_{uv},
    \qquad
    J^{[1]}(u,v)=\frac{J(u)-J(v)}{u-v},
    \qquad
    J^{[1]}(u,u)=J'(u).
    \label{eq:det_divided_difference}
\end{equation}
For $J(A)=A^n$, the variation
$\delta A^n=\sum_{j=0}^{n-1}A^jBA^{n-1-j}$ has matrix elements
$\sum_j u^jv^{n-1-j}B_{uv}=(u^n-v^n)B_{uv}/(u-v)$.
The same divided-difference rule applies to the smooth rational functions
used below, with continuous coincident limits. The first variation of
$\Tr L(A)$ is $\Tr[J(A)B]$, so the second variation is
\begin{equation}
    \left.\frac{\dd^2}{\dd\epsilon^2}\frac12\Tr L(A+\epsilon B)
    \right|_{\epsilon=0}
    =\frac12\sum_{u,v}J^{[1]}(u,v)B_{uv}B_{vu}.
    \label{eq:det_second_variation}
\end{equation}
Multiplying by the Taylor factor $1/2!$ gives the quadratic term
$\frac14\sum_{u,v}J^{[1]}(u,v)B_{uv}B_{vu}$, in the same normalization
as the fixed-kernel expansion.
Comparing it with the fixed-kernel term gives the mismatch kernel
\begin{equation}
    M(u,v)=-g(u)g(v)-J^{[1]}(u,v).
    \label{eq:det_mismatch_kernel}
\end{equation}
For a field-independent mass regulator, $r=\text{const}$ and $J=g$.
Then $J^{[1]}(u,v)=-g(u)g(v)$, so $M=0$.

\subsection{Reconstructed profiles and  divided differences}

For the reconstructed $m$-family, the condition
\eqref{eq:rm_ode_mixed_simplified}, $W=gs$ with
$s=2(r-xr')$, gives
\begin{equation*}
    \frac{xY'(x)}{Y(x)}=1-\frac{W_m(x)}{2},
    \qquad
    J_m(x)=\frac{1-W_m(x)/2}{x}.
\end{equation*}
Writing $z=m/(x+m)$, the geometric sum gives
$(1-z^m)/x=(x+m)^{-1}\sum_{j=0}^{m-1}z^j$. Thus the apparent
singularity at $x=0$ is removable, and for positive integer $m$,
\begin{equation}
    J_m(x)=\sum_{j=1}^{m}\frac{m^{\,j-1}}{(x+m)^j},
    \label{eq:det_J_profiles}
\end{equation}
so that
$J_2=1/(x+2)+2/(x+2)^2$ and
$J_3=1/(x+3)+3/(x+3)^2+9/(x+3)^3$.
Integrating $J=(\log Y)'$ with $Y(x)\sim x$ for large $x$
recovers \eqref{eq:r3_solution} for $m=3$.
For $a=u+m$ and $b=v+m$, use
$(a^{-j}-b^{-j})/(a-b)=-\sum_{n=1}^{j}a^{-(j-n+1)}b^{-n}$.
Collecting equal powers of $b^{-1}$ gives
\begin{equation}
    J_m^{[1]}(u,v)=-\sum_{n=1}^{m}\frac{c_n(u)}{b^n},
    \qquad
    c_n(u)=\sum_{j=n}^{m}\frac{m^{\,j-1}}{a^{\,j-n+1}},
    \label{eq:det_Jdiv}
\end{equation}
which remains regular at $u=v$ and avoids subtracting nearly equal
values of $J$ in numerical calculations.

\subsection{Subtracted form factor and outer projection}

For a four-point channel with loop momentum $\ell$ and external
momentum $P$, define
\begin{equation}
    y=\frac{P^2}{k^2},\qquad
    u=\frac{\ell^2}{k^2},\qquad
    v=\frac{(\ell+P)^2}{k^2}=u+y+2\sqrt{uy}\,c,
\end{equation}
where $c$ is the cosine of the angle between $\ell$ and $P$.
In four dimensions,
$\int\frac{\dd^4\ell}{(2\pi)^4}
 =\frac{k^4}{16\pi^2}\int_0^\infty\dd u\,u\avg{\cdots}$,
with the average defined in \eqref{eq:S3_average_new}.
Each insertion supplies $\lambda_k\phi^2/2$, so the quadratic term
carries $\lambda_k^2/16$ multiplying a product of two $\phi^2$ insertions.
For a specified pairing there are two assignments within each insertion
and two ways of exchanging the insertions, giving $2\cdot2\cdot2=8$.
The four field derivatives therefore supply $\lambda_k^2/2$ per channel.
The two factors $B=V/k^2$ contribute $k^{-4}$, which cancels
the $k^4$ in the loop measure. Extracting $\lambda_k^2/(16\pi^2)$ as in
\eqref{eq:D_difference_formfactor} leaves the factor $1/2$ below.

The momentum-independent part does not contribute to $\eta$.
Subtracting it inside the integral, with
$M(u,u)=-g(u)^2-J'(u)$, gives
\begin{equation}
    D_m(y)
    =\frac12\int_0^\infty\dd u\,u\,
    \avg{-g(u)g(v)-J^{[1]}(u,v)+g(u)^2+J'(u)},
    \qquad D_m(0)=0.
    \label{eq:det_D}
\end{equation}
This is the form factor $D_m=f_{\mathrm W}-f_{\mathrm{PT}}$
of \eqref{eq:D_difference_formfactor}, now obtained without the flow
sources. Subtracting before integration also removes the separate
logarithmic divergences of the two bubbles.

Inserting the bubble into the outer two-point loop and using
\eqref{eq:angular_projection_identity_new} gives
\begin{equation}
    C_{\mathrm{vert,det}}^{(m)}
    =\int_0^\infty\dd x\,x\,h_m(x)
    \left[D_m'(x)+\frac x2D_m''(x)\right].
    \label{eq:det_Cvert_D2}
\end{equation}
The second derivative is numerically inconvenient: its integrand
contains terms behaving as $y^{-3/2}$ near $y=0$.
Using
\begin{equation}
    x\left[D_m'+\frac x2D_m''\right]
    =\frac12\,\frac{\dd}{\dd x}\bigl(x^2D_m'\bigr),
\end{equation}
and integrating by parts yields
\begin{equation}
    C_{\mathrm{vert,det}}^{(m)}
    =-\frac12\int_0^\infty\dd x\,h_m'(x)\,x^2\,D_m'(x).
    \label{eq:det_Cvert}
\end{equation}
The boundary term is $\frac12[h_m(x)x^2D_m'(x)]_0^\infty$;
\eqref{eq:det_Cvert} requires its endpoint limits to vanish. At the origin,
$D_m'$ is finite for these smooth regulated profiles. At large $x$, the
reconstructed profiles have $h_m(x)=O(x^{-m-1})$; together with the
decaying derivative of the subtracted mismatch this removes the upper
endpoint for the profiles considered. For $m=1$ the mismatch vanishes
identically. This integration by parts acts on the outer integral only.

To compute the remaining derivative, the subtraction in \eqref{eq:det_D}
has no $y$ dependence. At fixed $u,c$ and $y>0$,
\begin{equation}
    v_y=1+c\sqrt{u/y},\qquad
    v_{yy}=-\frac{c\sqrt u}{2y^{3/2}},\qquad
    g'=-Jg,\qquad g''=(J^2-J')g.
    \label{eq:det_chain_derivatives}
\end{equation}
The rational expression \eqref{eq:det_Jdiv} gives
\begin{equation}
    \partial_vJ_m^{[1]}(u,v)
    =\sum_{n=1}^m\frac{n c_n(u)}{(v+m)^{n+1}},\qquad
    \partial_v^2J_m^{[1]}(u,v)
    =-\sum_{n=1}^m\frac{n(n+1)c_n(u)}{(v+m)^{n+2}}.
    \label{eq:det_Jdiv_derivatives}
\end{equation}
With $M_v=-g(u)g'(v)-\partial_vJ_m^{[1]}$ and
$M_{vv}=-g(u)g''(v)-\partial_v^2J_m^{[1]}$, differentiation under
the convergent subtracted integral yields
\begin{align}
    D_m'(y)&=\frac12\int_0^\infty\dd u\,u\,\avg{M_v v_y},
    \label{eq:det_Dprime_analytic}\\
    D_m''(y)&=\frac12\int_0^\infty\dd u\,u\,
    \avg{M_{vv}v_y^2+M_v\,v_{yy}}.
    \label{eq:det_Dsecond_analytic}
\end{align}
These formulas use only $g_m$, $J_m$ and their rational divided
differences. The positive quadrature nodes avoid evaluating the explicit
$y^{-1/2}$ and $y^{-3/2}$ expressions at the origin; their endpoint
cancellations belong to the combined angular integrand.
The by-parts form \eqref{eq:det_Cvert} needs only $D_m'$, avoiding
the second-derivative terms that slow convergence in
\eqref{eq:det_Cvert_D2}. Substituting
$D_m'=(I_m-I_{\mathrm W}^{(m)})/(2y)$ would instead recover
\eqref{eq:Cvert_byparts}, but that relation is used only as a comparison,
not to construct the determinant result.

The field derivatives of $Q^{(2)}$ give the four-point mismatch in
the crossed channels relevant here. The additional zero-channel-momentum
insertion does not contribute to the $p^2$ projection; see
Sec.~\ref{sec:restoring_universality}.

\subsection{Numerical checks}

We first compare the derivative of \eqref{eq:det_D} with
$(I_m-I_{\mathrm W}^{(m)})/(2y)$ at $y=0.1,1,5$.
For $m=2,3$, a five-point finite difference with step $10^{-3}$,
using 216 radial, 80 angular and 200 Feynman-parameter nodes,
agrees to better than $6\times10^{-14}$.
This tests the direct source independently: replacing $I_m$ by $1$
would leave the sum rule unchanged but give $D_2'(1)=0.1101$,
whereas the determinant gives $0.0377$.
At $m=1$, $C_{\mathrm{vert,det}}^{(1)}$ vanishes to $10^{-19}$.

Finally, evaluating \eqref{eq:det_Cvert} with 729 radial and
240 angular nodes gives the following differences from
$C_{\mathrm{vert}}^{(m)}$ in \eqref{eq:decomposition_table},
computed from the sources with 324 radial and 160 angular nodes:
\begin{equation}
    \begin{array}{c|ccccccc}
        m & 2 & 3 & 4 & 5 & 6 & 7 & 8\\[0.2em]
        \hline
        \rule{0pt}{1.1em}
        |C_{\mathrm{vert,det}}-C_{\mathrm{vert}}|
        & 3\times10^{-10} & 2\times10^{-12} & 2\times10^{-14}
        & 1\times10^{-14} & 1\times10^{-14} & 1\times10^{-14} & 1\times10^{-14}
    \end{array}
\end{equation}
Convergence is slowest for $m=2$, whose profile is widest.
The differences shown are observed values, not error bounds.
The determinant calculation thus confirms the sign and magnitude
of the projected vertical contribution independently of the flow sources.
It checks the  four-point mismatch in the crossed channels relevant
to $\eta$; it does not evaluate every component of $Q_1$.

\clearpage
\nocite{apsrev42Control}
\bibliographystyle{apsrev4-2}
\bibliography{horizontal_vertical
}

@article{Cornwall:1974vz,
    author = "Cornwall, John M. and Jackiw, R. and Tomboulis, E.",
    title = "{Effective action for composite operators}",
    journal = "Phys. Rev. D",
    volume = "10",
    pages = "2428--2445",
    year = "1974",
    doi = "10.1103/PhysRevD.10.2428"
}

@CONTROL{apsrev42Control,
  author = "08",
  editor = "1",
  pages  = "0",
  title  = "0",
  year   = "1"
}

@article{Wetterich:1992yh,
    author = "Wetterich, Christof",
    title = "{Exact evolution equation for the effective potential}",
    journal = "Phys. Lett. B",
    volume = "301",
    pages = "90--94",
    year = "1993",
    doi = "10.1016/0370-2693(93)90726-X"
}

@article{Reuter:1993kw,
    author = "Reuter, Martin and Wetterich, Christof",
    title = "{Effective average action for gauge theories and exact evolution equations}",
    journal = "Nucl. Phys. B",
    volume = "417",
    pages = "181--214",
    year = "1994",
    doi = "10.1016/0550-3213(94)90543-6"
}

@article{Liao:1992fm,
    author = "Liao, Sen-Ben and Polonyi, Janos",
    title = "{Blocking transformation in field theory}",
    journal = "Annals Phys.",
    volume = "222",
    pages = "122--156",
    year = "1993",
    doi = "10.1006/aphy.1993.1019"
}

@article{Bonanno:2000yp,
    author = "Bonanno, Alfio and Zappala, Dario",
    title = "{Towards an accurate determination of the critical exponents with the renormalization group flow equations}",
    eprint = "hep-th/0010095",
    archivePrefix = "arXiv",
    journal = "Phys. Lett. B",
    volume = "504",
    pages = "181--187",
    year = "2001",
    doi = "10.1016/S0370-2693(01)00273-8"
}

@article{Mazza:2001bp,
    author = "Mazza, Marco and Zappala, Dario",
    title = "{Proper time regulator and renormalization group flow}",
    eprint = "hep-th/0106230",
    archivePrefix = "arXiv",
    journal = "Phys. Rev. D",
    volume = "64",
    pages = "105013",
    year = "2001",
    doi = "10.1103/PhysRevD.64.105013"
}

@article{Litim:2001hk,
    author = "Litim, Daniel F. and Pawlowski, Jan M.",
    title = "{Predictive power of renormalisation group flows: a comparison}",
    eprint = "hep-th/0107020",
    archivePrefix = "arXiv",
    journal = "Phys. Lett. B",
    volume = "516",
    pages = "197--207",
    year = "2001",
    doi = "10.1016/S0370-2693(01)00909-2"
}

@article{Litim:2001ky,
    author = "Litim, Daniel F. and Pawlowski, Jan M.",
    title = "{Perturbation theory and renormalisation group equations}",
    eprint = "hep-th/0111191",
    archivePrefix = "arXiv",
    journal = "Phys. Rev. D",
    volume = "65",
    pages = "081701",
    year = "2002",
    doi = "10.1103/PhysRevD.65.081701"
}

@article{Litim:2002xm,
    author = "Litim, Daniel F. and Pawlowski, Jan M.",
    title = "{Completeness and consistency of renormalisation group flows}",
    eprint = "hep-th/0202188",
    archivePrefix = "arXiv",
    journal = "Phys. Rev. D",
    volume = "66",
    pages = "025030",
    year = "2002",
    doi = "10.1103/PhysRevD.66.025030"
}

@article{Papenbrock:1994kf,
    author = "Papenbrock, Thomas and Wetterich, Christof",
    title = "{Two loop results from one loop computations and nonperturbative solutions of exact evolution equations}",
    eprint = "hep-th/9403164",
    archivePrefix = "arXiv",
    journal = "Z. Phys. C",
    volume = "65",
    pages = "519--535",
    year = "1995",
    doi = "10.1007/BF01556140"
}

@article{Litim:2002hj,
    author = "Litim, Daniel F. and Pawlowski, Jan M.",
    title = "{Wilsonian flows and background fields}",
    eprint = "hep-th/0208216",
    archivePrefix = "arXiv",
    journal = "Phys. Lett. B",
    volume = "546",
    pages = "279--286",
    year = "2002",
    doi = "10.1016/S0370-2693(02)02693-X"
}

@article{Bonanno:2019ukb,
    author = "Bonanno, Alfio and Lippoldt, Stefan and Percacci, Roberto and Vacca, Gian Paolo",
    title = "{On exact proper time Wilsonian RG flows}",
    eprint = "1912.08135",
    archivePrefix = "arXiv",
    primaryClass = "hep-th",
    journal = "Eur. Phys. J. C",
    volume = "80",
    number = "3",
    pages = "249",
    year = "2020",
    doi = "10.1140/epjc/s10052-020-7798-9"
}

@article{Falls:2024noj,
    author = "Falls, Kevin and Ferrero, Renata",
    title = "{Asymptotic safety within on-shell perturbation theory}",
    eprint = "2411.00938",
    archivePrefix = "arXiv",
    primaryClass = "hep-th",
    doi = "10.1007/JHEP08(2025)173",
    journal = "JHEP",
    volume = "08",
    pages = "173",
    year = "2025"
}

@article{Falls:2020tmj,
    author = "Falls, Kevin",
    title = "{Background independent exact renormalisation}",
    eprint = "2004.11409",
    archivePrefix = "arXiv",
    primaryClass = "hep-th",
    doi = "10.1140/epjc/s10052-020-08803-0",
    journal = "Eur. Phys. J. C",
    volume = "81",
    number = "2",
    pages = "121",
    year = "2021"
}

@article{Falls:2017cze,
    author = "Falls, Kevin",
    title = "{Physical renormalization schemes and asymptotic safety in quantum gravity}",
    eprint = "1702.03577",
    archivePrefix = "arXiv",
    primaryClass = "hep-th",
    doi = "10.1103/PhysRevD.96.126016",
    journal = "Phys. Rev. D",
    volume = "96",
    number = "12",
    pages = "126016",
    year = "2017"
}

@article{deAlwis:2017ysy,
    author = "de Alwis, S. P.",
    title = "{Exact RG flow equations and quantum gravity}",
    eprint = "1707.09298",
    archivePrefix = "arXiv",
    primaryClass = "hep-th",
    journal = "JHEP",
    volume = "03",
    pages = "118",
    year = "2018",
    doi = "10.1007/JHEP03(2018)118"
}

@article{Wetterich:2024ivi,
    author = "Wetterich, Christof",
    title = "{Simplified functional flow equation}",
    eprint = "2403.17523",
    archivePrefix = "arXiv",
    primaryClass = "hep-th",
    journal = "Phys. Lett. B",
    volume = "864",
    pages = "139435",
    year = "2025",
    doi = "10.1016/j.physletb.2025.139435"
}

@article{Giacometti:2025qyy,
    author = "Giacometti, Gabriele and Rizzo, Daniele and Zappala, Dario",
    title = "{Universal content of the proper time flow in scalar and Yang--Mills theories}",
    eprint = "2510.04896",
    archivePrefix = "arXiv",
    primaryClass = "hep-th",
    journal = "Phys. Rev. D",
    volume = "113",
    number = "4",
    pages = "045020",
    year = "2026",
    doi = "10.1103/dsgp-mlkp"
}

@article{Morris:2005tv,
    author = "Morris, Tim R. and Rosten, Oliver J.",
    title = "{A manifestly gauge invariant, continuum calculation of the SU(N) Yang--Mills two-loop beta function}",
    eprint = "hep-th/0508026",
    archivePrefix = "arXiv",
    journal = "Phys. Rev. D",
    volume = "73",
    pages = "065003",
    year = "2006",
    doi = "10.1103/PhysRevD.73.065003"
}

@article{Nielsen:1975fs,
    author = "Nielsen, N. K.",
    title = "{On the gauge dependence of spontaneous symmetry breaking in gauge theories}",
    journal = "Nucl. Phys. B",
    volume = "101",
    pages = "173--188",
    year = "1975",
    doi = "10.1016/0550-3213(75)90301-6"
}

@article{Daletskii:1965,
    author = "Daletskii, Yu. L. and Krein, S. G.",
    title = "{Integration and differentiation of functions of Hermitian operators and applications to the theory of perturbations}",
    journal = "Amer. Math. Soc. Transl. (2)",
    volume = "47",
    pages = "1--30",
    year = "1965"
}

@article{Wilson:1973jj,
    author = "Wilson, Kenneth G. and Kogut, John B.",
    title = "{The Renormalization group and the epsilon expansion}",
    journal = "Phys. Rept.",
    volume = "12",
    pages = "75--199",
    year = "1974",
    doi = "10.1016/0370-1573(74)90023-4"
}

@article{Wegner:1972ih,
    author = "Wegner, Franz J. and Houghton, Anthony",
    title = "{Renormalization group equation for critical phenomena}",
    journal = "Phys. Rev. A",
    volume = "8",
    pages = "401--412",
    year = "1973",
    doi = "10.1103/PhysRevA.8.401"
}

@article{Falls:2026nuh,
    author = "Falls, Kevin and Ferrero, Renata and Oglialoro, Giovanni",
    title = "{Towards gauge independence in asymptotically safe quantum gravity}",
    eprint = "2607.06657",
    archivePrefix = "arXiv",
    primaryClass = "hep-th",
    doi = "10.1007/JHEP09(2026)194",
    journal = "JHEP",
    volume = "09",
    pages = "194",
    year = "2026"
}

@article{Manrique:2009uh,
    author = "Manrique, Elisa and Reuter, Martin",
    title = "{Bimetric Truncations for Quantum Einstein Gravity and Asymptotic Safety}",
    eprint = "0907.2617",
    archivePrefix = "arXiv",
    primaryClass = "gr-qc",
    reportNumber = "MZ-TH-09-17",
    doi = "10.1016/j.aop.2009.11.009",
    journal = "Annals Phys.",
    volume = "325",
    pages = "785--815",
    year = "2010"
}

@misc{Ihssen:2026ucr,
    author = "Ihssen, Friederike and Knorr, Benjamin and Mezger, Silas and Pawlowski, Jan M. and Sprenger, Paul P.",
    title = "{Diffeomorphism-invariant Approach to Asymptotically Safe Quantum Gravity}",
    eprint = "2609.07829",
    archivePrefix = "arXiv",
    primaryClass = "hep-th",
    month = "9",
    year = "2026"
}

@article{Polchinski:1983gv,
    author = "Polchinski, Joseph",
    title = "{Renormalization and Effective Lagrangians}",
    journal = "Nucl. Phys. B",
    volume = "231",
    pages = "269--295",
    year = "1984",
    doi = "10.1016/0550-3213(84)90287-6"
}

@article{Morris:1993qb,
    author = "Morris, Tim R.",
    title = "{The Exact renormalization group and approximate solutions}",
    eprint = "hep-ph/9308265",
    archivePrefix = "arXiv",
    journal = "Int. J. Mod. Phys. A",
    volume = "9",
    pages = "2411--2450",
    year = "1994",
    doi = "10.1142/S0217751X94000972"
}

@article{Morris:1994ie,
    author = "Morris, Tim R.",
    title = "{Derivative expansion of the exact renormalization group}",
    eprint = "hep-ph/9403340",
    archivePrefix = "arXiv",
    journal = "Phys. Lett. B",
    volume = "329",
    pages = "241--248",
    year = "1994",
    doi = "10.1016/0370-2693(94)90767-6"
}

@article{Morris:1994ki,
    author = "Morris, Tim R.",
    title = "{On truncations of the exact renormalization group}",
    eprint = "hep-th/9405190",
    archivePrefix = "arXiv",
    journal = "Phys. Lett. B",
    volume = "334",
    pages = "355--362",
    year = "1994",
    doi = "10.1016/0370-2693(94)90700-5"
}

@article{Morris:1998da,
    author = "Morris, Tim R.",
    title = "{Elements of the continuous renormalization group}",
    eprint = "hep-th/9802039",
    archivePrefix = "arXiv",
    journal = "Prog. Theor. Phys. Suppl.",
    volume = "131",
    pages = "395--414",
    year = "1998",
    doi = "10.1143/PTPS.131.395"
}

@article{Bagnuls:2000ae,
    author = "Bagnuls, C. and Bervillier, C.",
    title = "{Exact renormalization group equations. An Introductory review}",
    eprint = "hep-th/0002034",
    archivePrefix = "arXiv",
    journal = "Phys. Rept.",
    volume = "348",
    pages = "91--157",
    year = "2001",
    doi = "10.1016/S0370-1573(00)00137-X"
}

@article{Berges:2000ew,
    author = "Berges, Juergen and Tetradis, Nikolaos and Wetterich, Christof",
    title = "{Nonperturbative renormalization flow in quantum field theory and statistical physics}",
    eprint = "hep-ph/0005122",
    archivePrefix = "arXiv",
    journal = "Phys. Rept.",
    volume = "363",
    pages = "223--386",
    year = "2002",
    doi = "10.1016/S0370-1573(01)00098-9"
}

@article{Polonyi:2001se,
    author = "Polonyi, Janos",
    title = "{Lectures on the functional renormalization group method}",
    eprint = "hep-th/0110026",
    archivePrefix = "arXiv",
    journal = "Central Eur. J. Phys.",
    volume = "1",
    pages = "1--71",
    year = "2003",
    doi = "10.2478/BF02475552"
}

@article{Pawlowski:2005xe,
    author = "Pawlowski, Jan M.",
    title = "{Aspects of the functional renormalisation group}",
    eprint = "hep-th/0512261",
    archivePrefix = "arXiv",
    journal = "Annals Phys.",
    volume = "322",
    pages = "2831--2915",
    year = "2007",
    doi = "10.1016/j.aop.2007.01.007"
}

@article{Gies:2006wv,
    author = "Gies, Holger",
    title = "{Introduction to the functional RG and applications to gauge theories}",
    eprint = "hep-ph/0611146",
    archivePrefix = "arXiv",
    journal = "Lect. Notes Phys.",
    volume = "852",
    pages = "287--348",
    year = "2012",
    doi = "10.1007/978-3-642-27320-9_6"
}

@article{Rosten:2010vm,
    author = "Rosten, Oliver J.",
    title = "{Fundamentals of the Exact Renormalization Group}",
    eprint = "1003.1366",
    archivePrefix = "arXiv",
    primaryClass = "hep-th",
    journal = "Phys. Rept.",
    volume = "511",
    pages = "177--272",
    year = "2012",
    doi = "10.1016/j.physrep.2011.12.003"
}

@article{Delamotte:2007pf,
    author = "Delamotte, Bertrand",
    title = "{An Introduction to the nonperturbative renormalization group}",
    eprint = "cond-mat/0702365",
    archivePrefix = "arXiv",
    journal = "Lect. Notes Phys.",
    volume = "852",
    pages = "49--132",
    year = "2012",
    doi = "10.1007/978-3-642-27320-9_2"
}

@article{Dupuis:2020fhh,
    author = "Dupuis, N. and Canet, L. and Eichhorn, A. and Metzner, W. and Pawlowski, J. M. and Tissier, M. and Wschebor, N.",
    title = "{The nonperturbative functional renormalization group and its applications}",
    eprint = "2006.04853",
    archivePrefix = "arXiv",
    primaryClass = "cond-mat.stat-mech",
    journal = "Phys. Rept.",
    volume = "910",
    pages = "1--114",
    year = "2021",
    doi = "10.1016/j.physrep.2021.01.001"
}

@article{Litim:2000ci,
    author = "Litim, Daniel F.",
    title = "{Optimization of the exact renormalization group}",
    eprint = "hep-th/0005245",
    archivePrefix = "arXiv",
    journal = "Phys. Lett. B",
    volume = "486",
    pages = "92--99",
    year = "2000",
    doi = "10.1016/S0370-2693(00)00748-6"
}

@article{Litim:2001up,
    author = "Litim, Daniel F.",
    title = "{Optimized renormalization group flows}",
    eprint = "hep-th/0103195",
    archivePrefix = "arXiv",
    journal = "Phys. Rev. D",
    volume = "64",
    pages = "105007",
    year = "2001",
    doi = "10.1103/PhysRevD.64.105007"
}

@article{Litim:2001dt,
    author = "Litim, Daniel F.",
    title = "{Derivative expansion and renormalization group flows}",
    eprint = "hep-th/0111159",
    archivePrefix = "arXiv",
    journal = "JHEP",
    volume = "11",
    pages = "059",
    year = "2001",
    doi = "10.1088/1126-6708/2001/11/059"
}

@article{Canet:2003qd,
    author = "Canet, Leonie and Delamotte, Bertrand and Mouhanna, Dominique and Vidal, Julien",
    title = "{Nonperturbative renormalization group approach to the Ising model: A Derivative expansion at order $\partial^4$}",
    eprint = "hep-th/0302227",
    archivePrefix = "arXiv",
    journal = "Phys. Rev. B",
    volume = "68",
    pages = "064421",
    year = "2003",
    doi = "10.1103/PhysRevB.68.064421"
}

@article{Tetradis:1993ts,
    author = "Tetradis, N. and Wetterich, C.",
    title = "{Critical exponents from effective average action}",
    eprint = "hep-ph/9308214",
    archivePrefix = "arXiv",
    journal = "Nucl. Phys. B",
    volume = "422",
    pages = "541--592",
    year = "1994",
    doi = "10.1016/0550-3213(94)90446-4"
}

@article{Ellwanger:1993mw,
    author = "Ellwanger, Ulrich",
    title = "{Flow equations for N point functions and bound states}",
    eprint = "hep-ph/9308260",
    archivePrefix = "arXiv",
    journal = "Z. Phys. C",
    volume = "62",
    pages = "503--510",
    year = "1994",
    doi = "10.1007/BF01555911"
}

@article{Bonini:1992vh,
    author = "Bonini, M. and D'Attanasio, M. and Marchesini, G.",
    title = "{Perturbative renormalization and infrared finiteness in the Wilson renormalization group: The Massless scalar case}",
    eprint = "hep-th/9301114",
    archivePrefix = "arXiv",
    journal = "Nucl. Phys. B",
    volume = "409",
    pages = "441--464",
    year = "1993",
    doi = "10.1016/0550-3213(93)90588-G"
}

@article{Aoki:2000wm,
    author = "Aoki, K.-I.",
    title = "{Introduction to the nonperturbative renormalization group and its recent applications}",
    journal = "Int. J. Mod. Phys. B",
    volume = "14",
    pages = "1249--1326",
    year = "2000",
    doi = "10.1016/S0217-9792(00)00092-3"
}

@article{Schwinger:1951nm,
    author = "Schwinger, Julian S.",
    title = "{On gauge invariance and vacuum polarization}",
    journal = "Phys. Rev.",
    volume = "82",
    pages = "664--679",
    year = "1951",
    doi = "10.1103/PhysRev.82.664"
}

@article{Bonanno:2004sy,
    author = "Bonanno, Alfio and Reuter, Martin",
    title = "{Proper time flow equation for gravity}",
    eprint = "hep-th/0410191",
    archivePrefix = "arXiv",
    journal = "JHEP",
    volume = "02",
    pages = "035",
    year = "2005",
    doi = "10.1088/1126-6708/2005/02/035"
}

@article{Codello:2008vh,
    author = "Codello, Alessandro and Percacci, Roberto and Rahmede, Christoph",
    title = "{Investigating the Ultraviolet Properties of Gravity with a Wilsonian Renormalization Group Equation}",
    eprint = "0805.2909",
    archivePrefix = "arXiv",
    primaryClass = "hep-th",
    journal = "Annals Phys.",
    volume = "324",
    pages = "414--469",
    year = "2009",
    doi = "10.1016/j.aop.2008.08.008"
}

@article{Morris:1999px,
    author = "Morris, Tim R.",
    title = "{A Gauge invariant exact renormalization group. 1.}",
    eprint = "hep-th/9910058",
    archivePrefix = "arXiv",
    journal = "Nucl. Phys. B",
    volume = "573",
    pages = "97--126",
    year = "2000",
    doi = "10.1016/S0550-3213(99)00821-4"
}

@article{Morris:2000fs,
    author = "Morris, Tim R.",
    title = "{A Gauge invariant exact renormalization group. 2.}",
    eprint = "hep-th/0006064",
    archivePrefix = "arXiv",
    journal = "JHEP",
    volume = "12",
    pages = "012",
    year = "2000",
    doi = "10.1088/1126-6708/2000/12/012"
}

@article{Wetterich:2016ewc,
    author = "Wetterich, Christof",
    title = "{Gauge-invariant flow equation}",
    eprint = "1607.02989",
    archivePrefix = "arXiv",
    primaryClass = "hep-th",
    journal = "Nucl. Phys. B",
    volume = "931",
    pages = "262--282",
    year = "2018",
    doi = "10.1016/j.nuclphysb.2018.04.020"
}

@article{Pawlowski:2015mlf,
    author = "Pawlowski, Jan M. and Scherer, Michael M. and Schmidt, Richard and Wetzel, Sebastian J.",
    title = "{Physics and the choice of regulators in functional renormalisation group flows}",
    eprint = "1512.03598",
    archivePrefix = "arXiv",
    primaryClass = "hep-th",
    journal = "Annals Phys.",
    volume = "384",
    pages = "165--197",
    year = "2017",
    doi = "10.1016/j.aop.2017.06.017"
}

@article{Gies:2002af,
    author = "Gies, Holger",
    title = "{Running coupling in Yang-Mills theory: A flow equation study}",
    eprint = "hep-th/0202207",
    archivePrefix = "arXiv",
    reportNumber = "CERN-TH-2002-047",
    doi = "10.1103/PhysRevD.66.025006",
    journal = "Phys. Rev. D",
    volume = "66",
    pages = "025006",
    year = "2002"
}

@article{Reuter:1997gx,
    author = "Reuter, M. and Wetterich, C.",
    title = "{Gluon condensation in nonperturbative flow equations}",
    eprint = "hep-th/9708051",
    archivePrefix = "arXiv",
    doi = "10.1103/PhysRevD.56.7893",
    journal = "Phys. Rev. D",
    volume = "56",
    pages = "7893--7916",
    year = "1997"
}

@article{Wetterich:2020cxq,
    author = "Wetterich, C.",
    title = "{Fundamental scale invariance}",
    eprint = "2007.08805",
    archivePrefix = "arXiv",
    primaryClass = "hep-th",
    doi = "10.1016/j.nuclphysb.2021.115326",
    journal = "Nucl. Phys. B",
    volume = "964",
    pages = "115326",
    year = "2021"
}

@article{Dietz:2015owa,
    author = "Dietz, Juergen A. and Morris, Tim R.",
    title = "{Background independent exact renormalization group for conformally reduced gravity}",
    eprint = "1502.07396",
    archivePrefix = "arXiv",
    primaryClass = "hep-th",
    doi = "10.1007/JHEP04(2015)118",
    journal = "JHEP",
    volume = "04",
    pages = "118",
    year = "2015"
}

@article{Labus:2016lkh,
    author = {Labus, Peter and Morris, Tim R. and Slade, Zo{\"e} H.},
    title = "{Background independence in a background dependent renormalization group}",
    eprint = "1603.04772",
    archivePrefix = "arXiv",
    primaryClass = "hep-th",
    doi = "10.1103/PhysRevD.94.024007",
    journal = "Phys. Rev. D",
    volume = "94",
    number = "2",
    pages = "024007",
    year = "2016"
}

@article{Morris:2016spn,
    author = "Morris, Tim R.",
    title = "{Large curvature and background scale independence in single-metric approximations to asymptotic safety}",
    eprint = "1610.03081",
    archivePrefix = "arXiv",
    primaryClass = "hep-th",
    doi = "10.1007/JHEP11(2016)160",
    journal = "JHEP",
    volume = "11",
    pages = "160",
    year = "2016"
}

@article{Reuter:1996cp,
    author = "Reuter, M.",
    title = "{Nonperturbative evolution equation for quantum gravity}",
    eprint = "hep-th/9605030",
    archivePrefix = "arXiv",
    doi = "10.1103/PhysRevD.57.971",
    journal = "Phys. Rev. D",
    volume = "57",
    pages = "971--985",
    year = "1998"
}

@article{Freire:2000bq,
    author = "Freire, Filipe and Litim, Daniel F. and Pawlowski, Jan M.",
    title = "{Gauge invariance and background field formalism in the exact renormalization group}",
    eprint = "hep-th/0009110",
    archivePrefix = "arXiv",
    doi = "10.1016/S0370-2693(00)01231-4",
    journal = "Phys. Lett. B",
    volume = "495",
    pages = "256--262",
    year = "2000"
}

@article{Safari:2015dva,
    author = "Safari, Mahmoud",
    title = "{Splitting Ward identity}",
    eprint = "1508.06244",
    archivePrefix = "arXiv",
    primaryClass = "hep-th",
    doi = "10.1140/epjc/s10052-016-4036-6",
    journal = "Eur. Phys. J. C",
    volume = "76",
    number = "4",
    pages = "201",
    year = "2016"
}

@article{Manrique:2010am,
    author = "Manrique, Elisa and Reuter, Martin and Saueressig, Frank",
    title = "{Bimetric Renormalization Group Flows in Quantum Einstein Gravity}",
    eprint = "1006.0099",
    archivePrefix = "arXiv",
    primaryClass = "hep-th",
    doi = "10.1016/j.aop.2010.11.006",
    journal = "Annals Phys.",
    volume = "326",
    pages = "463--485",
    year = "2011"
}

@article{Kopietz:2000bh,
    author = "Kopietz, Peter",
    title = "{Two loop beta function from the exact renormalization group}",
    eprint = "hep-th/0007128",
    archivePrefix = "arXiv",
    doi = "10.1016/S0550-3213(00)00680-5",
    journal = "Nucl. Phys. B",
    volume = "595",
    pages = "493--518",
    year = "2001"
}

@article{Codello:2013bra,
    author = "Codello, Alessandro and Demmel, Maximilian and Zanusso, Omar",
    title = "{Scheme dependence and universality in the functional renormalization group}",
    eprint = "1310.7625",
    archivePrefix = "arXiv",
    primaryClass = "hep-th",
    doi = "10.1103/PhysRevD.90.027701",
    journal = "Phys. Rev. D",
    volume = "90",
    number = "2",
    pages = "027701",
    year = "2014"
}

@book{Zinn-Justin:2002ecy,
    author = "Zinn-Justin, Jean",
    title = "{Quantum field theory and critical phenomena}",
    publisher = "Clarendon Press",
    address = "Oxford",
    edition = "4th",
    series = "Int. Ser. Monogr. Phys.",
    volume = "113",
    year = "2002"
}

@article{Callan:1970yg,
    author = "Callan, Jr., Curtis G.",
    title = "{Broken scale invariance in scalar field theory}",
    doi = "10.1103/PhysRevD.2.1541",
    journal = "Phys. Rev. D",
    volume = "2",
    pages = "1541--1547",
    year = "1970"
}

@article{Symanzik:1970rt,
    author = "Symanzik, K.",
    title = "{Small distance behavior in field theory and power counting}",
    doi = "10.1007/BF01649434",
    journal = "Commun. Math. Phys.",
    volume = "18",
    pages = "227--246",
    year = "1970"
}

@article{Alexandre:2000eg,
    author = "Alexandre, Jean and Polonyi, Janos",
    title = "{Functional Callan-Symanzik equation}",
    eprint = "hep-th/0010128",
    archivePrefix = "arXiv",
    doi = "10.1006/aphy.2000.6109",
    journal = "Annals Phys.",
    volume = "288",
    pages = "37--51",
    year = "2001"
}
\end{document}